\documentclass[aps,prx,reprint,twocolumn,floatfix,superscriptaddress,amsmath,amssymb,amsfonts,nofootinbib,balancelastpage]{revtex4-2}

\usepackage{url}
\usepackage{bm}
\usepackage{graphicx}
\usepackage{amsmath}
\usepackage{amstext}
\usepackage{amssymb}
\usepackage{amsfonts}
\usepackage{amsbsy}
\usepackage{verbatim}
\usepackage{color}
\usepackage[colorlinks=true, urlcolor=blue, linkcolor=blue, citecolor=blue, pdftex]{hyperref}
\usepackage{floatrow}
\usepackage{float}
\usepackage{tikz}
\usepackage{gensymb}
\usepackage{textcomp}
\usepackage[version=3]{mhchem}
\usepackage{afterpage}
\usepackage{pbox}
\usepackage{dsfont}
\usepackage{siunitx}
\usepackage{stackengine,wasysym,scalerel}
\usepackage{latexsym}
\usepackage{cancel}
\usepackage{comment}
\usepackage{array, multirow, bigdelim, makecell, booktabs}
\usepackage[misc]{ifsym}
\usepackage[capitalise]{cleveref}
\usepackage{sidecap}

\usepackage{color}
\usepackage{pifont}

\usepackage[normalem]{ulem}
\newcommand{\pdagger}{\phantom{\dagger}}

\definecolor{darkblue}{HTML}{004D6B}
\definecolor{darkred}{HTML}{8c1515}
\definecolor{darkgreen}{HTML}{006400}
\usepackage{hyperref}
\hypersetup{
	colorlinks=true,
	urlcolor=blue,
	citecolor=blue,
	linkcolor=blue,
	breaklinks
}

\allowdisplaybreaks

\usepackage{tabularray}
\usepackage{bm}
\usepackage{slashed}
\usepackage{enumitem}
\usepackage[table,dvipsnames]{xcolor}

\newcommand{\mbf}[1]{\mathbf{#1}}

\newcommand{\Tr}{\mathop{\mathrm{Tr}}}
\newcommand{\vi}{{\boldsymbol{i}}}
\newcommand{\vj}{{\boldsymbol{j}}}
\newcommand{\vecr}{{\boldsymbol{r}}}

\newcommand{\haty}{{\mbf{\hat{y}}}}
\newcommand{\hatx}{{\mbf{\hat{x}}}}

\newcommand{\X}{{\mathcal{X}}}

\definecolor{colone}{HTML}{fec4de}
\definecolor{coltwo}{HTML}{cbf0f6}
\definecolor{colthree}{HTML}{c7ece0}

\newcommand{\SO}{\mathrm{SO}}
\newcommand{\SU}{\mathrm{SU}}
\newcommand{\U}{\mathrm{U}}
\newcommand{\Z}{\mathbb{Z}}

\newcommand{\eqnref}[1]{Eq.~\eqref{#1}}
\newcommand{\figref}[1]{Fig.~\ref{#1}}
\newcommand{\tabref}[1]{Table~\ref{#1}}
\newcommand{\secref}[1]{Sec.~\ref{#1}}
\newcommand{\appref}[1]{Appendix~\ref{#1}}

\begin{document}
\title{Unified gauge‑theory description of quantum spin liquids\\ on square‑based frustrated lattices}
\author{Atanu Maity}
\thanks{These authors contributed equally.}
\affiliation{Institut f\"ur Theoretische Physik und Astrophysik and W\"urzburg-Dresden Cluster of Excellence ctd.qmat, Julius-Maximilians-Universit\"at W\"urzburg, Am Hubland, Campus S\"ud, W\"urzburg 97074, Germany}
\author{Andreas Feuerpfeil}
\thanks{These authors contributed equally.}
\affiliation{Institut f\"ur Theoretische Physik und Astrophysik and W\"urzburg-Dresden Cluster of Excellence ctd.qmat, Julius-Maximilians-Universit\"at W\"urzburg, Am Hubland, Campus S\"ud, W\"urzburg 97074, Germany}
\affiliation{Center for Computational Quantum Physics, Flatiron Institute, 162 5th Avenue, New York, NY 10010, USA}
\author{Ronny Thomale}
\affiliation{Institut f\"ur Theoretische Physik und Astrophysik and W\"urzburg-Dresden Cluster of Excellence ctd.qmat, Julius-Maximilians-Universit\"at W\"urzburg, Am Hubland, Campus S\"ud, W\"urzburg 97074, Germany}
\affiliation{Department of Physics, Indian Institute of Technology Madras, Chennai 600036, India}
\author{Subir~Sachdev}
\affiliation{Department of Physics, Harvard University, Cambridge MA 02138, USA}
\affiliation{Center for Computational Quantum Physics, Flatiron Institute, 162 5th Avenue, New York, NY 10010, USA}
\author{Yasir Iqbal}
\email{yiqbal@physics.iitm.ac.in}
\affiliation{Department of Physics, Indian Institute of Technology Madras, Chennai 600036, India}


\begin{abstract}
Quantum spin liquids are commonly thought to be highly sensitive to lattice geometry, symmetry, and microscopic exchange patterns, leading to a proliferation of seemingly distinct phases across frustrated magnets. Here, we provide a framework that unifies phases that appear distinct from the viewpoint of this intuition. We postulate that the spin-$\tfrac{1}{2}$ Heisenberg antiferromagnets on the square, Shastry--Sutherland, and checkerboard lattices can realize a single unified quantum phase: a gapless $\mathbb{Z}_2$ Dirac quantum spin liquid, despite their markedly different lattice symmetries. Using a systematic projective symmetry group analysis, we identify a checkerboard spin-liquid state that completes a closed set of adiabatically connected phases linking the well-established square-lattice and Shastry--Sutherland spin liquids. Crucially, we show that this lattice-level unification is mirrored exactly in the continuum description. In all three cases, the spin liquids descend from a common SU(2) $\pi$-flux parent state and are governed by the same gauge theory, QED$_3$ with two Dirac fermion flavors coupled to two adjoint Higgs fields. As a result, we postulate that the surrounding N\'eel and valence-bond-solid phases and their confinement transitions admit a unified interpretation within the framework of deconfined quantum criticality. More broadly, our results suggest that quantum spin liquids are most fundamentally classified not by lattice geometry or microscopic couplings, but by the emergent gauge theory and its Higgs structure. Distinct frustrated lattices can thus host the same quantum phase and exhibit the same confinement mechanisms, despite substantial differences in their microscopic symmetries.\\
\end{abstract}

\maketitle

\section{Introduction}

Quantum spin liquids (QSLs) represent one of the most striking manifestations of strong correlations in condensed matter systems, where local magnetic moments evade conventional symmetry breaking and instead organize into highly entangled quantum states with emergent gauge fields and fractionalized excitations~\cite{Balents-2010,Savary-2016,Zhou-2017}. While a wide variety of lattice geometries are now known to support spin-liquid behavior, a central unresolved question is whether these realizations correspond to fundamentally distinct phases of matter or whether they reflect different microscopic routes to the same underlying quantum state.

This question is particularly pressing for frustrated antiferromagnets whose lattices are closely related but differ in symmetry and connectivity. The spin-$\tfrac{1}{2}$ Heisenberg antiferromagnets on the square, Shastry--Sutherland, and checkerboard lattices provide a paradigmatic example. The square lattice is bipartite and maximally symmetric, the Shastry--Sutherland lattice~\cite{Shastry-1981} introduces explicit dimerization and reduced translational symmetry, and the checkerboard lattice interpolates between the two with competing diagonal bonds and a distinct wallpaper group~\cite{Fouet-2003,Bernier-2004,Starykh-2005}. From a microscopic viewpoint, these differences would normally be expected to produce qualitatively different quantum disordered phases and phase transitions.

Surprisingly, however, a growing body of numerical and analytical work suggests a very different picture: all three models appear to host gapless $\mathbb{Z}_2$ quantum spin liquids~\cite{Wen-2002} with Dirac fermionic excitations~\cite{Hu-2013,Maity-2024,Feuerpfeil-2026}, and these phases can be smoothly connected as lattice couplings are tuned. This raises a fundamental question: \emph{does this apparent continuity merely reflect coincidences of variational wavefunctions and finite-size numerics, or a genuine unification at the level of emergent gauge structure, quantum order, and their continuum field theories?}

In this work, we show that this connection is neither accidental nor superficial. Instead, we demonstrate that the square, Shastry--Sutherland, and checkerboard Heisenberg antiferromagnets realize a unified family of quantum spin liquids governed by the same emergent gauge structure and the same low-energy quantum field theory (QFT), despite their distinct lattice symmetries. The central result of this paper is that all three systems support a gapless $\mathbb{Z}_2$ Dirac spin liquid that is adiabatically connected across lattices and descends from a common $\SU(2)$ $\pi$-flux parent state~\cite{Affleck1988} via Higgs condensation.

Our unification operates at a level stronger than similarity of variational wave functions or matching excitation spectra. Employing a systematic projective symmetry group (PSG) analysis~\cite{Wen-2002}, we identify a checkerboard spin-liquid state that completes a triangle of adiabatically connected phases linking the well-established square-lattice and Shastry--Sutherland spin liquids. Here, ``adiabatic connectivity'' does not refer to a continuous symmetrization of a fixed mean-field {\it Ansatz}, which is, by construction, a discrete operation, but rather to the existence of a continuous interpolation within the space of symmetry-allowed mean-field Hamiltonians for which the PSG and invariant gauge group remain unchanged. Additional lattice symmetries impose constraints on the allowed parameters without altering the underlying PSG class. Consequently, these states possess identical invariant gauge groups and projective symmetry implementations, establishing their equivalence at the level of quantum order.

Crucially, we then show that this lattice-level connectivity is mirrored exactly in the continuum description. For all three lattices, the Higgs transitions from the SU(2) $\pi$-flux state through an intermediate staggered-flux U(1) spin liquid to the stable $\mathbb{Z}_2$ Dirac spin liquid are governed by the same gauge theory---QED$_3$ with two Dirac fermion flavors coupled to two adjoint Higgs fields. We classify all symmetry-allowed fermion bilinears and Higgs couplings up to leading gradient order and demonstrate that lattice-specific terms either do not alter the scaling dimensions or correspond only to less relevant operators. As a result, scaling dimensions, critical behavior, and phase structure coincide across all three models within the controlled large-$N_f$, $N_b$ expansion of the resulting QED$_3$--Higgs theory.

This unified field-theoretic structure naturally places these systems within the broader framework of deconfined quantum criticality~\cite{Senthil_2004,Senthil2004a,Sandvik2007,Kaul_2008}, where conventional N\'eel and valence-bond-solid phases emerge as confined descendants of a common parent gauge theory. In this perspective, lattice geometry controls which Higgs fields condense, but not the fundamental structure of the emergent theory itself. The ordered phases surrounding the spin liquid are therefore expected to be adiabatically connected as well~\cite{Bernier-2004,Starykh-2005}, as indicated by recent numerical studies~\cite{LiuESS-2024,Qian-2025}.

While frustrated magnets have long been classified primarily by lattice geometry, the present results show that emergent gauge structure provides a more fundamental classification axis. Distinct frustrated lattices can thus realize the same quantum phase and the same confinement mechanisms, even when their microscopic symmetries differ substantially.

The square–Shastry–Sutherland–checkerboard family thus provides a controlled setting where microscopic diversity does not yield distinct quantum orders but instead converges onto a single universal gauge-theoretic description. Our results open a path toward uncovering hidden unifications among frustrated magnets and suggest that many seemingly disparate spin-liquid candidates may ultimately correspond to different realizations of a small number of fundamental quantum field theories. This demonstrates that microscopic space-group symmetry alone does not uniquely determine the quantum order~\cite{Wen-2002,Lu-2011}, and that distinct lattice geometries can share a common emergent gauge structure.

\begin{figure}
    \centering
    \includegraphics[width=1.0\linewidth]{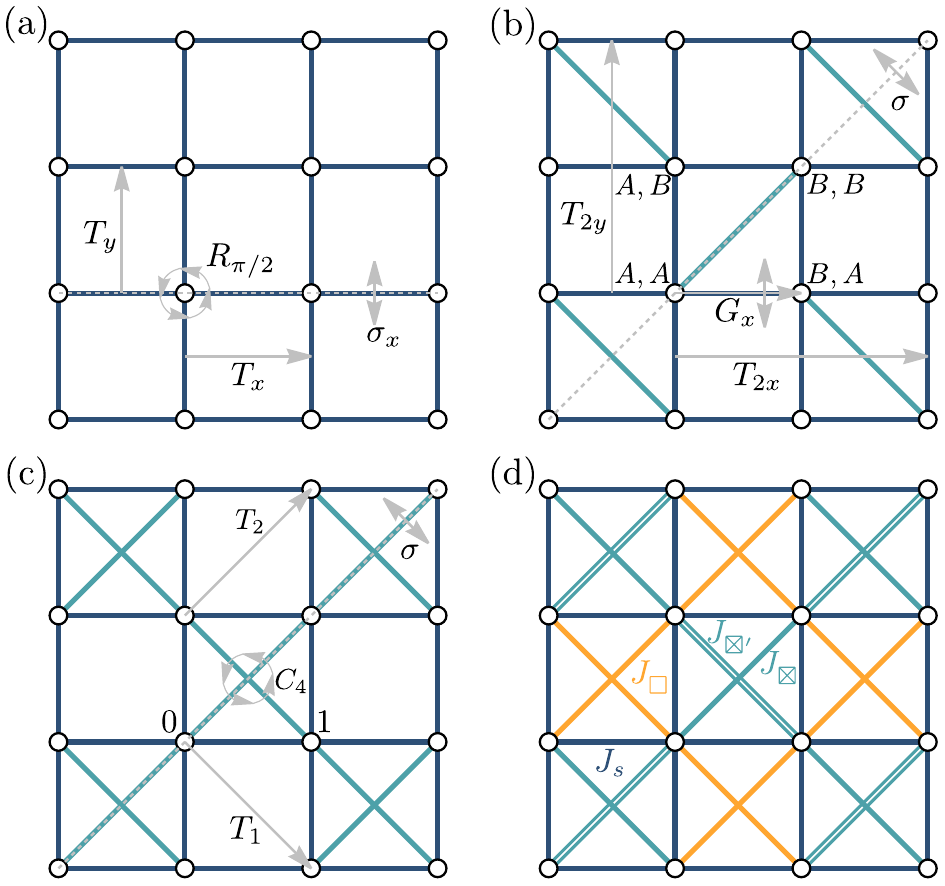}
    \caption{Lattice geometries and symmetry hierarchy of the square, Shastry--Sutherland (SS), and checkerboard models.
(a) Square lattice with minimal generators of the $p4mm$ wallpaper group.
(b) Shastry--Sutherland lattice ($p4g$), obtained by symmetry reduction via diagonal bond decoration.
(c) Checkerboard lattice ($p4gm$).
(d) Heisenberg exchange pattern at the lowest common symmetry level, corresponding to the SS geometry.
The limit $J_{\boxtimes'}=J_{\boxtimes}$ yields the checkerboard lattice, while $J_{\boxtimes'}=J_{\boxtimes}=J_{\square}$ yields the square lattice.}
    \label{fig:lattice}
\end{figure}

\begin{figure*}
    \centering    \includegraphics[width=1.0\linewidth]{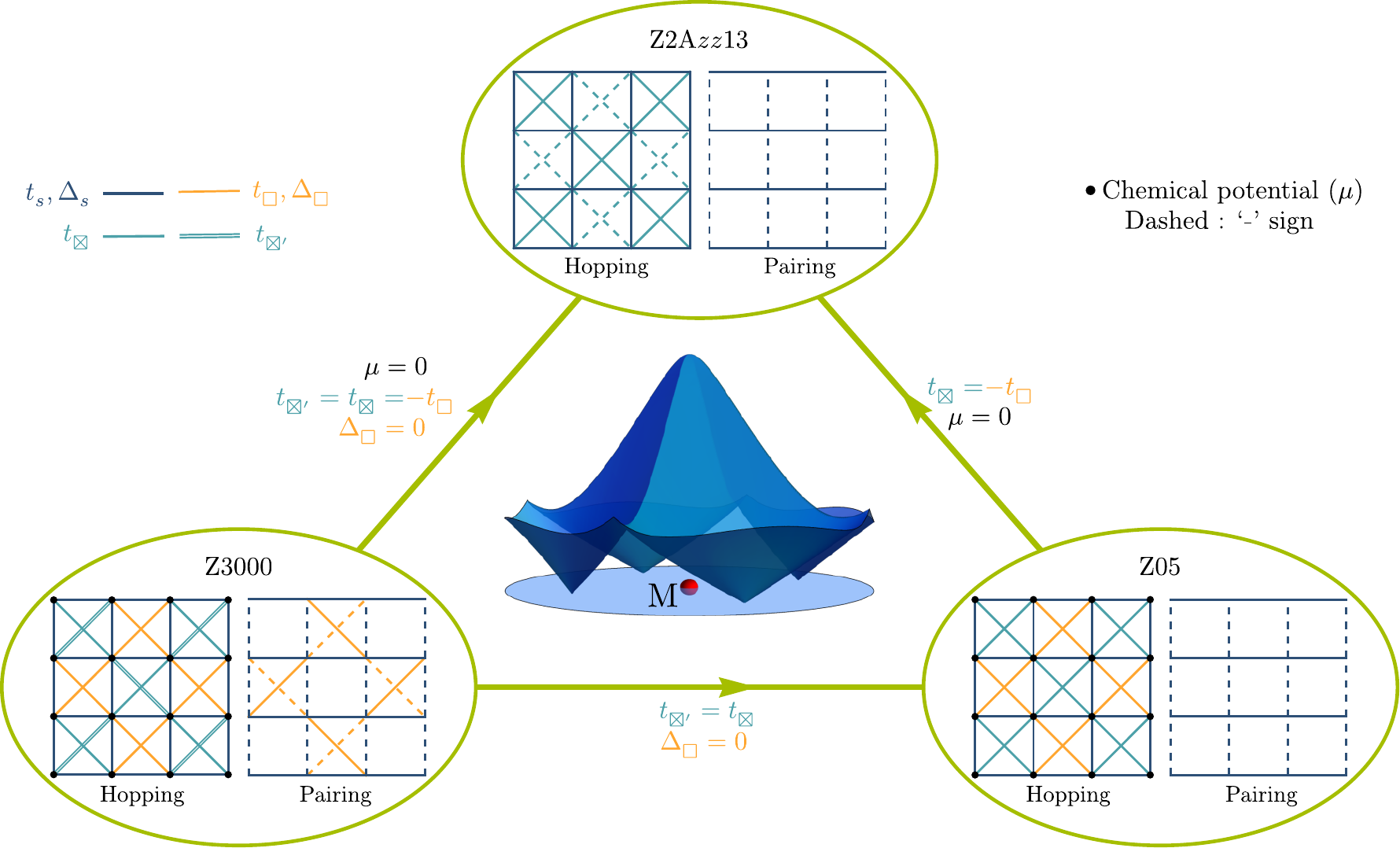}
    \caption{Triangle of adiabatically connected $\mathbb{Z}_2$ Dirac spin liquids on the square (Z2A$zz$13), Shastry--Sutherland ($\mathrm{Z3000}$), and checkerboard ($\mathrm{Z05}$) lattices. Continuous interpolation of symmetry-allowed hopping and pairing parameters preserves both the projective symmetry group (PSG) and the invariant gauge group (IGG), showing that all three states belong to the same class of quantum order. Representative Dirac dispersions are shown for selected parameter values along the interpolation. In all cases, the low-energy structure remains that of the same Dirac spin liquid.}
    \label{fig:z2_qsl_unification}
\end{figure*}

\emph{Summary of results.}
We show that the spin-$\tfrac{1}{2}$ Heisenberg antiferromagnets on the square, Shastry--Sutherland, and checkerboard lattices can realize the \emph{same} gapless $\mathbb{Z}_2$ Dirac spin liquid despite their distinct lattice geometries, and that these realizations are adiabatically connected as the lattice geometry and couplings are tuned within this family (Sec.~\ref{sec:adiabatic_connections}).

Our argument has two complementary parts. First, at the lattice level, we identify a fully symmetric checkerboard $\mathbb{Z}_2$ state (denoted $\mathrm{Z05}$) whose projective symmetry group matches those of the established square-lattice and Shastry--Sutherland candidates (Z2A$zz$13 and $\mathrm{Z3000}$ states, respectively), thereby completing a triangle of continuously connected spin liquids (Secs.~\ref{sec:adiabatic_connections}, \ref{sec:symmetry}, and \ref{sec:classification}).

Second, at the continuum level, we derive the low-energy gauge theory describing the Higgs route from a common $\mathrm{SU}(2)$ $\pi$-flux parent to a staggered-flux $\mathrm{U}(1)$ spin liquid and finally to the stable $\mathbb{Z}_2$ Dirac phase (Secs.~\ref{sec:continuum_theory}, \ref{sec:higgs_transitions}).

In all three cases, the Higgs transitions are governed by the same field theory---$\mathrm{QED}_3$ with two Dirac fermions coupled to two adjoint Higgs fields---and lattice-specific distinctions do not alter the scaling dimensions or critical behavior (Sec.~\ref{sec:qed_3}). Taken together, these results establish a unified gauge-theoretic framework for these models at both the level of quantum order (PSG/IGG) and continuum dynamics, situating their phase structures within a common paradigm of deconfined quantum criticality (Secs.~\ref{sec:dqcp} and \ref{sec:discussion}).

Figure~\ref{fig:z2_qsl_unification} summarizes the central result of the paper in a single view. The three corners correspond to the established square-lattice and Shastry--Sutherland $\mathbb{Z}_2$ Dirac spin liquids, Z2A$zz$13 and $\mathrm{Z3000}$, together with the checkerboard state $\mathrm{Z05}$ identified here. The edges represent continuous interpolations of symmetry-allowed hopping and pairing amplitudes that preserve both the projective symmetry group (PSG) and the invariant gauge group (IGG), thereby placing all three states in the same class of quantum order. The representative dispersions further illustrate that the low-energy Dirac structure remains unchanged along these interpolations. Sections~\ref{sec:adiabatic_connections} and \ref{sec:classification} establish the lattice-level part of this statement, while Sec.~\ref{sec:continuum} develops the corresponding continuum gauge-theory description.

\section{Unified $\mathbb{Z}_2$ Dirac spin liquid on square, Shastry--Sutherland and checkerboard lattices}
\label{sec:adiabatic_connections}
Recent work~\cite{Maity-2024} has demonstrated that QSL phases realized on closely related frustrated lattices, such as the square and Shastry–Sutherland (SS) geometries, can be continuously connected without encountering a phase transition. The checkerboard lattice naturally belongs to this same family, while exhibiting a symmetry that is lower (higher) than that of the square lattice (SS). This structural hierarchy suggests that a QSL phase, if stabilized on the checkerboard lattice, should be adiabatically connected to the QSL phases established on the SS and square lattices. Viewed together, these connections point to the existence of a triangular network of continuously connected spin-liquid phases spanning these three lattice geometries. 
By this ``triangle of spin liquids'' we mean the following: each lattice hosts a $\mathbb{Z}_2$ Dirac state whose projective symmetry group (PSG) allows the associated QSL mean-field parameters to be continuously deformed into that of the other two without closing the spinon gap or altering the invariant gauge group. 
Thus, the three states are adiabatically connected within the same PSG class.

To substantiate this triangle explicitly, we adopt the Abrikosov fermion framework [see \secref{sec:parton_classification}], which offers a transparent and widely used route to constructing mean-field descriptions of QSLs. Within this approach, the spin degrees of freedom are recast in terms of emergent spin-$1/2$ fermionic quasiparticles ($\hat{f}^{\pdagger}_{i\alpha}$ with $\alpha=\uparrow,\downarrow$), commonly referred to as {\it spinons}~\cite{Abrikosov-1965,Baskaran-1988}. This representation enables the formulation of an effective mean-field theory with quadratic Hamiltonians $\hat{H}_{\text{MF}}(u^{}_{ij},\mu)$ for quantum paramagnetic states in terms of a singlet link field $u^{}_{ij}$ composed of singlet hopping and pairing amplitudes, denoted by $t_{ij}=\sum_\alpha\langle\hat{f}^{\dagger}_{i\alpha}\hat{f}^{\pdagger}_{j\alpha}\rangle$ and $\Delta_{ij}=\epsilon^{\alpha\beta}\langle\hat{f}^{\pdagger}_{i\alpha}  \hat{f}_{j\beta}^{\pdagger}\rangle$, respectively, together with a chemical potential $\mu_i$ that enforces the physical Hilbert-space constraint. The resulting set of parameters $\{u^{}_{ij}(t_{ij},\Delta_{ij}),\mu_i\}$ defines a quantum spin-liquid {\it Ansatz}. In the following, we establish the proposed triangle of liquidity by introducing appropriate interpolating parameters between different lattice limits and by examining the nontrivial implementation of lattice symmetries within the PSG framework.

The {\it Ans\"atze} governing QSL phases which are already established to exist on square and SS are dubbed as `Z2A$zz$13' and `Z3000' in Ref.~\cite{Wen-2002,Hu-2013} and Ref.~\cite{Maity-2024} respectively. In this section, our aim is to identify the checkerboard QSL candidate that can be interpolated to Z2A$zz$13 (Z3000) in square lattice (SS). Our analysis establishes the symmetry and projective equivalence of the three $\mathbb{Z}_2$ Dirac states. The energetic realization of the proposed Z05 spin liquid is a separate dynamical question, whose numerical investigation will be presented in a forthcoming work by some of us. Our focus here is instead on demonstrating that, if realized, the three lattice geometries host symmetry-equivalent Z$_2$ Dirac spin liquids belonging to the same PSG class.

To do this, we first identify all $\mathbb{Z}_2$ {\it Ans\"atze} realizable up to next-nearest neighbors (NNN). Of these $\mathbb{Z}_2$ {\it Ans\"atze}, we identify a particular one, which we denote by Z05 [see Appendix~\ref{app:ansatz}] that can be adiabatically connected to the Z2A$zz$13 and Z3000 states. To make this connection explicit, we first write Z05 in a suitable gauge where all hopping and pairing parameters are real:
\begin{align}
&u^{\pdagger}_{(X,Y,0)(X,Y,1)}= t^{\pdagger}_s\tau^z+\Delta^{\pdagger}_s\tau^x\,,\nonumber\\
&u^{\pdagger}_{(X,Y,1)(X,Y+1,0)}=t^{\pdagger}_s\tau^z-\Delta^{\pdagger}_s\tau^x\,,\nonumber\\
&u^{\pdagger}_{(X,Y+1,0)(X-1,Y,1)}=t^{\pdagger}_s\tau^z+\Delta^{\pdagger}_s\tau^x\,,\nonumber\\
&u^{\pdagger}_{(X-1,Y,1)(X,Y,0)}=t^{\pdagger}_s\tau^z-\Delta^{\pdagger}_s\tau^x\,,\nonumber\\
&u^{\pdagger}_{(X,Y,0)(X,Y+1,0)}=u^{\pdagger}_{(X,Y,1)(X-1,Y,1)}=t^{\pdagger}_{\boxtimes}\,,\nonumber\\
&u^{\pdagger}_{(X,Y,0)(X+1,Y,0)}=u^{\pdagger}_{(X,Y,1)(X,Y-1,1)}=t^{\pdagger}_{\square}\,,\nonumber\\
&\mu_i=\mu\, .\label{eq:z05}
\end{align}

Here, we choose the coordinates as $\mathbf{r} \equiv (X,Y,s) \equiv X \mathbf{\hat{a}_{x\bar{y}}}+Y\mathbf{\hat{a}_{x{y}}}+sa\mathbf{\hat{x}}$ with $\mathbf{\hat{a}_{x\bar{y}}}=(\mathbf{\hat{x}}-\mathbf{\hat{y}})a$ and $\mathbf{\hat{a}_{x{y}}}=(\mathbf{\hat{x}}+\mathbf{\hat{y}})a$, $s=0,1$ denotes the sublattice index, and $a$ is the lattice spacing.
The sign structure of the Z05 {\it Ansatz} is shown in \figref{fig:z2_qsl_unification} along with that of the Z3000 and Z2A$zz$13 states~\footnote{This choice of gauge adopted for the Z2A$zz$13 {\it Ansatz} in Ref.~\cite{Wen-2002} is different from ours, the latter being expressed in terms of pairings \emph{only}.}. 

Since the SS has the lowest symmetry group within this lattice family, we begin the interpolation from its QSL {\it Ansatz}, namely the Z3000 state. The Z3000 {\it Ansatz} is composed of uniform hoppings $t^{\pdagger}_s$ (square edges), $t^{\pdagger}_\boxtimes$ (dimer bond diagonals), $t^{\pdagger}_{\boxtimes'}$ (diagonals perpendicular to the dimer bonds) and $t^{\pdagger}_\square$ (both diagonal bonds within the empty squares), along with a uniform chemical potential $\mu$. Real pairings include $\Delta^{\pdagger}_{s}$ with $d^{\pdagger}_{x^2-y^2}$ symmetry on the square edges and $\Delta^{\pdagger}_{\square}$ with $d^{\pdagger}_{xy}$ symmetry on the diagonals of the empty squares. With the {\it Ansatz} expressed in this gauge, the interpolation to the checkerboard Z05 {\it Ansatz} is manifest: it follows from tuning $t^{\pdagger}_{\boxtimes'}=t^{\pdagger}_{\boxtimes}$ and setting $\Delta^{\pdagger}_{\square}=0$. The further interpolation from Z05 to the square-lattice Z2A$zz$13 {\it Ansatz} is then achieved by additionally fixing $t^{\pdagger}_{\square}=-t^{\pdagger}_{\boxtimes}$ and $\mu=0$. 

However, the adiabatic connections among the aforementioned QSL states on the three lattices cannot be conclusively inferred from the continuous connectivity at the {\it Ans\"atze} level alone. This limitation arises because our analysis was restricted to NNN (diagonal) links only. Upon inclusion of further-neighbor links, the presumed interpolation may no longer hold. A conclusive establishment of such a connection requires an analysis at the level of PSGs, as it provides the fundamental characterization of the symmetry group, i.e., the quantum order of any {\it Ansatz}. We therefore establish the adiabatic connections between the Z05 and Z3000 {\it Ansatz}, as well as between the Z05 and Z2A$zz$13 states, complementing the prior PSG connection between Z3000 and Z2A$zz$13 established in Ref.~\cite{Maity-2024}. We emphasize that a QSL {\it Ansatz} transforms projectively under a symmetry element $\mathcal{O}$, i.e., it is $\mathcal{O}$-symmetric \textit{iff} it remains unchanged up to a local SU(2) gauge transformation $W(\mathbf{r})$ which defines the PSG~\cite{Wen-2002}.

From a methodological perspective, it is convenient to begin with the PSG of the lowest-symmetry lattice among the three, namely the Shastry--Sutherland lattice. Proceeding in this manner, we obtain the PSGs of the higher-symmetry square and checkerboard lattices by incorporating the projective actions of additional symmetry elements. 

We start with the {\it Ansatz} Z3000 which is invariant under the full symmetry group ($p4g$ wallper group $+$ $\mathcal{T}$) of the SS lattice. The $p4g$ wallpaper group is generated by translations $T_{2x}=T_x\circ T_x$, $T_{2y}=T_y\circ T_y$ ($T_x$ and $T_y$ denote the translations of the underlying square Bravais lattice), the glide $G_x=\sigma_x\circ T_x$ (where $\sigma_x$ generates reflection with respect to the $x$-axis), and reflection symmetry $\sigma$ about the line $x=y$ [\figref{fig:lattice}(b)]. 
The PSGs associated with Z3000 are given by 
\begin{align}
&W^{\pdagger}_{\mathcal{O}}(X,Y,s)=\tau^0,\;\mathcal{O}\in\{T_{2x},T_{2y},G_x\}\label{eq:z3000_psg_1}\\
&W^{\pdagger}_{\sigma}(X,Y,s)=\dot\iota\tau^z,\;W^{\pdagger}_{\mathcal{T}}(X,Y,s)=\dot\iota\tau^y\,.\label{eq:z3000_psg_2}
\end{align}
To establish the connectivity between Z3000 and Z05, we interpolate the wallpaper group $p4g$ to that of the checkerboard lattice, i.e., $p4gm$. This is achieved by incorporating $C_4=T_x \circ R_{\pi/2}$ which generates $\pi/2$ rotations about the center of the squares with diagonals. Here, $R_{\pi/2}$ denotes a $C_{4z}$ axis anchored at the origin), i.e., $\{p4g+C_4\}\equiv p4gm$. Now, we consider the following projective action of $C_4$
\begin{equation}
W^{\pdagger}_{C_4}(X,Y,s)=\dot\iota\tau^z\,.\label{eq:z3000_psg_3}    
\end{equation}
The PSGs given by Eqs.~\eqref{eq:z3000_psg_1}-\eqref{eq:z3000_psg_3} corresponds to the wallpaper group $p4gm$ of checkerboard lattice. Furthermore, these PSGs can be re-expressed in terms of a minimal set of generators [illustrated in Fig.~\ref{fig:lattice}(c) and discussed in Sec.~\ref{sec:symmetry}] by using the projective implementations of the algebraic relations $T_1=G_x\circ\sigma\circ C_4$ and $T_2=\sigma\circ C_4\circ G_x$ as follows
\begin{align}
&W^{\pdagger}_{T_1}(X,Y,s)=W^{\pdagger}_{T_2}(X,Y,s)=\tau^0,\label{eq:z05_psg_1}\\
&W^{\pdagger}_{\sigma}(X,Y,s)=W^{\pdagger}_{C_4}(X,Y,s)=\dot\iota\tau^z,\label{eq:z05_psg_2}\\
&W^{\pdagger}_{\mathcal{T}}(X,Y,s)=\dot\iota\tau^y\,,\label{eq:z05_psg_3}
\end{align}
which then yields the PSG for the checkerboard QSL candidate denoted as Z05 in the gauge choice of Eq.~\eqref{eq:z05}. This rigorously establishes the fact that the SS Z3000 and checkerboard Z05 QSLs are continuously connected at the level of projective symmetries.

Similarly, to establish the continuous connection between Z05 and Z2A$zz$13, we need to interpolate between the checkerboard wallpaper group $p4gm$ and $p4mm$ (the wallpaper group of square lattice) by including an additional symmetry element $\sigma_x$, i.e., $\{p4gm+\sigma_x\}\equiv p4mm$. Now, we consider the following projective action of $\sigma_x$:
\begin{equation}
W^{\pdagger}_{\sigma_x}(X,Y,s)=(-1)^{s}\dot\iota\tau^y\,.\label{eq:z05_psg_4}    
\end{equation}
 In terms of the symmetry group ($\mathcal{O}\in\{T_x,T_y,\sigma_x,\sigma_y,\sigma,\mathcal{T}\}$)~\footnote{$T_x$ and $T_y$ are the square lattice translations $a\mathbf{\hat{x}}$ and $a\mathbf{\hat{y}}$ respectively. $\sigma_y$ denotes reflection with respect to $y$ axis. In Ref.~\cite{Wen-2002} $\sigma_x$, $\sigma_y$ and $\sigma$ are denoted by $P_y$, $P_x$ and $P_{xy}$, respectively.} considered in Ref.~\cite{Wen-2002}, the PSG given by Eqs.~\eqref{eq:z05_psg_1}-\eqref{eq:z05_psg_4} can be re-expressed in the square lattice notation by using the projective implementations of the relations $T_x=G_x\circ\sigma_x$, $T_y=\sigma\circ  T_x\circ \sigma$ and $\sigma_y=\sigma\circ\sigma_x\circ\sigma$ in the following manner:
\begin{align}
&W^{\pdagger}_{\mathcal{O}}(i_x,i_y)=(-1)^{i_x+i_y}\dot\iota\tau^y,\,\mathcal{O}\in\{T_x,T_y,\sigma_x,\sigma_y\},\label{eq:z2azz13_psg_1}\\
&W^{\pdagger}_{\sigma}(i_x,i_y)=\dot\iota\tau^z,\;W^{\pdagger}_{\mathcal{T}}(i_x,i_y)=\dot\iota\tau^y\,\label{eq::z2azz13_psg_2}\,.
\end{align}
As expected, the PSG given by Eqs.~\eqref{eq:z05_psg_1}-\eqref{eq:z05_psg_4} (equivalently, Eqs.~\eqref{eq:z2azz13_psg_1} and~\eqref{eq::z2azz13_psg_2}) corresponds to the Z2A$zz$13 {\it Ansatz} in the gauge form illustrated in Fig.~\ref{fig:z2_qsl_unification}. 

Therefore, based on the above analysis, we infer that the trio of QSL candidates—Z2A$zz$13 on the square lattice, Z05 on the checkerboard lattice, and Z3000 on the Shastry–Sutherland lattice—forms a triangle of spin-liquid phases that can be continuously interpolated among one another. This trio is obtained from the same parent U(1) staggered ($\varphi,-\varphi$) flux  {\it Ansatz}~\cite{Lee2006} which further descends from the parent SU(2) $\pi$-flux {\it Ansatz} on the square lattice~\cite{Affleck1988}. 

\section{Unified continuum field theory and Higgs transitions}
\label{sec:continuum}
Having shown that the three $\mathbb{Z}_2$ Dirac spin liquids on the square, Shastry--Sutherland, and checkerboard lattices share identical quantum order, a crucial question remains: \emph{Why do these models exhibit similar phase diagrams, featuring transitions from a N\'eel ordered state through a $\mathbb{Z}_2$ Dirac spin liquid to VBS order?}

To address this, we demonstrate that their emergent gauge theories are identical up to additional terms allowed by the lower-symmetry lattices, which do not modify scaling dimensions or critical exponents. 
We propose deconfined quantum criticality as the mechanism producing adjacent N\'eel and VBS phases with shared confinement physics---strongly suggesting that not only the spin liquids, but their \emph{entire phase diagrams} can be adiabatically connected across these Heisenberg antiferromagnets.

\subsection[Common \texorpdfstring{$\SU(2)$}{SU(2)} \texorpdfstring{$\pi$}{pi}-flux parent and \texorpdfstring{$\U(1)$}{U(1)} staggered-flux state]{Common \texorpdfstring{$\mbf{\SU(2)}$}{SU(2)} \texorpdfstring{$\mbf{\pi}$}{pi}-flux parent and \texorpdfstring{$\mbf{\U(1)}$}{U(1)} staggered-flux state}\label{sec:continuum_theory}

Our result---that these models share the same QFT---stems from their identical projective symmetry groups (Sec.~\ref{sec:adiabatic_connections}) and common descent from the $\U(1)$ staggered-flux state~\cite{Lee2006} and ultimately the $\SU(2)$ $\pi$-flux state~\cite{Affleck1988,Wen-2002}. 
Thus, the low-energy modes and Higgs fields driving the $\U(1)$ and $\mathbb{Z}_2$ spin-liquid transitions share identical symmetries across all three lattices. 
This dictates the symmetry-allowed continuum terms, yielding QED$_3$ with two Dirac fermions coupled to two adjoint Higgs scalars as the universal gauge theory. 

The reduced symmetries of the checkerboard and Shastry--Sutherland lattices introduce additional terms for the Yukawa couplings between fermions and bosons, which preserve the scaling dimensions or are less relevant. Starting from this unified QFT, Higgs condensation drives deconfined phase transitions between $\SU(2)$, $\U(1)$, and $\mathbb{Z}_2$ spin liquids in all three models. 
While microscopic details differ between lattices, the confinement of the $\U(1)$ staggered-flux state via monopole proliferation obeys identical symmetry constraints and quantum numbers in all three models. Recent Monte Carlo studies~\cite{Shao2016Science,Nahum2015PRX,Zhou_2024,Takahashi2024}, alongside our field-theoretic analysis, indicate that the $\SU(2)$ $\pi$-flux is also unstable and likely confines to N\'eel or VBS orders. Consequently, these confinement transitions---and the resulting ordered phases---belong to the same universality class and should be adiabatically connected, a numerically testable prediction.

To make explicit the transition from $\SU(2)$ to $\U(1)$ and $\mathbb{Z}_2$ spin liquids and to connect with prior field-theoretic studies on the square~\cite{Shackleton2021,Shackleton:2022zzm} and Shastry--Sutherland~\cite{Feuerpfeil-2026} lattices we adopt the gauge of Ref.~\cite{Feuerpfeil-2026} [see \secref{sec:summary_classification}; full details in \appref{app:z2_ansatz_qft}]. 
To render the {\it Ansatz} translationally invariant in this gauge, we enlarge the unit cell to $2\times 2$, labeled by sublattice indices $m_x,m_y=A,B$, with Pauli matrices $\rho^a,\kappa^a$ acting on $m_x,m_y$ respectively [\figref{fig:lattice}(b)].

The $\SU(2)$ $\pi$-flux {\it Ansatz} is then given by $t^z_{\mathbf{i}\mathbf{j}}=\Delta_{\mathbf{i}\mathbf{j}}^{x/y}=0$ and
\begin{equation}
\begin{split}
    t^0_{\vi\vj}&=-t^0_{\vj\vi}\,, \quad t^0_{\vi,\vi+\hatx}=t, \quad t^0_{\vi,\vi+\haty}=(-1)^{i_x}t \,.
\end{split}
\end{equation}
This matches Refs.~\cite{Shackleton2021,Feuerpfeil-2026} and its mean-field Hamiltonian \eqnref{eq:mf_ham} has two Dirac points at $(0,0)$ and $(0,\pi)$, to which we associate a valley degree of freedom $v=0,1$ with Pauli matrices $\mu^a$. We choose the low-energy degrees of freedom as~\cite{Feuerpfeil-2026}
\begin{equation}\label{eq:low_energy_modes}
    \mathcal{F}_{\vecr}\sim\rho^x\X_{\vecr,0}+\kappa^z\X_{\vecr,1}\,.
\end{equation}
Expanding \eqnref{eq:mf_ham} around these yields the relativistic Dirac Lagrangian
\begin{equation}
\begin{split}
    \mathcal{L}_{\mathrm{MF}}&=i\Tr[\bar{\X}\gamma^\mu \partial_\mu\X]\,,
\end{split}
\end{equation}
where we absorbed $t$ into the speed of light and used the Lorentz signature $(+,-,-)$ with $\bar{\X}=\X^\dag\gamma^0$ and $\gamma^0=\rho^y,\gamma^x=i\rho^z,\gamma^y=i\rho^x$. This theory exhibits an emergent $\SO(5)$ symmetry from combined $\SU(2)$ spin $\times$ $\SU(2)$ valley symmetry~\cite{Tanaka2005,Ran2006,Wang_2017,Feuerpfeil-2026}.

The $\U(1)$ staggered-flux state emerges naturally as a Higgs descendant of this $\SU(2)$ $\pi$-flux parent by tuning the flux $\theta=\pi/2+\delta\theta$; we introduce it with the corresponding Higgs fields in the next section.

\subsection{Higgs transitions to the \texorpdfstring{$\mathbb{Z}_2$}{Z2} Dirac spin liquid}\label{sec:higgs_transitions}

At the level of the continuum description, the shared projective symmetry structure fixes the form of the low-energy effective theory. The spinons couple to Higgs fields transforming under the adjoint representation of the emergent gauge group; their condensation breaks the IGG from $\SU(2)$ to $\U(1)$ or $\mathbb{Z}_2$, driving a cascade of phase transitions from the $\SU(2)$ $\pi$-flux state to the $\U(1)$ staggered-flux state and Z05 state~\cite{Wen-2002}. Symmetry constrains the Lagrangian on the square lattice~\cite{Shackleton2021,Shackleton:2022zzm,Feuerpfeil-2026} to
\begin{equation}
\begin{split} \label{eq:full_lagrangian}
    \mathcal{L}&=i\Tr[\bar{\X}\gamma^\mu \partial_\mu \X]+\Phi_1^a\Tr[\tau^a\bar{\X}\gamma^x\mu^z \X]\\
    &+\Phi_2^a\Tr[\tau^a\bar{\X}\gamma^y\mu^x \X]\\
    &+\Phi_3^a\Tr[\tau^a \bar{\X}\mu^y(\gamma^yi\partial_x+\gamma^xi\partial_y)\X]+V(\Phi)\,,\\
\end{split}
\end{equation}
where $\Phi_{1,2,3}$ are the adjoint Higgs fields and $V(\Phi)$ is the most general symmetry-allowed Higgs potential (obtained by integrating out high-energy spinons). The microscopic origin of these Higgs fields and pairing channels is discussed in \appref{app:sym_couplings}. 

Crucially, this constitutes the entire effective Lagrangian for the checkerboard lattice as well. However, on the checkerboard lattice, additional terms are strictly symmetry-allowed and take the form
\begin{equation}
\begin{split}\label{eq:symm_allowed_terms}
    \mathcal{L}_{\mathrm{check.}} &\propto \Phi_3^a\Tr[\tau^a\bar{\X}(i\partial_0+\gamma^x\mu^yi\partial_y+\gamma^y\mu^yi\partial_x)\X]\\
    +\Phi_1^a&\Tr[\tau^a\bar{\X}\gamma^x\mu^x\X]+\Phi_2^a\Tr[\tau^a\bar{\X}\gamma^y\mu^z\X]\,,
\end{split}
\end{equation}
which are classified in \appref{app:sym_checkerboard}. In the specific lattice realization of the Z05 state truncated to next-nearest-neighbor hoppings and pairings, these terms do not appear. Because they are formally allowed by symmetry, they can naturally arise from higher-order microscopic couplings or be generated dynamically. 

On the Shastry--Sutherland lattice, we additionally get~\cite{Feuerpfeil-2026}
\begin{equation}
\begin{split}\label{eq:symm_allowed_terms_SS}
    \mathcal{L}_{\mathrm{SS}} \propto &\Phi_1^a\Tr[\tau^a\bar{\X}(\mu^y\gamma ^0i\partial_x+\mu^y\gamma^xi\partial_0+i\partial_y)\X]\\
    -&\Phi_2^a\Tr[\tau^a\bar{\X}(\mu^y\gamma^0i\partial_y+\mu^y\gamma^yi\partial_0+i\partial_x)\X]\,.
\end{split}
\end{equation}
Thus, up to the less relevant gradient terms, the checkerboard and Shastry--Sutherland lattices share identical continuum theories.

\begin{figure}
    \centering
    \includegraphics[width=1.0\linewidth]{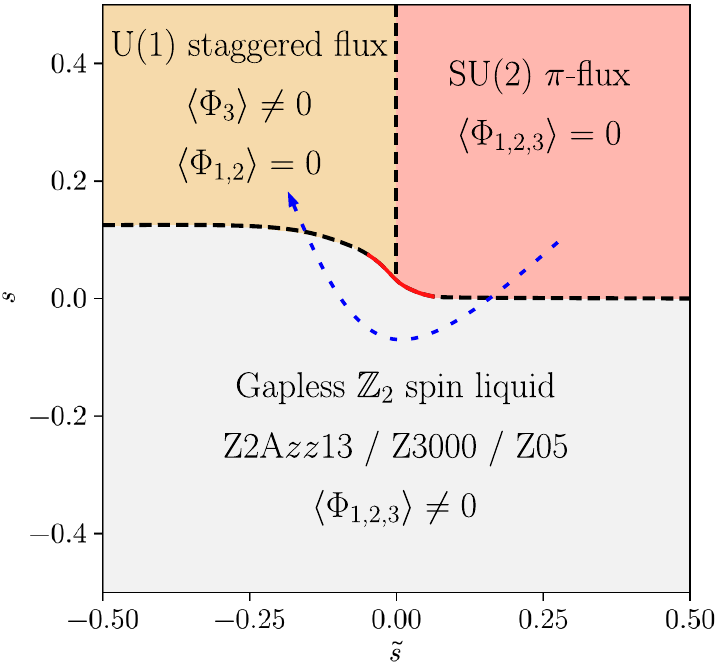}
    \caption{Mean-field phase diagram of the Higgs potential \eqnref{eq:Higgs_potential} for the square, Shastry--Sutherland, and checkerboard lattices. Dashed (solid red) lines indicate second (first) order transitions. We choose the mean-field parameters to be $w=u=1$, $v_2=-1$, $\tilde{u}=0.75$, and $v_4=0.5$ and employ the {\it Ansatz} $\Phi_1^a = c_1 \delta_{a,x}$, $\Phi_2^a = c_1 \delta_{a,y}$, and $\Phi_3^a = c_2 \delta_{a,z}$. Independent of the stable, gapless $\Z_2$ spin liquid---Z2A$zz$13 (square)~\cite{Wen-2002,Senthil-2000}, Z3000 (Shastry--Sutherland) or Z05 (checkerboard)---the phase diagram is qualitatively identical. We assume the $\SU(2)$ $\pi$-flux state confines to a N\'eel state whilst the $\U(1)$ staggered flux state confines to a columnar or plaquette VBS state, consistent with earlier checkerboard studies finding plaquette, staggered spin-Peierls, and crossed-dimer valence-bond orders in different regimes~\cite{Bernier-2004,Starykh-2005,Bishop-2012}. If confinements are reversed, the arrow direction (increasing $J_2$ for square, $J_d$ for Shastry--Sutherland, $J_\boxtimes$ for checkerboard) reverses. As the confinement mechanisms are the same across lattices, we postulate that also the ordered phases are adiabatically connected between the models.}
    \label{fig:Higgs_mean_field}
\end{figure}

The Higgs potential $V(\Phi)$ for the checkerboard lattice is deduced by constructing all gauge-invariant and symmetry-allowed terms up to quartic order:
\begin{equation}
\begin{split}\label{eq:Higgs_potential}
    V(\Phi)&=s(\Phi_1^a\Phi_1^a+\Phi_2^a\Phi_2^a)+\tilde{s}\Phi_3^a\Phi_3^a+w\epsilon_{abc}\Phi_1^a\Phi_2^b\Phi_3^c\\
    &+u(\Phi_1^a\Phi_1^a+\Phi_2^a\Phi_2^a)^2+\tilde{u}(\Phi_3^a\Phi_3^a)^2 +v_1 (\Phi_1^a\Phi_2^a)^2\\
    &+v_2 (\Phi_1^a\Phi_1^a)(\Phi_2^b\Phi_2^b)+ v_3[(\Phi_1^a\Phi_3^a)^2+(\Phi_2^a\Phi_3^a)^2] \\
    &+v_4(\Phi_1^a\Phi_1^a+\Phi_2^a\Phi_2^a)(\Phi_3^b\Phi_3^b)\,,
\end{split}
\end{equation}
where $\epsilon_{abc}$ is the Levi-Civita symbol. These terms are identical to those of the square and Shastry--Sutherland lattices and lead to a mean-field phase diagram as shown in \figref{fig:Higgs_mean_field}~\cite{Shackleton2021,Feuerpfeil-2026}.

The identical Higgs potentials to quartic order---and in fact to all orders---are no coincidence. 
As proven in Ref.~\cite{Feuerpfeil-2026}, any lattice whose wallpaper group contains the Shastry--Sutherland symmetries (p4g) as a subgroup and is contained within the square lattice symmetries (p4mm) yields identical symmetry-allowed Higgs potential terms for Higgs fields transforming as specified in \tabref{tab:higgs_transformations} under the square lattice symmetries. 

Since the Z05 state on checkerboard connects via PSG and the low-energy Dirac structure of the shared $\SU(2)$ $\pi$-flux state to the $\mathbb{Z}_2$ spin liquids on square (Z2A$zz$13) and Shastry--Sutherland (Z3000), the governing Higgs fields share identical symmetry properties. 
Thus, the Higgs potential is structurally identical, yielding qualitatively similar mean-field phase diagrams with matching phases and transitions. 
This constitutes a rare example where distinct microscopic symmetry groups yield identical symmetry-allowed continuum physics to arbitrary order.

\subsection[QED\texorpdfstring{$_3$}{3} description]{QED\texorpdfstring{$_\mbf{3}$}{3} description}\label{sec:qed_3}

We now analyze the quantum field theory corresponding to \eqnref{eq:full_lagrangian}.

It proves convenient to recast the theory in terms of $N_f=2$ flavors of complex Dirac fermions. These relate to the Majorana spinons via
\begin{equation}\label{eq:dirac_fermions}
    \psi_{a,m_x,v} = i \tau_{ab}^y[\X_{m_x,v}]_{1,b}\,.
\end{equation}
Introducing the $\SU(2)$ gauge field $A_\mu^\alpha$ via minimal coupling, and extending the two Higgs fields $\Phi_{1/2,\alpha}^a$ for a controlled large-$N_{f,b}$ expansion, \eqnref{eq:full_lagrangian} becomes
\begin{equation}
\begin{split}\label{eq:continuum_dirac_theory}
    \mathcal{L}&=i\bar{\psi}_v\gamma^\mu(\partial_\mu-iA_\mu^a\tau^a)\psi_v\\\
    &+\frac{1}{2g}\sum_{s=1,2}(\partial_s\Phi_{s\alpha}^a-2\epsilon_{abc}A_s^b\Phi_{s\alpha}^c)^2\\
    &+y\sum_{\alpha}(\Phi_{1\alpha}^a\bar{\psi}_v\mu^z\gamma^x\tau^a\psi_v+\Phi_{2\alpha}^a\bar{\psi}_v\mu^x\gamma^y\tau^a\psi_v)\,.
\end{split}
\end{equation}

This theory, known as QED$_3$ with two fundamental Dirac fermions and two adjoint Higgs scalars, matches those found for square~\cite{Shackleton2021,Shackleton:2022zzm} and Shastry--Sutherland~\cite{Feuerpfeil-2026} lattices---confirming all three models share identical low-energy physics. The lattice-specific terms Eqs.~\eqref{eq:symm_allowed_terms}--\eqref{eq:symm_allowed_terms_SS} yield additional fermion bilinears coupling to the Higgs fields: as the gradient terms are less relevant, they can be ignored. The leading zeroth-order bilinears alter the form of the Yukawa coupling and are given by
\begin{equation}
    y\sum_\alpha(\Phi_{1\alpha}^a\bar{\psi}\mu^x\gamma^x\tau^a\psi + \Phi_{2\alpha}^a\bar{\psi}\mu^z\gamma^y\tau^a\psi)\,.
\end{equation}

Importantly, as shown in Ref.~\cite{Feuerpfeil-2026}, the inclusion of these terms does not alter the vertex corrections of the Yukawa coupling. They differ from the Yukawa interaction in \eqnref{eq:continuum_dirac_theory} solely by the insertion of different valley matrices $\mu$. Since these matrices commute with the Dirac $\gamma$ matrices, the algebraic structure of the relevant Feynman diagrams remains unaffected, leaving the vertex corrections unchanged. Overall, the scaling dimension of the Yukawa coupling in the large-$N_{f,b}$ expansion is given by~\cite{Feuerpfeil-2026}
\begin{equation}
    \dim[y]=\frac{1}{2} - \frac{2}{9N_b\pi^2}+\frac{64}{3(N_f+8N_b)\pi^2}\,.
\end{equation}
For the physically relevant $N_f=2$ and $N_b=2$, the Yukawa coupling has dimension $\dim[y]\approx 0.609$ and is thus weakly relevant across all lattices.

All critical exponents and scaling dimensions thus match Ref.~\cite{Feuerpfeil-2026} for a weakly relevant Yukawa. Under the assumption of strong Yukawa coupling, one recovers the Lorentz-breaking theory of Refs.~\cite{Shackleton2021,Shackleton:2022zzm}.

\subsection{Relation to deconfined quantum criticality}
\label{sec:dqcp}
As established in the Secs.~\ref{sec:higgs_transitions} and~\ref{sec:qed_3}, both the QFT and Higgs potential structure--- including its mean-field phase diagram---are identical across all three models. This unifies not only the stable $\mathbb{Z}_2$ Dirac spin liquids, but also adjacent N\'eel/VBS ordered phases via deconfined quantum criticality~\cite{Senthil_2004}, as shown in Fig.~\ref{fig:Higgs_mean_field}.

The $\SU(2)$ $\pi$-flux state ($\langle\Phi_{1,2,3}^a\rangle=0$) retains $\SU(2)$ gauge and emergent $\SO(5)$ symmetries~\cite{Feuerpfeil-2026}, and is believed to confine to N\'eel or VBS order~\cite{Wang_2017,WangPRL2021}. Our analysis shows a lattice-independent weakly relevant Yukawa coupling, which breaks Lorentz-invariance and ultimately renders the theory unstable in the IR, consistent with the numerical observations of pseudo- or multi-criticality~\cite{Nahum2015,Wang_2017,WangPRL2021,Zhou_2024,Chen_2024,Meng24,Takahashi2024}. In the large-$y$ limit, it breaks $\SO(5)$ symmetry and favors N\'eel confinement in the large-$N_f$ expansion~\cite{Shackleton2021}.

The $\U(1)$ staggered-flux state ($\langle\Phi_3^a\rangle\propto\delta\theta\delta_{a,z}$) hosts a trivial monopole~\cite{Alicea-2008,Song2019} on the square lattice and thus also on the lower-symmetry checkerboard and Shastry--Sutherland lattices, which is expected to lead to confinement to the N\'eel or VBS phase across all lattices~\cite{SongPRX2020,Shackleton2021}.

The $\mathrm{Z05}$ state ($\langle\Phi_{1,2}^a\rangle\neq0$) has a single tuning parameter $s$ to condense the fields due to the symmetry transformations between the two Higgs fields [see \eqnref{eq:Higgs_potential}]. This gapless spin liquid is stable in 2D~\cite{Wen-2002} and connected through its PSG to the Z3000 and Z2A$zz$13 spin liquids. The cubic $w$-term induces $\langle \Phi_3^a\rangle\propto w\epsilon_{abc}\langle \Phi_{1}^b\rangle\langle \Phi_{2}^c\rangle$ and predicts a first-order transition near the triple point~\cite{Sachdev_2002}, which could reconcile deconfined quantum criticality predicting a continuous phase transition with the numerical observations of a weakly first-order transition~\cite{Nahum2015,Wang_2017,WangPRL2021,Zhou_2024,Chen_2024,Meng24,Takahashi2024,Shao2016Science,Zhou_2024}.

Shared PSGs, parent states, QED$_3$ theories, and confinement mechanisms thus provide strong evidence of full phase diagram connectivity---recently supported numerically~\cite{LiuESS-2024,Qian-2025}. This positions deconfined quantum criticality and QED$_3$ as a unifying framework for magnetically ordered phases across substantially different microscopic symmetries.

\section{Methodology}
\label{sec:parton_classification}

\subsection{Fermionic parton construction}
\label{sec:parton_mft}

We briefly review the Abrikosov-fermion construction for spin models~\cite{Abrikosov-1965,Wen-1991,Wen-2002}. In this approach, the spin-$\tfrac12$ operator at site $i$ is represented in terms of fermionic spinons \(f_{i\alpha}\) (\(\alpha=\uparrow,\downarrow\)) as
\begin{equation}
\hat S_i^\gamma
=
\frac12
\sum_{\alpha\beta}
\hat f^\dagger_{i\alpha}\,
\tau^\gamma_{\alpha\beta}\,
\hat f_{i\beta}\,,
\qquad
\gamma=x,y,z\,,
\end{equation}
where \(\tau^\gamma\) are Pauli matrices. This representation enlarges the Hilbert space, so the physical spin sector must be recovered by imposing the single-occupancy constraint
\begin{equation}
\sum_\alpha \hat f^\dagger_{i\alpha}\hat f_{i\alpha}=1,\,\hat f_{i\alpha}\hat f_{i\beta}\epsilon^{\alpha\beta}=0
\qquad
\text{on every site } i\,.
\end{equation}

The second constraint follows the first. We assemble spinons into the \(2\times2\) matrix
\begin{equation}
\hat{\mathcal F}_i =
\begin{pmatrix}
\hat f_{i\uparrow} & \hat f^\dagger_{i\downarrow} \\
\hat f_{i\downarrow} & -\hat f^\dagger_{i\uparrow}
\end{pmatrix}\,,
\end{equation}
yielding the compact spin representation
\begin{equation}
\hat S_i^\gamma
=
\frac14
\mathrm{Tr}
\!\left(
\hat{\mathcal F}_i^\dagger
\tau^\gamma
\hat{\mathcal F}_i
\right)\,.
\end{equation}
In this form the local SU(2) gauge redundancy is manifest:
\begin{equation}
\hat{\mathcal F}_i \rightarrow \hat{\mathcal F}_i W_i\,,
\qquad
W_i\in \mathrm{SU}(2)\,,
\end{equation}
leaving physical spin invariant. By contrast, left multiplication corresponds to physical spin rotations.

For spin-liquid mean-field states, one introduces singlet link fields on bonds \(\langle ij\rangle\),
\begin{equation}
u_{ij}
=
\bigl\langle
\hat{\mathcal F}_i^\dagger \hat{\mathcal F}_j
\bigr\rangle\,,
\end{equation}
which can be decomposed into singlet hopping and pairing amplitudes,
\begin{equation}
u_{ij}
=
\begin{pmatrix}
t_{ij} & -\Delta_{ij}^\ast \\
-\Delta_{ij} & -t_{ij}^\ast
\end{pmatrix}\,.
\end{equation}
Here, \(t_{ij}\) denotes the spinon hopping channel and \(\Delta_{ij}\) the singlet pairing channel.  A QSL {\it Ansatz} is specified by $\{u_{ij}\}$ and Lagrange multipliers enforcing the single-occupancy constraint.

The singlet mean-field Hamiltonian is
\begin{align}
\hat H_{\mathrm{MF}}
&=
\frac{3}{8}\sum_{\langle ij\rangle} J_{ij}
\left[
\frac12 \mathrm{Tr}(u_{ij}^\dagger u_{ij})
-
\mathrm{Tr}
\bigl(
\hat{\mathcal F}_i u_{ij}\hat{\mathcal F}_j^\dagger
+\mathrm{H.c.}
\bigr)
\right]
\notag\\
&\quad
+
\sum_i\sum_{\gamma} a_{\gamma,i}\,
\mathrm{Tr}
\!\left(
\hat{\mathcal F}_i \tau^\gamma \hat{\mathcal F}_i^\dagger
\right)\,,
\label{eq:mf_ham}
\end{align}
where $a_{\gamma,i}$ enforces the constraint on average.

\subsection{Projective symmetry group framework}
\label{sec:psg}

The mean-field Hamiltonian in Eq.~\eqref{eq:mf_ham} is invariant under local SU(2) gauge transformations. In matrix form, these act as
\begin{align}
\hat{\mathcal F}_i &\rightarrow \hat{\mathcal F}_i W_i,
\notag\\
u_{ij} &\rightarrow W_i^\dagger u_{ij} W_j,
\notag\\
a_{\gamma,i}\tau^\gamma &\rightarrow W_i^\dagger (a_{\gamma,i}\tau^\gamma) W_i,
\end{align}
with \(W_i\in \mathrm{SU}(2)\). As a result, two {\it Ans\"atze} related by such a local gauge transformation correspond to the same physical spin state.

This gauge redundancy implies that a symmetry operation \(\mathcal O\) need not leave an {\it Ansatz} invariant by itself; instead, it may be followed by a compensating gauge transformation. Symmetries therefore act projectively as \(W_{\mathcal O}\mathcal O\). The condition for symmetry invariance is then
\begin{equation}
u_{ij}
=
W_{\mathcal O}^\dagger(i)\,
u_{\mathcal O(i)\mathcal O(j)}\,
W_{\mathcal O}(j),
\end{equation}
i.e, the invariance of an {\it Ansatz} under the action of the spatial operation \(\mathcal O\) is defined up to a local SU(2) gauge transformation.

The full set of gauge transformations \(\{W_{\mathcal O}\}\) associated with all symmetry operations defines the projective symmetry group (PSG). The PSG plays for quantum spin liquids a role analogous to that of the ordinary symmetry group in Landau theory: it characterizes phases that cannot be distinguished solely by conventional symmetry breaking.

A particularly important subgroup of the PSG is the invariant gauge group (IGG), namely the set of pure gauge transformations that leave the {\it Ansatz} invariant. Equivalently, the IGG is the projective realization of the identity operation. Depending on the {\it Ansatz}, the IGG can be SU(2), U(1), or \(\mathbb Z_2\), and we use it to label the corresponding spin-liquid phase.

\section{Summary of Classification}
\label{sec:summary_classification}
In this section, we classify the quantum-spin-liquid mean-field {\it Ans\"atze} within the fermionic-parton PSG framework.

\subsection{Symmetry generators and projective realizations on the checkerboard lattice}
\label{sec:symmetry}

The checkerboard lattice is described by the wallpaper group \(p4gm\). Its symmetry group is generated by two translations \(T_1\) and \(T_2\), a fourfold rotation \(C_4\) about the center of a crossed square, and a diagonal reflection \(\sigma\), as illustrated in Fig.~\ref{fig:lattice}(c) [see also Appendix~\ref{app:lattice_symmetry}]. Together with time reversal \(\mathcal T\), these generate the full symmetry set
\begin{equation}
\mathcal O \in \{T_1,T_2,C_4,\sigma,\mathcal T\}.
\end{equation}

Within the PSG construction, each symmetry operation \(\mathcal O\) is implemented only up to an accompanying SU(2) gauge transformation. The corresponding projective realization is therefore specified by the set
\begin{equation}
\{W_{T_1},W_{T_2},W_{C_4},W_{\sigma},W_{\mathcal T}\}.
\end{equation}
The number of allowed PSG solutions depends on the invariant gauge group (IGG). For the checkerboard lattice, we obtain 386 gauge-inequivalent PSGs for the U(1) IGG and 400 gauge-inequivalent PSGs for the \(\mathbb Z_2\) IGG; details of the derivation are given in \appref{app:projective_symmetry}.

\subsection{Classification results}
\label{sec:classification}

Having obtained the full sets of gauge-inequivalent U(1) and $\mathbb{Z}_2$ PSGs, we next construct the corresponding mean-field {\it Ans\"atze}. Although the number of PSG solutions is large, the number of realizable U(1) and $\mathbb{Z}_2$ {\it Ans\"atze} depends on the range of bonds on which nonvanishing mean-field amplitudes are allowed. Here we first restrict the construction to the $J_{\boxtimes}$ bonds only [Fig.~\ref{fig:lattice}(d)]. Under this restriction, the 386 U(1) PSGs yield only 20 distinct U(1) {\it Ans\"atze}, shown graphically in Fig.~\ref{fig:u1_ansatz} and listed in Appendix~\ref{app:ansatz_u1}. Likewise, the 400 $\mathbb{Z}_2$ PSGs give rise to only 36 realizable $\mathbb{Z}_2$ {\it Ans\"atze}, which are tabulated in Table~\ref{table:z2_ansatz} of Appendix~\ref{app:ansatz_z2} and labeled Z01--Z36.

Throughout this construction, we choose a canonical gauge in which the IGG is represented by global gauge transformations,
\[
\{e^{\dot\iota\phi\tau^z}\}
\qquad \text{for U(1)}, \qquad
\{\pm\tau^0\}
\qquad \text{for } \mathbb{Z}_2.
\]
In this gauge, the U(1) {\it Ans\"atze} contain hopping terms only.

Our main interest is the $\mathbb{Z}_2$ {\it Ansatz} on the checkerboard lattice that can be continuously connected to the two {\it Ans\"atze} known to describe the QSL ground states on the square and Shastry--Sutherland lattices, namely Z2A$zz$13 and Z3000, respectively. To identify such a state, we extend the above 36 checkerboard $\mathbb{Z}_2$ {\it Ans\"atze} by allowing additional mean-field parameters on the diagonal $J_{\square}$ links connecting the vertices of the empty squares. We then find that the {\it Ansatz} labeled Z05 precisely provides the missing corner of the triangle of spin liquids in square-lattice-based geometries. The corresponding adiabatic connections are discussed explicitly in Sec.~\ref{sec:adiabatic_connections}.

The three spin liquids Z2A$zz$13, Z3000, and Z05 are all $\mathbb{Z}_2$ daughters of a common U(1) parent state, denoted U19 in the present classification and U800 in Refs.~\cite{Maity-2024,Feuerpfeil-2026}. This parent U(1) state carries a staggered $(\varphi,-\varphi)$ flux pattern through the $(\boxtimes,\square)$ plaquettes and is therefore referred to as the U(1) staggered-flux state. In turn, this U(1) state is itself a daughter of the SU(2) $\pi$-flux state. These relations become explicit when Z05 is written in the following gauge:
\begin{align}
&u^{}_s=\begin{bmatrix}
-t e^{-\dot\iota\theta} &0\\
0 & t e^{\dot\iota\theta}
\end{bmatrix}=\dot\iota{t}^{}_{s,0}\tau^0+{t}^{}_{s,z}\tau^z\,,\\
&u^{}_{(X,Y,0),(X,Y,1)}=u^{}_s\,,
\label{eq:U19_1}\\
&u^{}_{(X,Y,1),(X+1,Y+1,0)}=-u^{\dagger}_s\,,\\
&u^{}_{(X,Y,0),(X-1,Y,1)}=-(-1)^{X+Y}u^{\dagger}\,,\\
&u^{}_{(X,Y+1,0),(X,Y,1)}=-(-1)^{X+Y}u^{\dagger}_s\label{eq:U19_3}\,,\\
&u^{}_{(X,Y,0),(X,Y+1,0)}=\Delta^{}_{\boxtimes}\sigma^x\,,\\
&
u^{}_{(X,Y,1),(X-1,Y,1)}=\Delta^{}_{\boxtimes} \sigma^y \label{eq:Z05_01}\,,\\
&u^{}_{(X,Y,1),(X,Y-1,1)}=\Delta_{\square}\sigma^y\,,\\
&u^{}_{(X,Y,0),(X+1,Y,0)}=\Delta_{\square}\sigma^x \label{eq:Z05_02}\,,\\
&a^{}_{\gamma}(X,Y,0)\tau^\gamma=(-1)^{X+Y}a^{}_p\tau^y\,,\\
&a^{}_{\gamma}(X,Y,1)\tau^\gamma=(-1)^{X+Y}a^{}_p\tau^x\,.\label{eq:Z05_03}
\end{align}
Equations~\eqref{eq:U19_1}--\eqref{eq:U19_3} describe fermionic hopping amplitudes \(t^{}_{s,0}=t\sin\theta\) and \(t^{}_{s,z}=t\cos\theta\), whereas the remaining equations correspond to pairing terms.

This form makes it transparent how the $\mathbb{Z}_2$, U(1), and SU(2) spin-liquid states emerge in different parameter regimes:
\begin{enumerate}[label=\roman*.]
\item When all pairing terms are set to zero, \(\Delta=0\), and \(\theta=\pi/2\), one recovers the parent $\pi$-flux state with SU(2) invariant gauge structure. In the notation of Ref.~\cite{Wen-2002}, this state is denoted SU2B$n$0.
\item For vanishing pairings $\Delta=0$, but arbitrary \(\theta\), one obtains a U(1) state with staggered \((\varphi,-\varphi)\) flux through the square plaquettes, where \(\varphi=\pi+4\theta\).
\item Allowing the \(\Delta\)-terms breaks the U(1) IGG of U19/U800 down to \(\mathbb{Z}_2\), yielding the $\mathbb{Z}_2$ state Z05.
\end{enumerate}

All hopping and pairing amplitudes are illustrated schematically in Fig.~\ref{fig:u800_z3000_ansatz} [Appendix~\ref{app:z2_ansatz_qft}]. In Sec.~\ref{sec:adiabatic_connections}, we use a different gauge in which the {\it Ansatz} is written entirely in terms of real hoppings and pairings [Eqs.~\eqref{eq:z05}]. The parameters in the two gauges are related by
\[
\frac{t^{}_{s}+\Delta^{}_{s}}{\sqrt{2}}\rightarrow-t^{}_{s,0}\,,\qquad
\frac{t^{}_{s}-\Delta^{}_{s}}{\sqrt{2}}\rightarrow t^{}_{s,z}\,,\qquad
t^{}_{\boxtimes}\rightarrow\Delta^{}_{\boxtimes}\,,
\]
\[
t^{}_{\square}\rightarrow-\Delta^{}_{\square},\qquad
\mu\rightarrow a^{}_p\,.
\]
The PSG data in this gauge are collected in \appref{app:PSGs}. It is also worth noting that, in this gauge, the Dirac points appear around \(\Gamma(0,0)\), in contrast to the gauge used in Sec.~\ref{sec:adiabatic_connections}, where they appear around the \(M\) points.

\section{Discussion and outlook}
\label{sec:discussion}

The results presented here reveal a broader unifying structure beyond the prevailing view that microscopic lattice geometry and space-group structure provide the primary basis for classifying quantum spin liquids. We demonstrate this broader perspective explicitly within a well-defined family of models: the square, Shastry--Sutherland, and checkerboard Heisenberg antiferromagnets can realize the same $\mathbb{Z}_2$ Dirac spin liquid and the same Higgs-driven gauge-theory structure, despite distinct microscopic symmetry realizations. Crucially, this unification is not variational or heuristic, but follows from the projective symmetry structure (PSG/IGG) and the common continuum gauge theory.

Three ingredients make this unification structurally rigid. First, the checkerboard state Z05 shares the PSG class of the square-lattice Z2A$zz$13 and Shastry--Sutherland Z3000 states; adiabatic interpolation preserves both the invariant gauge group and projective symmetry implementation. Second, all three descend from a common SU(2) $\pi$-flux parent through a staggered-flux U(1) state, thereby fixing the symmetry structure of the low-energy Dirac fermions and Higgs fields. Third, the resulting continuum description is QED$_3$ with two Dirac fermions coupled to two adjoint Higgs fields, where reduced lattice symmetries only enter in terms that do not alter the scaling dimensions or the underlying gauge-theoretic backbone.

This alignment between lattice PSG structure and continuum Higgs theory provides a controlled notion of equivalence. The three models are not merely similar at the mean-field level. Rather, they share the same parent gauge theory and the same Higgs mechanism leading to the stable $\mathbb{Z}_2$ Dirac phase. Differences in microscopic symmetry constrain the accessible parameter subspaces, but they do not alter the emergent gauge structure. In this sense, lattice geometry selects realizations of a common gauge-theoretic scaffold rather than generating distinct scaffolds.

The energetic realization of the checkerboard candidate remains a quantitative question. The present work fixes its symmetry class and effective field theory, thereby providing a sharply defined target for future numerical tests. Within this perspective, any change of phase must correspond either to a change of PSG class or to confinement within the same gauge-theoretic framework.

The unified Higgs structure also governs the confinement mechanisms. The SU(2) and U(1) parent states determine which Higgs condensates and monopole operators are symmetry-allowed. Confinement into N\'eel or valence-bond-ordered phases can therefore be analyzed within a common field-theory description. Microscopic differences modify which symmetry-breaking patterns are favored, but the route to confinement is still set by the same underlying gauge structure.

The present unification also suggests a concrete experimental testbed. On the Shastry--Sutherland side, the canonical compound SrCu$_2$(BO$_3$)$_2$ already provides access to proximate plaquette-singlet and antiferromagnetic phases under pressure and field, making it a natural platform for testing the common confinement structure discussed here~\cite{Shi-2022,Cui-2023,Guo-2025}. Complementarily, the recently reported Shastry--Sutherland magnet Pr$_2$Ga$_2$BeO$_7$~\cite{Li-2024} exhibits gapless spin-liquid-like phenomenology, including power-law specific heat, gapless inelastic neutron spectra, anomalous low-energy spin dynamics, and a finite residual thermal conductivity. This indicates that Shastry--Sutherland materials can access fractionalized regimes directly relevant to the broader phenomenology of the present work, even though the microscopic anisotropy of that compound differs from the SU(2)-symmetric setting studied here. By contrast, on the checkerboard side, a comparably clean spin-$\tfrac{1}{2}$ material realization remains much less developed experimentally; checkerboard-related compounds such as PbCuTeO$_5$ appear better viewed as frustrated precursor systems than as established realizations of the present gapless $\mathbb{Z}_2$ Dirac phase~\cite{Koteswararao-2017}. More generally, the most direct experimental diagnostics of the unified scenario advanced here are the absence of magnetic Bragg peaks, broad continuum spectral weight in inelastic neutron scattering, and low-temperature power-law thermodynamics rather than activated behavior~\cite{Broholm-2020,Savary-2016}.

The triangle construction suggests a broader program. One may systematically search for hidden unifications by identifying PSG classes that persist under symmetry reduction or augmentation across related lattices. The procedure is concrete: (i) determine the PSG class on the lowest-symmetry member, (ii) identify higher-symmetry embeddings, and (iii) test whether the corresponding continuum Higgs theory closes consistently. This reframes classification in terms of gauge structure and Higgs content rather than geometry alone.

More broadly, we propose a hierarchical classification: microscopic lattice symmetry constrains the realization, while the emergent gauge theory and its Higgs structure define the quantum phase. Coincident gauge theories predict shared phase diagrams and confinement---even across substantially different lattices.

The square-Shastry-Sutherland-checkerboard family provides an explicit realization of this principle. 
It demonstrates that distinct microscopic settings can realize the same $\mathbb{Z}_2$ Dirac spin-liquid class and Higgs-driven QED$_3$ framework. 
Lattice diversity thus reveals a unified gauge-theoretic structure rather than proliferating distinct quantum orders.

\section{Acknowledgments}
This work is supported by the Deutsche Forschungsgemeinschaft (DFG, German Research Foundation) through Project-ID 258499086 - SFB 1170 and through the W\"urzburg-Dresden Cluster of Excellence on Complexity, Topology and Dynamics in Quantum Matter - ctd.qmat Project-ID 390858490 - EXC 2147. The Flatiron Institute is a division of the Simons Foundation. S.S. was supported by the U.S. National Science Foundation grant No. DMR 2245246 and by the Simons Collaboration on Ultra-Quantum Matter which is a grant from
the Simons Foundation (651440, S.S.). The work Y.I. was performed in part at the Aspen Center for Physics, which is supported by a grant from the Simons Foundation (1161654, Troyer). This research was also supported in part by grant NSF PHY-2309135 to the Kavli Institute for Theoretical Physics and by the International Centre for Theoretical Sciences (ICTS) for participating in the Discussion Meeting - Fractionalized Quantum Matter (code: ICTS/DMFQM2025/07). R.T. thanks IIT Madras for a Visiting Faculty Fellow position under the IoE program.  Y.I. acknowledges support from the Abdus Salam International Centre for Theoretical Physics through the Associates Programme, from the Simons Foundation through Grant No.~284558FY19, from IIT Madras through the Institute of Eminence program for establishing QuCenDiEM (Project No. SP22231244CPETWOQCDHOC).

\clearpage
\appendix

\begin{table*}
	\caption{The 386 U(1) PSG classes, characterized by the different gauge-inequivalent choices of $w^{}_\mathcal{O}$, $\bar{\phi}^{}_{\mathcal{O},s}$, and the $\xi_{\ldots}$ parameters appearing in Eqs.~\eqref{eq:g_translation_u}--\eqref{eq:g_time_u}. The parameters $n_{\ldots}$ are binary variables taking the values 0 or 1, while $p^{\pdagger}_{C_4}$ takes the values $0,1,2,3$ and $\bar{\phi}^{\pdagger}_{C_4}\in[0,2\pi)$.}
	\begin{ruledtabular}
		\begin{tabular}{ccccccccccccc}
$w^{\pdagger}_{T_1}$&$w^{\pdagger}_{\mathcal{T}}$&$w^{\pdagger}_{C_4}$&$w^{\pdagger}_{\sigma}$&$\xi^{\pdagger}_{\mathcal{T}T_2}$&$\xi^{\pdagger}_T$&$\xi^{\pdagger}_{C_4T_2}$&$\xi^{\pdagger}_{\sigma T_1}$&$\xi^{\pdagger}_{\sigma T_2}$&$\bar{\phi}^{\pdagger}_{\mathcal{T},s}$&$\bar{\phi}^{\pdagger}_{C_4,s}$&$\bar{\phi}^{\pdagger}_{\sigma,s}$ & \# of PSG\\
			\hline	
$0$ & $0$ & $0$ & $0$ & $0$ & $n^{\pdagger}_{T}\pi$& $0$ & $n^{\pdagger}_{\sigma T_1}\pi$ & $n^{\pdagger}_{\sigma T_1}\pi$ & $\{0,\pi\}$ & $0$ & $\{0,\frac{n^{\pdagger}_{\sigma T_1}\pi}{2}+n^{\pdagger}_{\sigma}\pi\}$ & $8$ \\
$0$ & $0$ & $0$ & $1$ & $0$ & $\xi^{\pdagger}_{T}\pi$& $0$ & $0$ & $0$ & $\{0,\pi\}$ & $0$ & $\{0,n^{\pdagger}_{\sigma}\pi\}$ & $2$ \\
$0$ & $0$ & $1$ & $0$ & $0$ & $n^{\pdagger}_{T}\pi$& $0$ & $n^{\pdagger}_{\sigma T_1}\pi$ & $n^{\pdagger}_{\sigma T_1}\pi$ & $\{0,\pi\}$ & $\{0,\frac{p^{\pdagger}_{C_4}\pi}{2}\}$ & $\{0,n^{\pdagger}_{\sigma}\pi-\frac{p^{\pdagger}_{C_4}\pi}{2}\}$ & $32$ \\
$0$ & $0$ & $1$ & $1$ & $0$ & $n^{\pdagger}_{T}\pi$& $0$ & $0$ & $0$ & $\{0,\pi\}$ & $\{0,\frac{p^{\pdagger}_{C_4}\pi}{2}\}$ & $0$ & $8$ \\
\hline
$0$ & $1$ & $0$ & $0$ & $0$ & $n^{\pdagger}_{T}\pi$& $n^{\pdagger}_{\sigma T_2}\pi$ & $0$ & $n^{\pdagger}_{\sigma T_2}\pi$ & $0$ & $\{0,n^{\pdagger}_{C_4}\pi\}$ & $\{0,n^{\pdagger}_{\sigma}\pi\}$ & $16$ \\
$0$ & $1$ & $0$ & $1$ & $0$ & $n^{\pdagger}_{T}\pi$& $n^{\pdagger}_{\sigma T_2}\pi$ & $0$ & $n^{\pdagger}_{\sigma T_2}\pi$ & $0$ & $\{0,n^{\pdagger}_{C_4}\pi\}$ & $\{0,n^{\pdagger}_{\sigma}\pi\}$ & $16$ \\
$0$ & $1$ & $1$ & $0$ & $0$ & $n^{\pdagger}_{T}\pi$& $(n^{\pdagger}_{\sigma T_1}+n^{\pdagger}_{\sigma T_2})\pi$ & $n^{\pdagger}_{\sigma T_1}\pi$ & $n^{\pdagger}_{\sigma T_2}\pi$ & $0$ & $0$ & $\{0,n^{\pdagger}_{\sigma}\pi\}$ & $16$ \\
$0$ & $1$ & $1$ & $1$ & $0$ & $n^{\pdagger}_{T}\pi$& $n^{\pdagger}_{\sigma T_2}\pi$ & $0$ & $n^{\pdagger}_{\sigma T_2}\pi$ & $0$ & $0$ & $\{0,n^{\pdagger}_{\sigma}\pi\}$ & $8$ \\
\hline
$1$ & $0$ & $0$ & $0$ & $0$ & $0$& $n^{\pdagger}_{C_4 T_2}\pi$ & $0$ & $n^{\pdagger}_{\sigma T_2}\pi$ & $\{0,\pi\}$ & $\{\bar{\phi}^{\pdagger}_{C_4},\frac{(n^{\pdagger}_{C_4 T_2}+n^{\pdagger}_{\sigma T_2})\pi}{2}-\bar{\phi}^{\pdagger}_{C_4}\}$ & $\{0,n^{\pdagger}_{\sigma}\pi\}$ & $8$ \\
$1$ & $0$ & $0$ & $1$ & $0$ & $0$& $n^{\pdagger}_{\sigma T_2}\pi$ & $0$ & $n^{\pdagger}_{\sigma T_2}\pi$ & $\{0,\pi\}$ & $\{\bar{\phi}^{\pdagger}_{C_4},\bar{\phi}^{\pdagger}_{C_4}\}$ & $\{0,n^{\pdagger}_{\sigma}\pi\}$ & $4$ \\
$1$ & $0$ & $1$ & $0$ & $0$ & $0$& $n^{\pdagger}_{\sigma T_2}\pi$ & $0$ & $n^{\pdagger}_{\sigma T_2}\pi$ & $\{0,\pi\}$ & $\{\bar{\phi}^{\pdagger}_{C_4},\bar{\phi}^{\pdagger}_{C_4}\}$ & $\{0,n^{\pdagger}_{\sigma}\pi\}$ & $4$ \\
$1$ & $0$ & $1$ & $1$ & $0$ & $0$& $n^{\pdagger}_{C_4 T_2}\pi$ & $0$ & $n^{\pdagger}_{\sigma T_2}\pi$ & $\{0,\pi\}$ & $\{\bar{\phi}^{\pdagger}_{C_4},\frac{(n^{\pdagger}_{C_4 T_2}+n^{\pdagger}_{\sigma T_2})\pi}{2}-\bar{\phi}^{\pdagger}_{C_4}\}$ & $\{0,n^{\pdagger}_{\sigma}\pi\}$ & $8$ \\
\hline
$1$ & $1$ & $0$ & $0$ & $n^{\pdagger}_{\mathcal{T} T_2}\pi$ & $0$& $(n^{\pdagger}_{\mathcal{T} T_2}+n^{\pdagger}_{\sigma T_2})\pi$ & $0$ & $n^{\pdagger}_{\sigma T_2}\pi$ & $0$ & $\frac{n^{\pdagger}_{\mathcal{T} T_2}\pi}{4}+\frac{p^{\pdagger}_{C_4}\pi}{2}+\{0,n^{\pdagger}_{C_4}\pi\}$ & $\{0,n^{\pdagger}_{\sigma}\pi\}$ & $64$ \\
$1$ & $1$ & $0$ & $1$ & $n^{\pdagger}_{\mathcal{T}T_2}\pi$ & $0$& $n^{\pdagger}_{\sigma T_2}\pi$ & $0$ & $n^{\pdagger}_{\sigma T_2}\pi$ & $0$ & $\frac{n^{\pdagger}_{\mathcal{T}T_2}\pi}{4}+\frac{p^{\pdagger}_{C_4}\pi}{2}+\{0,n^{\pdagger}_{C_4}\pi\}$ & $\{0,n^{\pdagger}_{\sigma}\pi\}$ & $64$ \\
$1$ & $1$ & $1$ & $0$ & $n^{\pdagger}_{\mathcal{T}T_2}\pi$ & $0$& $n^{\pdagger}_{\sigma T_2}\pi$ & $0$ & $n^{\pdagger}_{\sigma T_2}\pi$ & $0$ & $\frac{n^{\pdagger}_{\mathcal{T}T_2}\pi}{4}+\frac{p^{\pdagger}_{C_4}\pi}{2}+\{0,n^{\pdagger}_{C_4}\pi\}$ & $\{0,n^{\pdagger}_{\sigma}\pi\}$ & $64$ \\
$1$ & $1$ & $0$ & $0$ & $n^{\pdagger}_{\mathcal{T} T_2}\pi$ & $0$& $(n^{\pdagger}_{\mathcal{T} T_2}+n^{\pdagger}_{\sigma T_2})\pi$ & $0$ & $n^{\pdagger}_{\sigma T_2}\pi$ & $0$ & $\frac{n^{\pdagger}_{\mathcal{T} T_2}\pi}{4}+\frac{p^{\pdagger}_{C_4}\pi}{2}+\{0,n^{\pdagger}_{C_4}\pi\}$ & $\{0,n^{\pdagger}_{\sigma}\pi\}$ & $64$ \\
		\end{tabular}
	\end{ruledtabular}
	\label{table:u1_psg}
\end{table*}

\begin{table}[h]
\caption{Gauge-inequivalent choices of the matrices $\bar{W}_{\mathcal{O}}$ appearing in Eqs.~\eqref{eq:solution_c4_z2}, \eqref{eq:solution_sigma_z2}, and \eqref{eq:solution_time_z2}.}
\begin{ruledtabular}
\begin{tabular}{cccccc}
No. & $\eta^{\pdagger}_{\sigma T_1}$ & $\eta^{\pdagger}_{\mathcal{T} T_1}$& $\bar{W}^{\pdagger}_{C_4,s}$ &$\bar{W}^{\pdagger}_{\sigma,s}$&$\bar{W}^{\pdagger}_{\mathcal{T},s}$ \\
			\hline
1& $+1$& $+1$&$\{\tau^0,\eta^{\pdagger}_{C_4,1}\tau^0\}$&$\{\tau^0,\eta^{\pdagger}_{\sigma,1}\tau^0\}$&$\{\dot\iota\tau^y,\eta^{\pdagger}_{\mathcal{T},1}\dot\iota\tau^y\}$ \\
2& $+1$& $+1$&$\{\tau^0,\eta^{\pdagger}_{C_4,1}\tau^0\}$&$\{\tau^0,\eta^{\pdagger}_{\sigma,1}\tau^0\}$&$\{\tau^0,-\tau^0\}$ \\
3& $+1$& $+1$&$\{\tau^0,\eta^{\pdagger}_{C_4,1}\tau^0\}$&$\{\dot\iota\tau^z,\eta^{\pdagger}_{\sigma,1}\dot\iota\tau^z\}$&$\{\dot\iota\tau^y,\eta^{\pdagger}_{\mathcal{T},1}\dot\iota\tau^y\}$ \\
4& $+1$& $+1$&$\{\tau^0,\eta^{\pdagger}_{C_4,1}\tau^0\}$&$\{\dot\iota\tau^z,\eta^{\pdagger}_{\sigma,1}\dot\iota\tau^z\}$&$\{\dot\iota\tau^z,\eta^{\pdagger}_{\mathcal{T},1}\dot\iota\tau^z\}$ \\
5& $+1$& $+1$&$\{\tau^0,\eta^{\pdagger}_{C_4,1}\tau^0\}$&$\{\dot\iota\tau^z,\eta^{\pdagger}_{\sigma,1}\dot\iota\tau^z\}$&$\{\tau^0,-\tau^0\}$ \\
6& $-1$& $-1$&$\{\tau^0,\eta^{\pdagger}_{C_4,1}\dot\iota\tau^z\}$&$\{\tau^0,\eta^{\pdagger}_{\sigma,1}\dot\iota\tau^z\}$&$\{\dot\iota\tau^y,\eta^{\pdagger}_{\mathcal{T},1}\dot\iota\tau^y\}$ \\
7& $-1$& $+1$&$\{\tau^0,\eta^{\pdagger}_{C_4,1}\dot\iota\tau^z\}$&$\{\tau^0,\eta^{\pdagger}_{\sigma,1}\dot\iota\tau^z\}$&$\{\dot\iota\tau^z,\eta^{\pdagger}_{\mathcal{T},1}\dot\iota\tau^z\}$ \\
8& $-1$& $+1$&$\{\tau^0,\eta^{\pdagger}_{C_4,1}\dot\iota\tau^z\}$&$\{\tau^0,\eta^{\pdagger}_{\sigma,1}\dot\iota\tau^z\}$&$\{\tau^0,-\tau^0\}$ \\
9& $-1$& $-1$&$\{\tau^0,\eta^{\pdagger}_{C_4,1}\dot\iota\tau^z\}$&$\{\dot\iota\tau^z,\eta^{\pdagger}_{\sigma,1}\tau^0\}$&$\{\dot\iota\tau^y,\eta^{\pdagger}_{\mathcal{T},1}\dot\iota\tau^y\}$ \\
10& $-1$& $+1$&$\{\tau^0,\eta^{\pdagger}_{C_4,1}\dot\iota\tau^z\}$&$\{\dot\iota\tau^z,\eta^{\pdagger}_{\sigma,1}\tau^0\}$&$\{\dot\iota\tau^z,\eta^{\pdagger}_{\mathcal{T},1}\dot\iota\tau^z\}$ \\
11& $-1$& $+1$&$\{\tau^0,\eta^{\pdagger}_{C_4,1}\dot\iota\tau^z\}$&$\{\dot\iota\tau^z,\eta^{\pdagger}_{\sigma,1}\tau^0\}$&$\{\tau^0,-\tau^0\}$ \\
12& $+1$& $+1$&$\{\tau^0,\eta^{\pdagger}_{C_4,1}\dot\iota\tau^z\}$&$\{\dot\iota\tau^x,\eta^{\pdagger}_{\sigma,1}\dot\iota\tau^y\}$&$\{\dot\iota\tau^z,\eta^{\pdagger}_{\mathcal{T},1}\dot\iota\tau^z\}$ \\
13& $+1$& $-1$&$\{\tau^0,\eta^{\pdagger}_{C_4,1}\dot\iota\tau^z\}$&$\{\dot\iota\tau^x,\eta^{\pdagger}_{\sigma,1}\dot\iota\tau^y\}$&$\{\dot\iota\tau^y,\eta^{\pdagger}_{\mathcal{T},1}\dot\iota\tau^y\}$ \\
14& $+1$& $-1$&$\{\tau^0,\eta^{\pdagger}_{C_4,1}\dot\iota\tau^z\}$&$\{\dot\iota\tau^x,\eta^{\pdagger}_{\sigma,1}\dot\iota\tau^y\}$&$\{\dot\iota\tau^x,\eta^{\pdagger}_{\mathcal{T},1}\dot\iota\tau^x\}$ \\
15& $+1$& $+1$&$\{\tau^0,\eta^{\pdagger}_{C_4,1}\dot\iota\tau^z\}$&$\{\dot\iota\tau^x,\eta^{\pdagger}_{\sigma,1}\dot\iota\tau^y\}$&$\{\tau^0,-\tau^0\}$ \\
		\end{tabular}
	\end{ruledtabular}
	\label{table:z2_psg}
\end{table}

\section{Symmetries of the checkerboard model}
\label{app:symmetry}

\subsection{Lattice symmetries}
\label{app:lattice_symmetry}
The symmetry group of the checkerboard lattice [see Fig.~\ref{fig:lattice}(c)] is the wallpaper group $p4gm$. To construct its symmetry generators, it is convenient to begin from the full square-lattice wallpaper group, $p4mm$. Since $p4gm$ is a lower-symmetry subgroup of $p4mm$, the generators of the former can be expressed in terms of those of the latter. We denote the positions of the lattice sites in Cartesian coordinates by $r \equiv (i_x,i_y) \equiv (i_x\hat{x}+i_y\hat{y})a$, where $a$ is the lattice spacing. In the following, we list the generators of $p4mm$ (i.e., the square lattice) and their actions on the lattice-site coordinates:
\allowdisplaybreaks
\begin{equation}
\begin{aligned}\label{eq:square_symmetries}
    &T^{}_x(i_x,i_y)  \rightarrow (i_x+1, i_y),\,\\
    &T^{}_y(i_x,i_y) \rightarrow (i_x, i_y+1)\,, \\
    &\sigma^{}_x(i_x,i_y) \rightarrow (i_x, -i_y)\,, \\
    &\sigma^{}_y(i_x,i_y) \rightarrow (-i_x, i_y)\,, \\
    &R^{}_{\pi/2}(i_x,i_y) \rightarrow (-i_y, i_x)\,.
\end{aligned}
\end{equation}
Here, $T_x$ and $T_y$ generate translations by $a\hat{x}$ and $a\hat{y}$, respectively. The reflections $\sigma_x$ and $\sigma_y$ act about the lines $y=0$ and $x=0$, respectively, while $R_{\pi/2}$ denotes a $\pi/2$ rotation about the $C_4$ axis centered at $(0,0)$. We can now express all symmetry elements of the checkerboard wallpaper group, $p4gm$, in terms of the above generators of $p4mm$. The operations $T_1 = T_x \circ T_y^{-1}$ and $T_2 = T_x \circ T_y$ generate translations by $a(\hat{x}-\hat{y})$ and $a(\hat{x}+\hat{y})$, respectively, on the underlying square Bravais lattice. The symmetries $G_x = T_x \circ \sigma_x$ and $G_y = T_y \circ \sigma_y$ generate glide reflections, combining a translation by $a\hat{x}$ or $a\hat{y}$ with a reflection about the lines $y=0$ or $x=0$, respectively. The operations $\sigma_{xy}=R_{\pi/2}\circ\sigma_x$ and $\sigma_{\bar{x}y}=T_x\circ T_y\circ R_{\pi/2}\circ\sigma_y$ generate reflections about the lines $x=y$ and $x=-y$, respectively. Two further reflections, $\sigma_h=T_x\circ\sigma_y$ and $\sigma_v=T_y\circ\sigma_x$, act about the lines $y=a/2$ and $x=a/2$, respectively. Finally, $C_4^{\square}=T_x^2\circ T_y^{-1}\circ R_{\pi/2}$ and $C_4^{\boxtimes}=T_x\circ R_{\pi/2}$ generate fourfold rotations about the centers of the empty and crossed squares, respectively. However, the operations $\sigma_h$, $\sigma_v$, $G_x$, $G_y$, $\sigma_{\bar{x}y}$, and $C_4^{\square}$ can all be expressed in terms of the minimal generating set $\{T_1,T_2,C_4^{\boxtimes},\sigma_{xy}\}$ as
\begin{equation}
\begin{aligned}
    \sigma^{}_h & = \sigma^{}_{xy}\circ (C^{\boxtimes}_4)^3,\;\sigma^{}_v = \sigma^{}_{xy}\circ C^{\boxtimes}_4, \\
    G^{}_x & = \sigma^{}_{xy}\circ C^{\boxtimes}_4\circ T_2 ,\; G^{}_y = \sigma^{}_{xy}\circ G^{}_x\circ \sigma_{xy}, \\
    \sigma^{}_{x\bar{y}} & = \sigma^{}_{xy}\circ G^{}_x\circ G_y^{-1}, \\
    C^{\square}_4 & = G^{}_x\circ \sigma^{}_{xy}\circ G^{}_x\circ G_y^{-1}.
\end{aligned}
\end{equation}
Therefore, a minimal generating set for the full $p4gm$ wallpaper group is $\{T_1,T_2,C_4^{\boxtimes},\sigma_{xy}\}$. In the following, we denote $C_4^{\boxtimes}$ and $\sigma_{xy}$ simply by $C_4$ and $\sigma$, respectively. 

Now, we introduce a new notation $(X,Y,s)$ for a lattice site, where $(X,Y)$ denotes the unit-cell position and $s=0,1$ denotes the sublattice index, such that
\begin{equation}\label{eq:coordinate}
\mathbf{r} \equiv (r, s) \equiv (X,Y,s) \equiv X \mathbf{\hat{a}_{x\bar{y}}}+Y\mathbf{\hat{a}_{x{y}}}+sa\mathbf{\hat{x}},
\end{equation}
where $\mathbf{\hat{a}_{x\bar{y}}}=(\mathbf{\hat{x}}-\mathbf{\hat{y}})a$ and $\mathbf{\hat{a}_{x{y}}}=(\mathbf{\hat{x}}+\mathbf{\hat{y}})a$.
In this notation, the symmetry operations act as follows:
\begin{equation}
\begin{aligned}
    T^{}_1: (X, Y, s) & \rightarrow (X+1, Y, s), \\
    T^{}_2: (X, Y, s) & \rightarrow (X, Y+1, s), \\
    C^{}_4: (X, Y, s) & \rightarrow (-Y, X + s, 1-s), \\
    \sigma: (X, Y, s) & \rightarrow (-X-s, Y, s).
\end{aligned}
\end{equation}
Furthermore, constructing fully symmetric {\it Ans\"atze} requires the inclusion of time-reversal symmetry $\mathcal{T}$, which is an internal symmetry that leaves $(X, Y, s)$ unchanged and commutes with all space-group operations. The action of $\mathcal{T}$ on the mean fields is nontrivial, namely $\mathcal{T}(u_{ij}, a_\gamma) = -(u_{ij}, a_\gamma)$~\cite{Wen-2002}. Thus, its projective action is given by
\begin{equation}
W^\dagger_{\mathcal{T}}(i) u^{\pdagger}_{ij} W^{\pdagger}_{\mathcal{T}}(j) = -u^{\pdagger}_{ij}, \;W^{\pdagger}_{\mathcal{T}} \in\text{ SU(2)}.
\end{equation}
Combining $\mathcal{T}$ with the wallpaper-group operations $\{T_1,T_2,C_4,\sigma\}$, the symmetry group of the checkerboard lattice is fully specified by the following identity relations:
\allowdisplaybreaks
\begin{equation}
\label{eq:id_relation}
\begin{aligned}
    T_1^{-1} T_2^{-1} T^{}_1 T^{}_2 & = \mathds{1}, \\
    C^{}_4 T^{-1}_2C_4^{-1} T_1^{-1}  & = \mathds{1}, \\
    C^{}_4 T^{}_1C_4^{-1} T_2^{-1}  & = \mathds{1}, \\
    C_4^4 & = \mathds{1}, \\
    \sigma T^{-1}_1\sigma^{-1} T^{-1}_1& = \mathds{1}, \\
    \sigma T^{}_2\sigma^{-1} T^{-1}_2 & = \mathds{1}, \\
    \sigma^2 & = \mathds{1}, \\
    C_4\sigma C_4\sigma & = \mathds{1}, \\    
    \mathcal{T}^2 & = \mathds{1}, \\
    \mathcal{T} \mathcal{O} \mathcal{T}^{-1} \mathcal{O}^{-1} & = \mathds{1}, \quad \mathcal{O} \in \{T^{}_1, T^{}_2, C^{}_4, \sigma\}.
\end{aligned}
\end{equation}

\subsection{PSG solutions}
\label{app:projective_symmetry}

We now proceed to obtain the projective actions of all the above-mentioned symmetries. We start with the case in which the IGG is U(1). The generic form of a U(1) PSG solution for any symmetry operator $\mathcal{O}$, in the canonical gauge (i.e., the {\it Ans\"atze} in this gauge consist only of hopping terms), is given by
\begin{equation}
    \label{eq:canonical_u1_ansatz}
    W^{}_\mathcal{O}(\mathbf{r},s)=e^{\dot\iota \Phi^{\pdagger}_\mathcal{O}(X,Y,s)\tau^z}(i\tau^x)^{w^{\pdagger}_\mathcal{O}},
\end{equation}
where $w^{}_\mathcal{O}$ is a binary parameter that takes the values $0$ or $1$. However, upon considering the projective representations of the relations~\eqref{eq:id_relation}, we find that solutions exist only for $w^{\pdagger}_{T_1}=w^{\pdagger}_{T_2}$ for $\mathcal{O}\in\{T_x,T_y\}$. All other binary parameters remain unconstrained. The solutions for $\Phi_\mathcal{O}(X,Y,s)$ can be written as
\begin{equation}
 \Phi^{\pdagger}_{\mathcal{O}}(X,Y,s)=(-1)^{w^{\pdagger}_{T_1}(X+Y)} \phi^{\pdagger}_{\mathcal{O}}(X,Y,s) \, ,   
\end{equation}
where
\begin{align}
 \phi^{\pdagger}_{T_1}(X,Y,s)&=Y\,\xi^{\pdagger}_T,\;\phi^{\pdagger}_{T_2}(X,Y,s)=-\frac{\xi^{\pdagger}_{\mathcal{T}T_2}\pi}{2},\label{eq:g_translation_u}\\
\phi^{\pdagger}_{C_4}(X,Y,s)&=Y \xi^{\pdagger}_{C_4T_2}+XY\xi^{\pdagger}_T+\bar{\phi}^{\pdagger}_{C_4,s}\notag\\
&+((-1)^{w^{\pdagger}_{C_4}+u}X-Y)\frac{\xi^{\pdagger}_{\mathcal{T}T_2}}{2},\label{eq:g_C4}\\
 \phi^{\pdagger}_{\sigma}(X,Y,s)&=X \xi^{\pdagger}_{\sigma T_1}+ Y \xi^{\pdagger}_{\sigma T_2}+\bar{\phi}^{\pdagger}_{\sigma,s}\notag\\
 &+((-1)^{w^{\pdagger}_{\sigma}+u}-1)\frac{Y\xi^{\pdagger}_{\mathcal{T}T_2}}{2},\label{eq:g_sigma}\\
 \phi^{\pdagger}_{\mathcal{T}}(X,Y,s)&=\bar{\phi}^{\pdagger}_{\mathcal{T},s}\, .
\label{eq:g_time_u}
\end{align}
Here, $\bar{\phi}^{\pdagger}_{\mathcal{O},s}$\,$\equiv$\,$\phi^{\pdagger}_{\mathcal{O}}(0,0,s)$ and $\xi^{\pdagger}_T,\,\xi^{\pdagger}_{C_4T_2},\,\xi^{\pdagger}_{\sigma T_i},\,\xi^{\pdagger}_{\mathcal{T}T_2},\,\bar{\phi}^{\pdagger}_{\mathcal{O},s}\in[0,2\pi)$. For details, we refer to Appendix~\ref{app:u1_psg_derivation}. All gauge-inequivalent choices of the $\xi$ and $\bar{\phi}$ parameters are listed in Table~\ref{table:u1_psg}. We find a total of $386$ U(1) PSG solutions that can be realized on the checkerboard lattice.

Similarly, one can proceed to obtain the solutions for $\mathbb{Z}_2$ IGG, where the identity can be defined up to a global sign in the canonical gauge. The resulting $\mathbb{Z}_2$ PSG solutions (for details, see Appendix~\ref{app:z2_psg_derivation}) can be expressed as
\begin{align}
  W^{\pdagger}_{T_1}(X,Y,s)&=\eta^{Y}_{T}\tau^0\ , \;W^{\pdagger}_{T_2}(X,Y,s)=\tau^0\ , \; \label{eq:solution_translationj_z2}\\ 
  W^{\pdagger}_{C_4}(X,Y,s)&=(\eta^{\pdagger}_{\sigma T_1}\eta^{\pdagger}_{\sigma T_2})^Y\eta^{XY}_{T}\bar{W}^{\pdagger}_{C_4,s}\ , \; \label{eq:solution_c4_z2}\\
 W^{\pdagger}_{\sigma}(X,Y,s)&=\eta^{X}_{\sigma T_1}\eta^{Y}_{\sigma T_2}\bar{W}^{\pdagger}_{\sigma,s}\
, \; \label{eq:solution_sigma_z2}\\  
 W^{\pdagger}_{\mathcal{T}}(X,Y,s)&=\eta^{X+Y}_{\mathcal{T} T_1}\bar{W}^{\pdagger}_{\mathcal{T},s}\, ,\label{eq:solution_time_z2}
\end{align}
where $\eta^{\pdagger}_{T},\eta^{\pdagger}_{\sigma T_1},\eta^{\pdagger}_{\sigma T_2},\eta^{\pdagger}_{\mathcal{T}T_1}=\pm1$ and the matrices $\bar{W}_{\mathcal{O},s}=W^{\pdagger}_{\mathcal{O}}(0,0,s)$ can be read from Table~\ref{table:z2_psg}. There are a total of $400$ $\mathbb{Z}_2$ PSGs.

\begin{figure}
\includegraphics[width=1.0\linewidth]{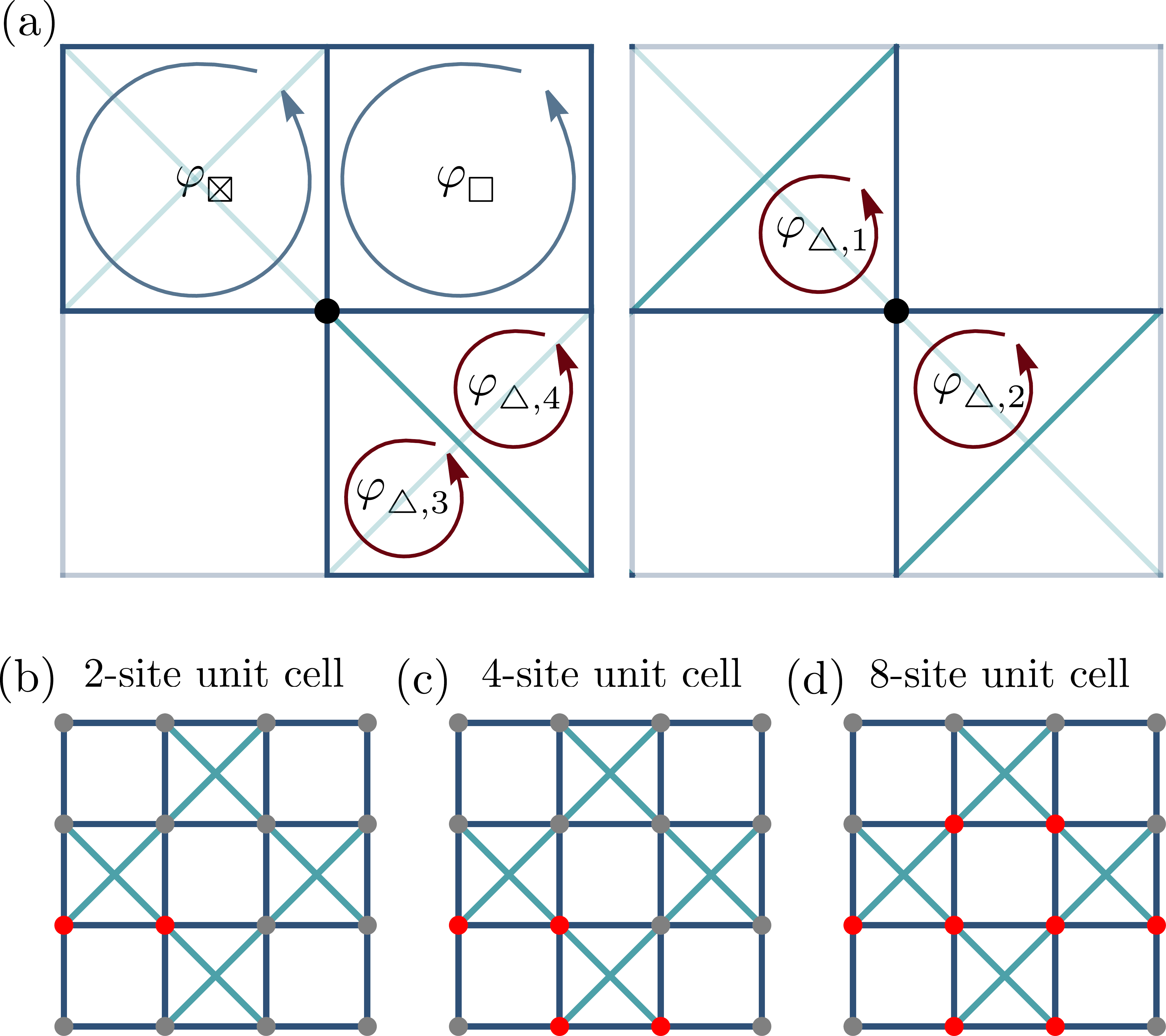}
	\caption{(a) SU(2) fluxes characterizing the U(1) {\it Ans\"atze}, defined with respect to a base site indicated by the black circle. (b)–(d) Illustrations of the 2-, 4-, and 8-site unit cells required to realize different classes of QSL {\it Ans\"atze}, respectively.}
	\label{fig:flux_unit_cell}
\end{figure}

\begin{figure*}[tb]	\includegraphics[width=1.0\linewidth]{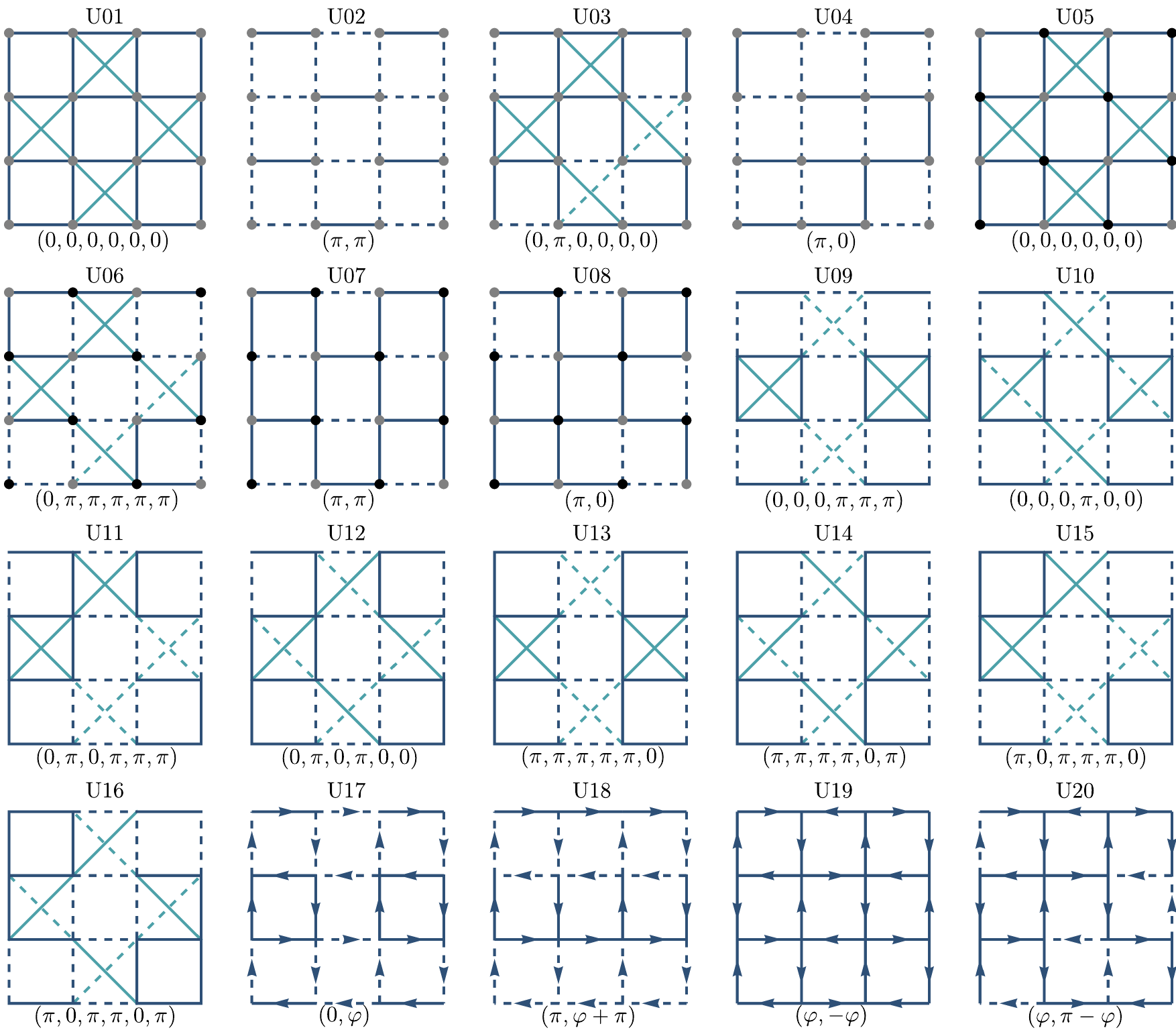}
	\caption{Sign structure of all U(1) {\it Ans\"atze}. Gray (black) dots on the lattice sites denote positive (negative) signs of the chemical potential, while solid and dashed lines or arrows represent positive and negative hopping amplitudes, respectively. Among these {\it Ans\"atze}, U01--U16 consist only of real hoppings, whereas the remaining states involve complex hoppings. The directions of the complex link fields $u^{\pdagger}_{ij}=\dot\iota t^{0}_{ij}\tau^0+t^{z}_{ij}\tau^z$ are indicated by arrows. The fluxes $(\varphi^{}_{\boxtimes},\varphi^{}_{\square},\varphi^{}_{\triangle,1},\varphi^{}_{\triangle,2},\varphi^{}_{\triangle,3},\varphi^{}_{\triangle,4})$ are also shown for each {\it Ansatz}.}
	\label{fig:u1_ansatz}
\end{figure*}

\begin{table}[h]
\caption{List of the 36 gauge inequivalent $\mathbb{Z}_2$ {\it Ans\"atze} when the symmetry-allowed link fields $u^{}_{ij}$ are allowed on $J_s$ and $J_\boxtimes$ links only. The gauge parameters specified by $(\eta^{\pdagger}_T,\eta^{\pdagger}_{\sigma T_2},\eta^{\pdagger}_{C_4,1}\eta^{\pdagger}_{\sigma,1},\eta^{\pdagger}_{\mathcal{T},1})$.}
\begin{ruledtabular}
\begin{tabular}{ccccc}
\multirow{ 2}{*}{Label} & PSG no & \multirow{ 2}{*}{$u^{\pdagger}_s$} & \multirow{ 2}{*}{$u^{\pdagger}_{\boxtimes}$} & \multirow{ 2}{*}{Onsite}\\
&Gauge parameters&&&\\
			\hline
Z01 & $1(+,+,+,+)$& $\tau^{z,x}$&$\tau^{z,x}$&$\tau^{z,x}$\\
Z02 & $1(-,+,+,+)$& $\tau^{z,x}$&$\tau^{z,x}$&$\tau^{z,x}$\\
Z03 & $1(+,+,-,-)$& $\tau^{0}$&$\tau^{z,x}$&$\tau^{z,x}$\\
Z04 & $1(-,+,-,-)$& $\tau^{0}$&$\tau^{z,x}$&$\tau^{z,x}$\\
Z05 & $3(+,+,-,-)$& $\tau^{0,y}$&$\tau^{z}$&$\tau^{z}$\\
Z06 & $3(-,+,-,-)$& $\tau^{0,y}$&$\tau^{z}$&$\tau^{z}$\\
Z07 & $3(+,-,-,-)$& $\tau^{0,y}$&$0$&$\tau^{z}$\\
Z08 & $3(-,-,-,-)$& $\tau^{0,y}$&$0$&$\tau^{z}$\\
Z09 & $6(+,+,+,+)$& $\tau^{x}$&$\tau^{0,y}$&$\tau^{z}$\\
Z10 & $6(-,+,+,+)$& $\tau^{x}$&$\tau^{0,y}$&$\tau^{z}$\\
Z11 & $6(+,+,-,+)$& $\tau^{z}$&$\tau^{0,y}$&$\tau^{z}$\\
Z12 & $6(-,+,-,+)$& $\tau^{z}$&$\tau^{0,y}$&$\tau^{z}$\\
Z13 & $6(+,+,+,-)$& $\tau^{0,y}$&$\tau^{0,y}$&$\tau^{z}$\\
Z14 & $6(-,+,+,-)$& $\tau^{0,y}$&$\tau^{0,y}$&$\tau^{z}$\\
Z15 & $6(+,-,+,-)$& $\tau^{0,y}$&$0$&$\tau^{z}$\\
Z16 & $6(-,-,+,-)$& $\tau^{0,y}$&$0$&$\tau^{z}$\\
Z17 & $7(+,+,+,+)$& $\tau^{x,y}$&$\tau^{x,y}$&$\tau^{z}$\\
Z18 & $7(-,+,+,+)$& $\tau^{x,y}$&$\tau^{x,y}$&$\tau^{z}$\\
Z19 & $7(+,+,+,-)$& 
$\tau^{0}$&$\tau^{x,y}$&$\tau^{z}$\\
Z20 & $7(-,+,+,-)$& 
$\tau^{0}$&$\tau^{x,y}$&$\tau^{z}$\\
Z21 & $7(+,+,-,-)$& 
$\tau^{z}$&$\tau^{x,y}$&$\tau^{z}$\\
Z22 & $7(-,+,-,-)$& 
$\tau^{z}$&$\tau^{x,y}$&$\tau^{z}$\\
Z23 & $9(+,+,+,+)$& $\tau^{z,x}$&$\tau^{0}$&$\tau^{z}$\\
Z24 & $9(-,+,+,+)$& $\tau^{z,x}$&$\tau^{0}$&$\tau^{z}$\\
Z25 & $9(+,-,+,+)$& $\tau^{z,x}$&$0$&$\tau^{z}$\\
Z26 & $9(-,-,+,+)$& $\tau^{z,x}$&$0$&$\tau^{z}$\\
Z27 & $12(+,-,+,-)$& $\tau^{0,z}$&$\tau^{y}$&$0$\\
Z28 & $12(-,-,+,-)$& $\tau^{0,z}$&$\tau^{y}$&$0$\\
Z29 & $13(+,-,+,+)$& $\tau^{z,x}$&$\tau^{y}$&$0$\\
Z30 & $13(-,-,+,+)$& $\tau^{z,x}$&$\tau^{y}$&$0$\\
Z31 & $15(+,+,+,-)$& $\tau^{0,z,x}$&$0$&$0$\\
Z32 & $15(-,+,+,-)$& $\tau^{0,z,x}$&$0$&$0$\\
Z33 & $15(+,-,+,-)$& $\tau^{0,z,x}$&$0$&$0$\\
Z34 & $15(-,-,+,-)$& $\tau^{0,z,x}$&$0$&$0$\\
Z35 & $1(+,+,+,-)$& $\tau^{y}$&$\tau^{z,x}$&$\tau^{z,x}$\\
Z36 & $1(-,+,+,-)$& $\tau^{y}$&$\tau^{z,x}$&$\tau^{z,x}$\\
		\end{tabular}
	\end{ruledtabular}
	\label{table:z2_ansatz}
\end{table}

\begin{table}[h]
\caption{Relisting of the 36 gauge-inequivalent $\mathbb{Z}_2$ {\it Ans\"atze} originally presented in Table~\ref{table:z2_ansatz}, after incorporating the symmetry-allowed link field $u_{ij}$ on the $J_\square$ bonds.}
\begin{ruledtabular}
\begin{tabular}{ccccc}
\multirow{ 2}{*}{Label} & \multirow{ 2}{*}{$u^{\pdagger}_{s}$} & \multirow{ 2}{*}{$u^{\pdagger}_{\boxtimes}$} & \multirow{ 2}{*}{Onsite} & \multirow{ 2}{*}{$u^{\pdagger}_{\square}$} \\
&&&&\\
			\hline
Z01& $\tau^{z,x}$&$\tau^{z,x}$&$\tau^{z,x}$&$\tau^{z,x}$\\
Z02 & $\tau^{z,x}$&$\tau^{z,x}$&$\tau^{z,x}$& $0$\\
Z03 & $\tau^{0}$&$\tau^{z,x}$&$\tau^{z,x}$&$\tau^{z,x}$\\
Z04 &  $\tau^{0}$&$\tau^{z,x}$&$\tau^{z,x}$&$0$\\
Z05 &$\tau^{0,y}$&$\tau^{z}$&$\tau^{z}$&$\tau^{z}$\\
Z06 & $\tau^{0,y}$&$\tau^{z}$&$\tau^{z}$&$0$\\
Z07 &$\tau^{0,y}$&$0$&$\tau^{z}$&$0$\\
Z08 & $\tau^{0,y}$&$0$&$\tau^{z}$&$\tau^{z}$\\
Z09 &$\tau^{x}$&$\tau^{0,y}$&$\tau^{z}$&$\tau^{0}$\\
Z10 & $\tau^{x}$&$\tau^{0,y}$&$\tau^{z}$&$0$\\
Z11 & $\tau^{z}$&$\tau^{0,y}$&$\tau^{z}$&$\tau^{0}$\\
Z12 & $\tau^{z}$&$\tau^{0,y}$&$\tau^{z}$&$0$\\
Z13 & $\tau^{0,y}$&$\tau^{0,y}$&$\tau^{z}$&$\tau^{0}$\\
Z14 & $\tau^{0,y}$&$\tau^{0,y}$&$\tau^{z}$&$0$\\
Z15 & $\tau^{0,y}$&$0$&$\tau^{z}$&$0$\\
Z16 &$\tau^{0,y}$&$0$&$\tau^{z}$&$\tau^{0}$\\
Z17 & $\tau^{x,y}$&$\tau^{x,y}$&$\tau^{z}$&$0$\\
Z18 &  $\tau^{x,y}$&$\tau^{x,y}$&$\tau^{z}$&$0$\\
Z19 & 
$\tau^{0}$&$\tau^{x,y}$&$\tau^{z}$&$0$\\
Z20 & 
$\tau^{0}$&$\tau^{x,y}$&$\tau^{z}$&$0$\\
Z21 &
$\tau^{z}$&$\tau^{x,y}$&$\tau^{z}$&$0$\\
Z22 &  
$\tau^{z}$&$\tau^{x,y}$&$\tau^{z}$&$0$\\
Z23 &  $\tau^{z,x}$&$\tau^{0}$&$\tau^{z}$&$\tau^{0,y}$\\
Z24 & $\tau^{z,x}$&$\tau^{0}$&$\tau^{z}$&$0$\\
Z25 & $\tau^{z,x}$&$0$&$\tau^{z}$&$0$\\
Z26 & $\tau^{z,x}$&$0$&$\tau^{z}$&$\tau^{0,y}$\\
Z27 & $\tau^{0,z}$&$\tau^{y}$&$0$&$\tau^{x}$\\
Z28 &  $\tau^{0,z}$&$\tau^{y}$&$0$&$0$\\
Z29 & $\tau^{z,x}$&$\tau^{y}$&$0$&$0$\\
Z30 & $\tau^{z,x}$&$\tau^{y}$&$0$&$0$\\
Z31 & $\tau^{0,z,x}$&$0$&$0$&$0$\\
Z32 &  $\tau^{0,z,x}$&$0$&$0$&$0$\\
Z33 & $\tau^{0,z,x}$&$0$&$0$&$0$\\
Z34 & $\tau^{0,z,x}$&$0$&$0$&$0$\\
Z35 & $\tau^{y}$&$\tau^{z,x}$&$\tau^{z,x}$&$\tau^{z,x}$\\
Z36 & $\tau^{y}$&$\tau^{z,x}$&$\tau^{z,x}$&$0$\\
		\end{tabular}
	\end{ruledtabular}
\label{table:z2_ansatz_extended}
\end{table}

\section[Mean-field {\it Ans\"atze}]{Mean-field {\textbf{Ans\"atze}}}
\label{app:ansatz}
In this section, we derive the different gauge-independent symmetric mean-field {\it Ans\"atze} from the PSG solutions obtained in the previous section. We restrict the mean-field amplitudes to square-lattice bonds [nearest neighbor (NN)] and diagonal bonds [next-nearest neighbor (NNN)].

In addition to their PSG characterization, {\it Ans\"atze} may also be distinguished through the definition of SU(2) flux operators~\cite{Wen-2002}, which are associated with closed loops emanating from a fixed base site $i$:
\begin{align}
P^{\pdagger}_{\mathcal{C}_{i}} &= u^{}_{ij} u^{}_{jk} \cdots u^{}_{l i},
\end{align}
where $j,k,l,\ldots$ label the successive sites along the contour $\mathcal{C}_{i}$. Under a local SU(2) gauge transformation \(W(i)\), the flux operator transforms as \(P_{\mathcal{C}_{i}} \rightarrow W^\dagger(i) P_{\mathcal{C}_{i}} W(i)\). As a consequence, all loop operators sharing the same base point are modified only by a common global rotation in Pauli space. This gauge-covariant behavior makes flux operators a useful diagnostic for distinguishing between different QSL {\it Ans\"atze}. For a loop consisting of $q$ links, the corresponding flux operator $P_{\mathcal{C}{i}}$ generally assumes the form
\begin{equation}
P^{\pdagger}_{\mathcal{C}_{i}}(\varphi^{\pdagger}_{\mathcal{C}_{i,l}}) \propto g^{\pdagger}_i e^{i\varphi^{}_{\mathcal{C}_{i}} \tau^z} (\tau^z)^q g_i^\dagger, \quad g^{\pdagger}_i \in \mathrm{SU(2)},
\end{equation}
where the phase $\varphi^{}_{\mathcal{C}_{i}}$ admits an interpretation as an effective magnetic flux threading the loop $\mathcal{C}_{i}$. Moreover, the invariant gauge group (IGG) associated with an {\it Ansatz} may not be immediately apparent when the {\it Ansatz} is expressed in a noncanonical gauge. In such situations, the hidden IGG corresponding to an {\it Ansatz} written in a generic gauge can be identified through the condition~\cite{Bieri-2016}
\begin{equation}
[\mathcal{G}, P_{\mathcal{C}i}] = 0,
\end{equation}
which must be satisfied for all the loop operators ${P_{\mathcal{C}_i}}$ and for all $\mathcal{G} \in \text{IGG}$.

\subsection[U(1) QSL {\it Ans\"atze}]{U(1) QSL {\textbf{Ans\"atze}}}
\label{app:ansatz_u1}
We begin by introducing the fluxes required to characterize the U(1) {\it Ans\"atze}. To this end, we first define two square plaquettes and four triangular plaquettes, as illustrated in Fig.~\ref{fig:flux_unit_cell}(a). The fluxes threading the diagonal and empty square plaquettes are denoted by $\varphi^{}_{\boxtimes}$ and $\varphi^{}_{\square}$, respectively. In addition, we define four fluxes, $\varphi^{}_{\triangle,1}$, $\varphi^{}_{\triangle,2}$, $\varphi^{}_{\triangle,3}$, and $\varphi^{}_{\triangle,4}$, which thread the four triangular plaquettes shown in Fig.~\ref{fig:flux_unit_cell}(a).

By imposing projective invariance under lattice space-group symmetries and time-reversal symmetry, we obtain a total of 20 U(1) QSL {\it Ans\"atze}, which are graphically illustrated in Fig.~\ref{fig:u1_ansatz} together with their fluxes and are labeled U01--U20. Among these, and according to our gauge choice, U01--U16 consist exclusively of real hopping amplitudes. Solid and dashed bond lines denote positive and negative hopping parameters, respectively. The remaining {\it Ans\"atze} involve generic complex hopping amplitudes and therefore require the definition of a link orientation, which we indicate by arrows. The final two {\it Ans\"atze}, namely U19 and U20, constitute special cases of a broader class of {\it Ans\"atze} given by:
\begin{align}
&u^{\pdagger}_{(X,Y,0)(X,Y,1)}= \dot\iota t^{0}_{s}\tau^0+t^{z}_{s}\tau^z,\\
&u^{\pdagger}_{(X-1,Y,1)(X,Y+1,0)}=e^{-\dot\iota X\xi^{\pdagger}_T\tau^z} (\dot\iota t^{0}_{s}\tau^0+t^{z}_{s}\tau^z),\\
&u^{\pdagger}_{(X-1,Y,1)(X,Y,0)}= \dot\iota t^{0}_{s}\tau^0+t^{z}_{s}\tau^z,\\
&u^{\pdagger}_{(X,Y,1)(X,Y+1,0)}=e^{-\dot\iota X\xi^{\pdagger}_T\tau^z} (\dot\iota t^{0}_{s}\tau^0+t^{z}_{s}\tau^z)\, .
\end{align}
The corresponding PSG is given by the second row of Table~\ref{table:u1_psg} with $n^{\pdagger}_{\sigma}=-1$. However, such a structure of the {\it Ans\"atze} is compatible with the lattice only if one imposes the constraint
\begin{equation}\label{eq:u_ansatze_row2}
\xi^{\pdagger}_{T} = \frac{m}{n}\pi, \quad \text{with } m, n \in \mathbb{Z}.
\end{equation}
In Fig.~\ref{fig:u1_ansatz}, U19 and U20 correspond to $m=0$ and $m=n$, respectively.

Within the chosen gauge, the {\it Ans\"atze} U01, U02, U05, U07, and U19 exhibit translational invariance with non-projective action ($\eta^{}_T=e^{\dot\iota n^{}_T\pi}=+1$). These can be accommodated within a unit cell matching the crystallographic lattice, namely the two-site cell depicted in Fig.~\ref{fig:flux_unit_cell}(b). By contrast, the U(1) {\it Ans\"atze} U03, U04, U06, U08, and U20 realize projective translations ($\eta^{}_T=e^{\dot\iota n^{}_T\pi}=-1$), giving rise to the following structure:
\begin{align}
&u^{\pdagger}_{(X,Y,0)(X,Y,1)}= u^{\pdagger}_{(0,0,0)(0,0,1)},\\
&u^{\pdagger}_{(X,Y,1)(X,Y+1,0)}=(-1)^{X}u^{\pdagger}_{(0,0,1)(0,1,0)},\\
&u^{\pdagger}_{(X,Y+1,0)(X-1,Y,1)}= (-1)^{X}u^{\pdagger}_{(0,1,0)(-1,0,1)},\\
&u^{\pdagger}_{(X-1,Y,1)(X,Y,0)}=u^{\pdagger}_{(-1,0,1)(0,0,0)},\\
&u^{\pdagger}_{(X,Y,0)(X,Y+1,0)}=(-1)^{X}u^{\pdagger}_{(0,0,0)(0,1,0)},\\
&u^{\pdagger}_{(X,Y,1)(X-1,Y,1)}=u^{\pdagger}_{(0,0,1)(-1,0,1)}\, .
\end{align}
Consequently, the link fields on specific bonds change sign under the translation $T_1$. This sign reversal requires the unit cell to double along $T_1$, resulting in a four-site unit cell [Fig.~\ref{fig:flux_unit_cell}(c)]. In the remaining {\it Ans\"atze}, $w^{}_{T_1}=1$ holds, and the $(i_x,i_y)$ dependence exhibits the following structure:
\begin{align}
&u^{\pdagger}_{(X,Y,0)(X,Y,1)}=g^{X+Y} u^{\pdagger}_{(0,0,0)(0,0,1)}g^{X+Y},\\
&u^{\pdagger}_{(X,Y,1)(X,Y+1,0)}=\eta_{\mathcal{T}T_2}^{Y}g^{X+Y}u^{\pdagger}_{(0,0,1)(0,1,0)}g^{X+Y},\\
&u^{\pdagger}_{(X,Y+1,0)(X-1,Y,1)}= g^{X+Y}u^{\pdagger}_{(0,1,0)(-1,0,1)}g^{X+Y},\\
&u^{\pdagger}_{(X-1,Y,1)(X,Y,0)}=\eta_{\mathcal{T}T_2}^{Y}g^{X+Y}u^{\pdagger}_{(-1,0,1)(0,0,0)}g^{X+Y},\\
&u^{\pdagger}_{(X,Y,0)(X,Y+1,0)}=\eta_{\mathcal{T}T_2}^{Y}g^{X+Y}u^{\pdagger}_{(0,0,0)(0,1,0)}g^{X+Y},\\
&u^{\pdagger}_{(X,Y,1)(X-1,Y,1)}=\eta_{\mathcal{T}T_2}^{Y}g^{X+Y}u^{\pdagger}_{(0,0,1)(-1,0,1)}g^{X+Y}\, 
\end{align}
where $g=\tau^x$ and $\eta_{\mathcal{T}T_2}=e^{\dot\iota n^{}_{\mathcal{T}T_2}}$. Therefore, for $\eta_{\mathcal{T}T_2}=+1$, the real hopping term changes sign under translations along both the $T_1$ and $T_2$ directions, leading to an {\it Ansatz} that can be realized within a four-site unit cell, as shown in Fig.~\ref{fig:flux_unit_cell}(c), with periodicity $2a\mathbf{\hat{x}}\equiv T_1T_2$ and $2a\mathbf{\hat{y}}\equiv T^{-1}_1T_2$. The U(1) {\it Ans\"atze} labeled by U09, U10, U13, U14, U17, and U18 belong to this family.

On the other hand, for $\eta_{\mathcal{T}T_2}=-1$, in addition to the real hopping term changing sign under translations along both $T_1$ and $T_2$, some bonds also experience an overall sign change under translation along $T_2$. This results in an {\it Ansatz} that can be realized within an eight-site unit cell, as shown in Fig.~\ref{fig:flux_unit_cell}(d), with periodicity $2a(\mathbf{\hat{x}-\hat{y}})\equiv 2T_1$ and $2a(\mathbf{\hat{x}+\hat{y}})\equiv 2T_2$. The U(1) {\it Ans\"atze} labeled by U11, U12, U15, and U16 belong to this family. The PSGs of all these U(1) {\it Ans\"atze} can be found in Appendix~\ref{app:psg_ansatz}.
\subsection[\texorpdfstring{$\mathbb{Z}_2$}{Z2} QSL {\it Ans\"atze}]{\texorpdfstring{$\mathbb{Z}_2$}{Z2} QSL {\textbf{Ans\"atze}}}
In this section, we tabulate all the {\it Ans\"atze} realized with IGG $\mathbb{Z}_2$. Restricting to $J_\boxtimes$ links, we obtain a total of 36 $\mathbb{Z}_2$ {\it Ans\"atze}. The structure of the {\it Ans\"atze} is given by:
\label{app:ansatz_z2}
\begin{align}
&u^{\pdagger}_{(X,Y,0)(X,Y,1)}= u^{\pdagger}_s,\\
&u^{\pdagger}_{(X,Y,1)(X,Y+1,0)}=\eta^{X}_{T}\eta^{\pdagger}_{\sigma T_1}\eta^{\pdagger}_{\sigma T_2}\bar{W}^{\pdagger}_{C_4,1} u^{\pdagger}_s,\\
&u^{\pdagger}_{(X,Y+1,0)(X-1,Y,1)}= \eta^{X}_{T}\bar{W}^{\pdagger}_{C_4,1} u^{\pdagger}_s\bar{W}^{\dagger}_{C_4,1},\\
&u^{\pdagger}_{(X-1,Y,1)(X,Y,0)}= u^{\pdagger}_s\bar{W}^{\pdagger}_{C_4,1},\\
&u^{\pdagger}_{(X,Y,0)(X,Y+1,0)}=\eta^{X}_{T}u^{\pdagger}_{\boxtimes},\\
&u^{\pdagger}_{(X,Y,1)(X-1,Y,1)}=\bar{W}^{\pdagger}_{C_4,1} u^{\pdagger}_{\boxtimes}\bar{W}^{\dagger}_{C_4,1}\,,\\
&a^{}_{\gamma}(X,Y,s)\tau^\gamma=a^{}_{\gamma}(0,0,s)\tau^\gamma\, .
\end{align}
where, in general, $u^{\pdagger}_s=\dot\iota t^{0}_{s}\tau^0+t^{3}_{z}\tau^z+\Delta^{x}_{s}\tau^x+\Delta^{y}_{s}\tau^y$ and $u^{\pdagger}_{\boxtimes}=\dot\iota t^{0}_{\boxtimes}\tau^0+t^{z}_{\boxtimes}\tau^z+\Delta^{x}_{\boxtimes}\tau^x+\Delta^{y}_{\boxtimes}\tau^y$. However, the symmetry-allowed parameters for the 36 $\mathbb{Z}_2$ {\it Ans\"atze} are listed in Table~\ref{table:z2_ansatz}. Furthermore, as a first step to identify the $\mathbb{Z}_2$ {\it Ansatz} that is adiabatically connected to QSL ground state {\it Ans\"atze} on SS and square lattice, we evaluate the symmetry-allowed mean-field parameters on the empty diagonals as well (i.e., $J_\square$ links) for this 36 $\mathbb{Z}_2$ {\it Ans\"atze}:
\begin{align}
&u^{\pdagger}_{(X,Y,0)(X+1,Y,0)}=\eta^{X}_{T}u^{\pdagger}_{\square},\\
&u^{\pdagger}_{(X,Y,1)(X,Y-1,1)}=\eta^{X+1}_{T}u^{\pdagger}_{\square}\,     
\end{align}
where, in general, $u^{\pdagger}_{\square}=\dot\iota t^{0}_{\square}\tau^0+t^{z}_{\square}\tau^z+\Delta^{x}_{\square}\tau^x+\Delta^{y}_{\square}\tau^y$ and the terms allowed on the $J_\square$ links for all the 36 $\mathbb{Z}_2$ {\it Ans\"atze} can be found from Table~\ref{table:z2_ansatz_extended}. We find the $\mathbb{Z}_2$ {\it Ansatz} denoted by Z05 is the desired one and discuss the connectivity in main article [Sec.~\ref{sec:adiabatic_connections}].

\section{Gauge-enriched symmetry relations}
\label{sec:genric_gauge_con}
The projective extensions of the algebraic relations in Eq.~\eqref{eq:id_relation} read as follows:
\begin{widetext}
\begin{subequations}
\label{eq:id_gauge_relation}
\begin{alignat}{2}
 W^{-1}_{T_1}(T^{\pdagger}_1(\mathbf{r}))W^{-1}_{T_{2}}(T^{\pdagger}_1T^{\pdagger}_2(\mathbf{r}))W^{\pdagger}_{T_1}(T^{\pdagger}_1T^{\pdagger}_2(\mathbf{r}))W^{\pdagger}_{T_{2}}(T^{\pdagger}_2(\mathbf{r}))&=e^{\dot\iota\xi^{}_{T}\tau^z}&&\mbox{ or }\,\,\eta^{\pdagger}_{T}\tau^0, \\ 
W^{\pdagger}_{C_4}(\mathbf{r})W^{-1}_{T_2}(C^{-1}_4T^{-1}_1(\mathbf{r}))W^{-1}_{C_4}(T^{-1}_1(\mathbf{r}))W^{-1}_{T_1}(\mathbf{r})&=e^{\dot\iota\xi^{}_{C_{4}T_1}\tau^z}&&\mbox{ or }\,\,\eta^{\pdagger}_{C_{4}T_1}\tau^0, \\ 
W^{\pdagger}_{C_4}(\mathbf{r})W^{\pdagger}_{T_1}(C^{-1}_4(\mathbf{r}))W^{-1}_{C_4}(T^{-1}_2(\mathbf{r}))W^{-1}_{T_2}(\mathbf{r})&=e^{\dot\iota\xi^{}_{C_{4}T_2}\tau^z}&&\mbox{ or }\,\,\eta^{\pdagger}_{C_{4}T_2}\tau^0, \\
W^{\pdagger}_{C_4}(\mathbf{r})W^{\pdagger}_{C_4}(C^3_4(\mathbf{r}))W^{\pdagger}_{C_4}(C^2_4(\mathbf{r}))W^{\pdagger}_{C_4}(C^{\pdagger}_4(\mathbf{r}))&=e^{\dot\iota\xi^{}_{C_{4}}\tau^z}&&\mbox{ or }\,\,\eta^{\pdagger}_{C_{4}}\tau^0, \\ 
W^{\pdagger}_{\sigma}(\mathbf{r})W^{-1}_{T_1}(\sigma^{-1}T^{-1}_1(\mathbf{r}))W^{-1}_{\sigma}(T^{-1}_1(\mathbf{r}))W^{-1}_{T_1}(\mathbf{r})&=e^{\dot\iota\xi^{}_{\sigma T_1}\tau^z}&&\mbox{ or }\,\,\eta^{\pdagger}_{\sigma T_1}\tau^0, \\ 
W^{\pdagger}_{\sigma}(\mathbf{r})W^{\pdagger}_{T_2}(\sigma^{-1}(\mathbf{r}))W^{-1}_{\sigma}(T^{-1}_2(\mathbf{r}))W^{-1}_{T_2}(\mathbf{r})&=e^{\dot\iota\xi^{}_{\sigma T_2}\tau^z}&&\mbox{ or }\,\,\eta^{\pdagger}_{\sigma T_2}\tau^0, \\
W^{\pdagger}_{\sigma}(\mathbf{r})W^{\pdagger}_{\sigma}(\sigma(\mathbf{r}))&=e^{\dot\iota\xi^{}_{\sigma }\tau^z}&&\mbox{ or }\,\,\eta^{\pdagger}_{\sigma}\tau^0, \\ 
W^{\pdagger}_{C_4}(\mathbf{r})W^{\pdagger}_{\sigma}(C^{-1}_4(\mathbf{r}))W^{\pdagger}_{C_4}(C^{\pdagger}_4\sigma(\mathbf{r}))W^{\pdagger}_{\sigma}(\sigma(\mathbf{r}))&=e^{\dot\iota\xi^{}_{C_{4}\sigma}\tau^z}&&\mbox{ or }\,\,\eta^{\pdagger}_{C_{4}\sigma}\tau^0, \\ 
W^{\pdagger}_\mathcal{T}(\mathbf{r})W^{\pdagger}_\mathcal{T}(\mathbf{r})&=e^{\dot\iota\xi^{}_{\mathcal{T}}\tau^z}&&\mbox{ or }\,\,\eta^{\pdagger}_{\mathcal{T}}\tau^0 , \\ 
W^{\pdagger}_\mathcal{T}(\mathbf{r}) W^{}_\mathcal{O}(\mathbf{r}) W^{-1}_\mathcal{T}(\mathcal{O}^{-1}(\mathbf{r}))  W^{-1}_\mathcal{O}(\mathbf{r})&=e^{\dot\iota\xi^{}_{\mathcal{TO}}\tau^z}&&\mbox{ or }\,\,\eta^{\pdagger}_{\mathcal{TO}}\tau^0\
.  
\end{alignat}
\end{subequations}    
\end{widetext}
Here, all the $\xi=[0,2\pi)$ and $\eta=\pm1$ appearing on the right-hand sides denote the IGG parameters for the U(1) and $\mathbb{Z}_2$ IGGs, respectively.

\section{\textbf{U(1)} PSG derivation}
\label{app:u1_psg_derivation}
We choose a structure of the U(1) PSG that leads to a form of {\it Ans\"atze} containing hopping terms only, i.e., $u^{\pdagger}_{ij}=\dot{\iota}t^0_{ij} \tau^0 + t^z_{ij} \tau^z$. Such a structure is known as the canonical gauge form. The choice of projective gauges that keeps this canonical form intact is given by:
\begin{equation}
    \label{eq:canonical_u1_gauge structure}
    W^{}_\mathcal{O}(\mathbf{r},s)=e^{\dot\iota \phi^{\pdagger}_\mathcal{O}(X,Y,s)\tau^z}(i\tau^x)^{w^{\pdagger}_\mathcal{O}},
\end{equation}
where $\mathcal{O}\in\{T_1,T_2,C_4,\sigma,\mathcal{T}\}$, and $w^{}_\mathcal{O}$ is a $\mathbb{Z}_2$ parameter taking the values 0 or 1. We now work out the possibilities for $w^{}_{\mathcal{O}}$ and $\phi_{\mathcal{O}}$ systematically, one by one, for the different symmetries. Let us begin with translations. Equation~\eqref{eq:canonical_u1_gauge structure} for $\mathcal{O}\in\{T_1,T_2\}$ can be written explicitly as:
\begin{equation}
\left.\begin{aligned}
&W^{\pdagger}_{T_1}(X,Y,s)=F^{\pdagger}_3(\phi^{\pdagger}_{T_1}(X,Y,s))(i\tau^x)^{w^{\pdagger}_{T_1}},\\
&W^{\pdagger}_{T_2}(X,Y,s)=F^{\pdagger}_3(\phi^{\pdagger}_{T_2}(X,Y,s))(i\tau^x)^{w^{\pdagger}_{T_2}}.\\
\end{aligned}\right.
\end{equation}
For the two binary parameters $w^{}_{T_{1}}$ and $w^{}_{T_{2}}$, there are four possible cases:
\begin{enumerate}
    \item $w^{\pdagger}_{T_1}=0$, $w^{\pdagger}_{T_2}=0$\,,
    \item $w^{\pdagger}_{T_{1}}=1$, $w^{\pdagger}_{T_{2}}=1$\,,
    \item $w^{\pdagger}_{T_{1}}=0$, $w^{\pdagger}_{T_{2}}=1$\,,
    \item $w^{\pdagger}_{T_{1}}=1$, $w^{\pdagger}_{T_{2}}=0$\,.
\end{enumerate}
Now, conditions~\eqref{eq:id_gauge_relation}(b,c) readily discard cases 3 and 4. Hence, we need to consider only the first two cases. We discuss these two cases separately below. Before proceeding further, we define the following notation, which will be used in the derivations below:
\begin{itemize}
    \item $\Omega^{}_i\phi^{}_\mathcal{O}(X,Y,s)\equiv\phi^{}_\mathcal{O}(X,Y,s)-\phi^{}_\mathcal{O}[T^{-1}_i(X,Y,s)]$ for $i=1,2$.
    \item $n$-parameters should be read as binary parameters taking the values $0$ or $1$.
    \item All the $p$ parameters can take the values $0,1,2,3$.
\end{itemize}

\subsection{$w^{\pdagger}_{T_1}=0$, $w^{\pdagger}_{T_2}=0$}
\label{sec:translation_case_1}
One can always choose a gauge transformation $W(X,Y,s)=e^{\dot\iota \theta(X,Y,s)\tau^z}$ to arrive at the following simplifications:
\begin{equation}
\label{eq:B4}
W^{\pdagger}_{T_1}(X,0,s)=W^{\pdagger}_{T_2}(X,Y,s)=\tau^0.\\
\end{equation}
With this gauge choice, the symmetry condition~\eqref{eq:id_gauge_relation}(a) leads to the following solutions for the projective gauges of translations:
\begin{equation}
\label{eq:tran_sol_1_1}
\left.\begin{aligned}
&W^{\pdagger}_{T_1}(X,Y,s)=e^{\dot\iota Y\,\xi^{}_T\tau^z},\\
&W^{\pdagger}_{T_2}(X,Y,s)=\tau^0.\\
\end{aligned}\right.
\end{equation}
The next step is to consider the other two wallpaper-group symmetries, $C_4$ and $\sigma$. After substituting the solutions~\eqref{eq:tran_sol_1_1}, Eqs.~\eqref{eq:id_gauge_relation}(b,c,e,f) can be recast as:
\begin{equation}
\label{eq:c4_1_1}
\left.\begin{aligned}
\Omega^{\pdagger}_1\phi^{\pdagger}_{C_4}(X,Y,s)&=\xi^{\pdagger}_{C_{4}T_1}+Y\xi^{\pdagger}_T \\
\Omega^{\pdagger}_2\phi^{\pdagger}_{C_4}(X,Y,s)&= \xi^{\pdagger}_{C_{4}T_2}+X(-1)^{w^{\pdagger}_{C_4}}\xi^{\pdagger}_T,
\end{aligned}\right.
\end{equation}
\begin{equation}
\label{eq:s_1_1}
\left.\begin{aligned}
\Omega^{\pdagger}_1\phi^{\pdagger}_{\sigma}(X,Y,s)&=\xi^{\pdagger}_{\sigma T_1}+(1+(-1)^{w^{\pdagger}_\sigma})Y\xi^{\pdagger}_T \\
\Omega^{\pdagger}_2\phi^{\pdagger}_{\sigma}(X,Y,s)&= \xi^{\pdagger}_{\sigma T_2}\,.
\end{aligned}\right.
\end{equation}
Now, all $\phi^{}_\mathcal{O}(X,Y,s)$ must satisfy the following consistency condition:
\begin{equation}
\label{eq:consistent}
\left.\begin{aligned}
\Omega^{\pdagger}_1&\phi^{\pdagger}_{\mathcal{O}}(X,Y,s)+\Omega^{\pdagger}_2\phi^{\pdagger}_{\mathcal{O}}[T^{-1}_1(X,Y,s)]\\
&=\Omega^{\pdagger}_2\phi^{\pdagger}_{\mathcal{O}}(X,Y,s)+\Omega^{\pdagger}_1\phi^{\pdagger}_{\mathcal{O}}[T^{-1}_2(X,Y,s)],\\
\end{aligned}\right.
\end{equation}
which, for $\mathcal{O}=C_4$, imposes the constraint $(1-(-1)^{w^{\pdagger}_{C_4}})\xi^{\pdagger}_T=(1+(-1)^{w^{\pdagger}_\sigma})\xi^{\pdagger}_T=0$ implying:
\begin{align}\label{eq:wc4ws_cond_1}
\xi^{\pdagger}_T&\in[0,2\pi)\text{ for }w^{\pdagger}_{C_4}=0,\,w^{\pdagger}_{\sigma}=1\notag\\
&=n^{\pdagger}_T\pi\text{ otherwise. }
\end{align}
Consequently, the solutions for $\phi^{\pdagger}_{C_4}$ and $\phi^{\pdagger}_{\sigma}$ can be obtained from relations~\eqref{eq:c4_1_1} and~\eqref{eq:s_1_1}, respectively.
\begin{align}
 \phi^{\pdagger}_{C_4}(X,Y,s)&=X\xi^{\pdagger}_{C_{4}T_1}+Y\xi^{\pdagger}_{C_{4}T_2}+XY\xi^{\pdagger}_{T}+\bar{\phi}^{\pdagger}_{C_4,s}\label{eq:c4_1_2}\\ 
\phi^{\pdagger}_{\sigma}(X,Y,s)&=X\xi^{\pdagger}_{\sigma T_1}+Y\xi^{\pdagger}_{\sigma T_2}+\bar{\phi}^{\pdagger}_{\sigma,s}\,.\label{eq:s_1_2}
\end{align}

Now, we employ a gauge transformation of the form $W(X,Y,s)=e^{\dot\iota{(X\theta_x+Y\theta_y+\theta_s)\tau^z}}$. This does not change $W_{T_1}$ and $W_{T_2}$, except for a global phase that can be set to zero using IGG freedom. However, such a transformation modifies $W_{C_4}$ and $W_{\sigma}$. With an appropriate choice of the parameters $\{\theta_x,\theta_y,\theta_s\}$, one can set:
\allowdisplaybreaks
\begin{align} &\xi^{\pdagger}_{C_{4}T_1}=\xi^{\pdagger}_{C_{4}T_2}=0\label{eq:u1_fixing_1_1} \\ &w^{\pdagger}_{C_4}=0,\,w^{\pdagger}_{\sigma}=0,1:\,\bar{\phi}^{\pdagger}_{C_4,s}=0\label{eq:u1_fixing_1_2}\\ &w^{\pdagger}_{C_4}=1,\,w^{\pdagger}_{\sigma}=0:\,\bar{\phi}^{\pdagger}_{C_4,0}=0\label{eq:u1_fixing_1_3} \\ &w^{\pdagger}_{C_4}=1,\,w^{\pdagger}_{\sigma}=1:\,\bar{\phi}^{\pdagger}_{\sigma,s}=0\,.\label{eq:u1_fixing_1_4}  
\end{align}
We are still left with three more relations,~\eqref{eq:id_gauge_relation}(d), (g), and (h), imposed by the wallpaper group, which can lead to further simplifications.
Upon substitution of Eqs.~\eqref{eq:c4_1_2}-\eqref{eq:u1_fixing_1_4} into these equations, we obtain the following choices.
\allowdisplaybreaks
\begin{align}
&w^{\pdagger}_{C_4}=0,\,w^{\pdagger}_{\sigma}=0:\notag\\
 &\xi^{\pdagger}_{\sigma T_1}=\xi^{\pdagger}_{\sigma T_2}=n^{\pdagger}_{\sigma T_1}\pi,\,\bar{\phi}^{\pdagger}_{C_4,s}=0,\\
 &\bar{\phi}^{\pdagger}_{\sigma,s}=\{0,\frac{n^{\pdagger}_{\sigma T_1}\pi}{2}+n^{\pdagger}_{\sigma,1}\pi\}\\ &w^{\pdagger}_{C_4}=0,\,w^{\pdagger}_{\sigma}=1:\notag\\
 &\xi^{\pdagger}_{\sigma T_1}=\xi^{\pdagger}_{\sigma T_2}=0,\,\bar{\phi}^{\pdagger}_{C_4,s}=0,\,\bar{\phi}^{\pdagger}_{\sigma,s}=\{0,n^{\pdagger}_{\sigma,1}\pi\}\\ 
&w^{\pdagger}_{C_4}=1,\,w^{\pdagger}_{\sigma}=0:\notag\\
 &\xi^{\pdagger}_{\sigma T_1}=\xi^{\pdagger}_{\sigma T_2}=n^{\pdagger}_{\sigma T_1}\pi,\,\bar{\phi}^{\pdagger}_{C_4,s}=\{0,\frac{p^{\pdagger}_{C_4,1}\pi}{2}\},\\
 &\bar{\phi}^{\pdagger}_{\sigma,s}=\{0,-\frac{p^{\pdagger}_{C_4,1}\pi}{2}+n^{\pdagger}_{\sigma,1}\pi\}\\  &w^{\pdagger}_{C_4}=1,\,w^{\pdagger}_{\sigma}=1:\notag\\
 &\xi^{\pdagger}_{\sigma T_1}=\xi^{\pdagger}_{\sigma T_2}=0,\,\bar{\phi}^{\pdagger}_{C_4,s}=\{0,\frac{p^{\pdagger}_{C_4,1}\pi}{2}\},\,\bar{\phi}^{\pdagger}_{\sigma,s}=0\,.
\end{align}
Now, we proceed to obtain the PSG solutions for time-reversal symmetry. From Eq.~\eqref{eq:id_gauge_relation}(j) for $\mathcal{O}\in T_i$, we have
\begin{equation}
\label{eq:time_u1_sol_1}
\left.\begin{aligned}
&\Omega_1\phi^{}_{\mathcal{T}}(X,Y,s)=\xi^{}_{\mathcal{T}T_1}+[1-(-1)^{w^{}_{\mathcal{T}}}]Y\,\,\xi^{\pdagger}_T,\\
&\Omega_2\phi^{}_{\mathcal{T}}(X,Y,s)=\xi^{}_{\mathcal{T}T_2}.\\
\end{aligned}\right.
\end{equation}
The consistency condition for $\mathcal{O}\in\mathcal{T}$ yields
\begin{equation}
    [1-(-1)^{w^{}_{\mathcal{T}}}]\xi^{\pdagger}_T=0,
\end{equation}
implying that for $w^{}_{\mathcal{T}}=1$, $2\xi^{\pdagger}_T=0$. The solution for $W_\mathcal{T}$ can be obtained from Eq.~\eqref{eq:time_u1_sol_1} as
\begin{equation}\label{eq:time_u1_sol_2}
    \phi^{}_\mathcal{T}(X,Y,s)=X\,\xi^{}_{\mathcal{T}T_1}+Y\,\xi^{}_{\mathcal{T}T_2}+\bar{\phi}^{\pdagger}_{\mathcal{T},s}\, .
\end{equation}
Now, let us consider the remaining conditions separately for $w^{\pdagger}_\mathcal{T}=0$ and $w^{\pdagger}_\mathcal{T}=1$.

For $w^{\pdagger}_\mathcal{T}=0$, Eq.~\eqref{eq:id_gauge_relation}(i) yields:
\begin{equation}\label{eq:time_u1_fixing_1}
 2\xi^{\pdagger}_{\mathcal{T}T_i}=0,\, 2\bar{\phi}^{\pdagger}_{\mathcal{T},0}=2\bar{\phi}^{\pdagger}_{\mathcal{T},1}\,. 
\end{equation}
To obtain a nonvanishing mean-field amplitude on the NN bonds, we need to choose $\bar{\phi}^{\pdagger}_{\mathcal{T},s}=\{0,\pi\}$. With this choice, using Eq.~\eqref{eq:id_gauge_relation}(j), one can conclude $\xi^{\pdagger}_{\mathcal{T}T_i}=0$ and thus,
\begin{equation}\label{eq:time_u1_sol_3}
    \phi^{}_\mathcal{T}(X,Y,s)=s\,\pi\, .
\end{equation}
For $w^{\pdagger}_\mathcal{T}=1$, we first perform a gauge transformation $W(X,Y,s)=e^{-\dot\iota\frac{\phi^{\pdagger}_{\mathcal{T}}(X,Y,s)}{2}\tau^z}$ to set
\begin{equation}
 \phi^{\pdagger}_{\mathcal{T}}=0\,.   
\end{equation}
However, such a gauge transformation also modifies $\phi^{\pdagger}_{C_4}$ and $\phi^{\pdagger}_{\sigma}$. Finally, we use Eqs.~\eqref{eq:id_gauge_relation}(j) and obtain the following.
\begin{align}
&w^{\pdagger}_{C_4}=0,\,w^{\pdagger}_{\sigma}=0:\notag\\
 &\phi^{\pdagger}_{C_4}(X,Y,s)=-X n^{\pdagger}_{\sigma T_1}\pi+XYn^{\pdagger}_{T}\pi+s\,n^{\pdagger}_{C_4}\pi,\\
 &\phi^{\pdagger}_{\sigma}(X,Y,s)=Y n^{\pdagger}_{\sigma T_1}\pi+s\,n^{\pdagger}_{\sigma}\pi,\\
 &w^{\pdagger}_{C_4}=0,\,w^{\pdagger}_{\sigma}=1:\notag\\
 &\phi^{\pdagger}_{C_4}(X,Y,s)=-X n^{\pdagger}_{\mathcal{T} T_1}\pi+XYn^{\pdagger}_{T}\pi+s\,n^{\pdagger}_{C_4}\pi,\\
 &\phi^{\pdagger}_{\sigma}(X,Y,s)=-Y n^{\pdagger}_{\mathcal{T} T_1}\pi+s\,n^{\pdagger}_{\sigma}\pi,\\
&w^{\pdagger}_{C_4}=1,\,w^{\pdagger}_{\sigma}=0:\notag\\
 &\phi^{\pdagger}_{C_4}(X,Y,s)=-Y n^{\pdagger}_{\mathcal{T} T_1}\pi+XYn^{\pdagger}_{T}\pi,\\
 &\phi^{\pdagger}_{\sigma}(X,Y,s)=(X+Y) n^{\pdagger}_{\sigma T_1}\pi-X n^{\pdagger}_{\mathcal{T} T_1}\pi+s\,n^{\pdagger}_{\sigma}\pi,\\
 &w^{\pdagger}_{C_4}=1,\,w^{\pdagger}_{\sigma}=1:\notag\\
 &\phi^{\pdagger}_{C_4}(X,Y,s)=-Y n^{\pdagger}_{\mathcal{T} T_1}\pi+XYn^{\pdagger}_{T}\pi,\\
 &\phi^{\pdagger}_{\sigma}(X,Y,s)=-Y n^{\pdagger}_{\mathcal{T}T_1}\pi+s\,n^{\pdagger}_{\sigma}\pi\,.
\end{align}
where $\xi^{\pdagger}_{\mathcal{T} T_1}=n^{\pdagger}_{\mathcal{T} T_1}\pi$.
Now perform a gauge transformation $W(X,Y,s)=e^{-\dot\iota Yn^{\pdagger}_{\sigma T_1}\pi\tau^z}$ and $W(X,Y,s)=e^{-\dot\iota Yn^{\pdagger}_{\mathcal{T} T_1}\pi\tau^z}$ for $(w^{\pdagger}_{C_4},\,w^{\pdagger}_{\sigma})=(0,0)$ and $(w^{\pdagger}_{C_4},\,w^{\pdagger}_{\sigma})=(0,1)$, respectively, to transfer the $X$ dependence of $\phi^{\pdagger}_{C_4}$ to $Y$ dependence, similar to the rest of the cases, as follows.
 \begin{align}
&w^{\pdagger}_{C_4}=0,\,w^{\pdagger}_{\sigma}=0:\notag\\
 &\phi^{\pdagger}_{C_4}(X,Y,s)=-Y n^{\pdagger}_{\sigma T_1}\pi+XYn^{\pdagger}_{T}\pi+s\,n^{\pdagger}_{C_4}\pi,\\
 &\phi^{\pdagger}_{\sigma}(X,Y,s)=Y n^{\pdagger}_{\sigma T_1}\pi+s\,n^{\pdagger}_{\sigma}\pi,\\
 &w^{\pdagger}_{C_4}=0,\,w^{\pdagger}_{\sigma}=1:\notag\\
 &\phi^{\pdagger}_{C_4}(X,Y,s)=-Y n^{\pdagger}_{\mathcal{T} T_1}\pi+XYn^{\pdagger}_{T}\pi+s\,n^{\pdagger}_{C_4}\pi,\\
 &\phi^{\pdagger}_{\sigma}(X,Y,s)=-Y n^{\pdagger}_{\mathcal{T} T_1}\pi+s\,n^{\pdagger}_{\sigma}\pi\,.
 \end{align}
\subsection{$w^{\pdagger}_{T_1}=1$, $w^{\pdagger}_{T_2}=1$}
\label{sec:translation_case_2}
In this class, the PSGs corresponding to translations have the following form:
\begin{align}
    W^{\pdagger}_{T_1}(X,Y,s)
    &=e^{\dot\iota\phi^{\pdagger}_{T_1}(X,Y,s)\tau^z}\dot\iota\tau^x,\\
    W^{\pdagger}_{T_2}(X,Y,s)
    &=e^{\dot\iota\phi^{\pdagger}_{T_2}(X,Y,s)\tau^z}\dot\iota\tau^x\,.
\end{align}
Now, similar to the class $w^{\pdagger}_{T_1}=w^{\pdagger}_{T_2}=0$, we begin with the gauge choice $\phi^{\pdagger}_{T_2}(X,Y,s)=\phi^{\pdagger}_{T_1}(X,0,s)=0$. Using this and Eq.~\eqref{eq:id_gauge_relation}(a), one can obtain the following solutions:
\begin{equation}
    \phi^{\pdagger}_{T_1}(X,Y,s) = \frac{1}{2}\bigl(1-(-1)^{y}\bigr)\xi^{\pdagger}_T\,.
\end{equation}
Now, we use a gauge transformation $W(X,Y,s)=e^{\dot\iota(-1)^y\frac{\xi^{\pdagger}_T}{4}\tau^z}$ to obtain $ \phi^{\pdagger}_{T_1}(X,Y,s) =\frac{\xi^{\pdagger}_T}{2}$. Furthermore, with the use of a global phase shift, we end up with the following solutions for the PSGs of translations:
\begin{equation}\label{eq:trans_2_1}
\phi^{\pdagger}_{T_1}(X,Y,s)=\phi^{\pdagger}_{T_2}(X,Y,s)=0\,.    
\end{equation}
Now we consider $\mathcal{O}=C_4$. After substituting Eq.~\eqref{eq:trans_2_1}, Eqs.~\eqref{eq:id_gauge_relation}(b) and (c) yield:
\begin{align}
&W^{\pdagger}_{C_4}(X,Y,s)\tau^xW^{-1}_{C_4}(X-1,Y,s)\tau^x=e^{\dot\iota\xi^{\pdagger}_{C_4T_1}\tau^z},\\
&W^{\pdagger}_{C_4}(X,Y,s)\tau^xW^{-1}_{C_4}(X,Y-1,s)\tau^x=e^{\dot\iota\xi^{\pdagger}_{C_4T_2}\tau^z}\,.
\end{align}
This can have a solution of the following form:
\begin{equation}
\label{eq:c4_2_1}
W^{\pdagger}_{C_4}(X,Y,s)=e^{\dot\iota(-1)^{X+Y}\phi^{\pdagger}_{C_4}(X,Y,s)\tau^z}(\dot\iota\tau^x)^{w^{\pdagger}_{C_4}} \,,  
\end{equation}
where $\phi^{\pdagger}_{C_4}(X,Y,s)$ satisfies:
\begin{align}
&\Omega^{\pdagger}_1\phi^{\pdagger}_{C_4}(X,Y,s)=(-1)^{m+n}\xi^{\pdagger}_{C_4T_1},\\
&\Omega^{\pdagger}_2\phi^{\pdagger}_{C_4}(X,Y,s)=(-1)^{m+n}\xi^{\pdagger}_{C_4T_2}\,.
\end{align}
Now, the consistency condition implies $2\xi^{\pdagger}_{C_4T_1}=2\xi^{\pdagger}_{C_4T_2}$ $\implies$ $\xi^{\pdagger}_{C_4T_1}=2\xi^{\pdagger}_{C_4T}$ (say) and $\xi^{\pdagger}_{C_4T_2}=2\xi^{\pdagger}_{C_4T}+n^{\pdagger}_{C_4T}\pi$. Consequently, the solutions, by considering $e^{\dot\iota(-1)^{X+Y}n^{\pdagger}_{C_4T_i}\pi\tau^z}=e^{\dot\iota n^{\pdagger}_{C_4T_i}\pi\tau^z}$, read as:
\begin{align}
\phi^{\pdagger}_{C_4}(X,Y,s)=&(-1)^{X+Y}\xi^{\pdagger}_{C_4T}+ Y n^{\pdagger}_{C_4T}\pi+\bar{\phi}^{\pdagger}_{C_4,s}\,.  
\end{align}
When this is substituted into Eq.~\eqref{eq:c4_2_1}, the parameter $\xi^{\pdagger}_{C_4T}$ appears only as a constant phase, which can be dropped using IGG freedom, i.e.,
\begin{equation}\label{eq:c4_2_2}
\phi^{\pdagger}_{C_4}(X,Y,s)=Y n^{\pdagger}_{C_4T}\pi+\bar{\phi}^{\pdagger}_{C_4,s}\,.      
\end{equation}
Similarly, the solution for $\mathcal{O}=\sigma$ reads as:
\begin{equation}\label{eq:s_2_1}
W^{\pdagger}_{\sigma}(X,Y,s)=e^{\dot\iota(-1)^{X+Y}\phi^{\pdagger}_{\sigma}(X,Y,s)\tau^z}(\dot\iota\tau^x)^{w^{\pdagger}_{\sigma}} \,,  
\end{equation}
where
\begin{equation}\label{eq:s_2_2}
\phi^{\pdagger}_{\sigma}(X,Y,s)= Y n^{\pdagger}_{\sigma T}\pi+\bar{\phi}^{\pdagger}_{\sigma,s}\,.  
\end{equation}

Now, we apply a gauge transformation of the form $W(X,Y,s)=e^{\dot\iota(-1)^{X+Y}\theta\tau^z}$ that does not affect the solutions in Eq.~\eqref{eq:trans_2_1}. However, one can choose $\theta$ such that we can set:
\begin{align}
&w^{\pdagger}_{\sigma}=0:\bar{\phi}^{\pdagger}_{\sigma,1}=0,\,\\
&w^{\pdagger}_{\sigma}=1:\bar{\phi}^{\pdagger}_{\sigma,0}=0\,.\label{eq:u1_fixing_2_1}
\end{align}

Furthermore, one can exploit the remaining relations~\eqref{eq:id_gauge_relation}(d), (g) and (h) imposed by the wallpaper group, which can offer further simplifications. After substituting Eqs.~\eqref{eq:c4_2_2}-\eqref{eq:u1_fixing_2_1} into these equations, we arrive at the following choices.
\allowdisplaybreaks
\begin{align}
&w^{\pdagger}_{C_4}=0,\,w^{\pdagger}_{\sigma}=0:\notag\\
 &\bar{\phi}^{\pdagger}_{\sigma,s}=\{n^{\pdagger}_{\sigma}\pi,0\},\\
 &\bar{\phi}^{\pdagger}_{C_4,s}=\{\bar{\phi}^{\pdagger}_{C_4,0},\frac{n\pi}{2}+n^{\prime}\pi-\bar{\phi}^{\pdagger}_{C_4,0}\},\\
&w^{\pdagger}_{C_4}=0,\,w^{\pdagger}_{\sigma}=1:\notag\\
 &\bar{\phi}^{\pdagger}_{\sigma,s}=\{0,n^{\pdagger}_{\sigma}\pi\},\,n^{\pdagger}_{\sigma T}=n^{\pdagger}_{C_4 T},\\
 &\bar{\phi}^{\pdagger}_{C_4,s}=\{\bar{\phi}^{\pdagger}_{C_4,0},n^{\prime}\pi+\bar{\phi}^{\pdagger}_{C_4,0}\},\\
&w^{\pdagger}_{C_4}=1,\,w^{\pdagger}_{\sigma}=0:\notag\\
  &\bar{\phi}^{\pdagger}_{\sigma,s}=\{n^{\pdagger}_{\sigma}\pi,0\},\,n^{\pdagger}_{\sigma T}=n^{\pdagger}_{C_4 T},\\
 &\bar{\phi}^{\pdagger}_{C_4,s}=\{\bar{\phi}^{\pdagger}_{C_4,0},n^{\prime}\pi+\bar{\phi}^{\pdagger}_{C_4,0}\},\\
 &w^{\pdagger}_{C_4}=1,\,w^{\pdagger}_{\sigma}=1:\notag\\
 &\bar{\phi}^{\pdagger}_{\sigma,s}=\{0,n^{\pdagger}_{\sigma}\pi\},\\
 &\bar{\phi}^{\pdagger}_{C_4,s}=\{\bar{\phi}^{\pdagger}_{C_4,0},\frac{n\pi}{2}+n^{\prime}\pi-\bar{\phi}^{\pdagger}_{C_4,0}\},
\end{align}
where we have defined $n\pi=(n^{\pdagger}_{\sigma T}+n^{\pdagger}_{C_4 T})\pi$. Furthermore, we perform a gauge transformation $W(X,Y,s)=e^{\dot\iota(-1)^{X+Y}(X+Y)n^\prime\pi\tau^z}$ to set $n^\prime=0$ without affecting anything else. In addition, one can use a global sign shift to set $\bar{\phi}^{\pdagger}_{\sigma,s}=\{0,n^{\pdagger}_{\sigma}\pi\}$ for all the above cases. Now, we consider Eq.~\eqref{eq:id_gauge_relation}(j) for $\mathcal{O}=T_i$ to obtain the following solution for the projective representation of time-reversal symmetry:
\begin{equation}\label{eq:t_2_1}
W^{\pdagger}_{\mathcal{T}}(X,Y,s)=e^{\dot\iota(-1)^{X+Y}\phi^{\pdagger}_{\mathcal{T}}(X,Y,s)\tau^z}(\dot\iota\tau^x)^{w^{\pdagger}_{\mathcal{T}}} \,,  
\end{equation}
where
\begin{equation}\label{eq:t_2_2}
\phi^{\pdagger}_{\mathcal{T}}(X,Y,s)= Y n^{\pdagger}_{\mathcal{T} T}\pi+\bar{\phi}^{\pdagger}_{\mathcal{T},s}\,.  
\end{equation}
Next, one can use Eq.~\eqref{eq:id_gauge_relation}(i) and Eq.~\eqref{eq:id_gauge_relation}(j) for $\mathcal{O}=C_4,\sigma$ to obtain $n^{\pdagger}_{\mathcal{T} T}=0$ and $\bar{\phi}^{\pdagger}_{\mathcal{T},s}=\{0, n^{\pdagger}_{\mathcal{T}}\pi\}$ for $w^{\pdagger}_\mathcal{T}=0$. Now, nonvanishing mean-field parameters on NN bonds require $n^{\pdagger}_{\mathcal{T}}=1$ and thus, for $w^{\pdagger}_\mathcal{T}=0$:
\begin{equation}
\phi^{\pdagger}_{\mathcal{T}}(X,Y,s)= s\, \pi\,.
\end{equation}
For $w^{\pdagger}_\mathcal{T}=1$, we perform a gauge transformation $W(X,Y,s)=e^{\dot\iota(-1)^{X+Y+1}\frac{\bar{\phi}^{\pdagger}_{\mathcal{T},s}}{2}\tau^z}$ to set
\begin{equation}\label{eq:t_2_2_1}
\phi^{\pdagger}_{\mathcal{T}}(X,Y,s)=Y n^{\pdagger}_{\mathcal{T} T}\pi\,.  
\end{equation}
However, this modifies $\phi^{\pdagger}_{C_4}$ and $\phi^{\pdagger}_{\sigma}$ as follows.
\begin{align}
&w^{\pdagger}_{C_4}=0,\,w^{\pdagger}_{\sigma}=0:\notag\\
 &\bar{\phi}^{\pdagger}_{\sigma,s}=\{0,\bar{\phi}^{\pdagger}_{\sigma}+n^{\pdagger}_{\sigma}\pi\},\\
 &\bar{\phi}^{\pdagger}_{C_4,s}=\{\bar{\phi}^{\pdagger}_{C_4,0},\frac{n\pi}{2}+\bar{\phi}^{\pdagger}_{\sigma}-\bar{\phi}^{\pdagger}_{C_4,0}\},\\
&w^{\pdagger}_{C_4}=0,\,w^{\pdagger}_{\sigma}=1:\notag\\
 &\bar{\phi}^{\pdagger}_{\sigma,s}=\{\bar{\phi}^{\pdagger}_{\sigma},n^{\pdagger}_{\sigma}\pi\},\,n^{\pdagger}_{\sigma T}=n^{\pdagger}_{C_4 T},\\
 &\bar{\phi}^{\pdagger}_{C_4,s}=\{\bar{\phi}^{\pdagger}_{C_4,0},\bar{\phi}^{\pdagger}_{C_4,0}-\bar{\phi}^{\pdagger}_{\sigma}\},\\
&w^{\pdagger}_{C_4}=1,\,w^{\pdagger}_{\sigma}=0:\notag\\
 &\bar{\phi}^{\pdagger}_{\sigma,s}=\{0,\bar{\phi}^{\pdagger}_{\sigma}+n^{\pdagger}_{\sigma}\pi\},\,n^{\pdagger}_{\sigma T}=n^{\pdagger}_{C_4 T},\\
 &\bar{\phi}^{\pdagger}_{C_4,s}=\{\bar{\phi}^{\pdagger}_{C_4,0},\bar{\phi}^{\pdagger}_{C_4,0}+\bar{\phi}^{\pdagger}_{\sigma}\},\\
 &w^{\pdagger}_{C_4}=1,\,w^{\pdagger}_{\sigma}=1:\notag\\
  &\bar{\phi}^{\pdagger}_{\sigma,s}=\{\bar{\phi}^{\pdagger}_{\sigma},n^{\pdagger}_{\sigma}\pi\},\\
 &\bar{\phi}^{\pdagger}_{C_4,s}=\{\bar{\phi}^{\pdagger}_{C_4,0},\frac{n\pi}{2}+\bar{\phi}^{\pdagger}_{\sigma}-\bar{\phi}^{\pdagger}_{C_4,0}\},
\end{align}
where we define the parameters as follows:
\begin{align}
&w^{\pdagger}_{C_4}=0,\,w^{\pdagger}_{\sigma}=0:\notag\\
&\bar{\phi}^{\pdagger}_{\mathcal{T},1}\rightarrow-\bar{\phi}^{\pdagger}_{\sigma},\,\bar{\phi}^{\pdagger}_{C_4,0}-\frac{\bar{\phi}^{\pdagger}_{\mathcal{T},0}}{2}-\frac{\bar{\phi}^{\pdagger}_{\mathcal{T},1}}{2}\rightarrow\bar{\phi}^{\pdagger}_{C_4,0},\\
&w^{\pdagger}_{C_4}=0,\,w^{\pdagger}_{\sigma}=1:\notag\\
&\bar{\phi}^{\pdagger}_{\mathcal{T},0}\rightarrow-\bar{\phi}^{\pdagger}_{\sigma},\,\bar{\phi}^{\pdagger}_{C_4,0}-\frac{\bar{\phi}^{\pdagger}_{\mathcal{T},0}}{2}-\frac{\bar{\phi}^{\pdagger}_{\mathcal{T},1}}{2}\rightarrow\bar{\phi}^{\pdagger}_{C_4,0},\\
&w^{\pdagger}_{C_4}=1,\,w^{\pdagger}_{\sigma}=0:\notag\\
&\bar{\phi}^{\pdagger}_{\mathcal{T},1}\rightarrow-\bar{\phi}^{\pdagger}_{\sigma},\,\bar{\phi}^{\pdagger}_{C_4,0}-\frac{\bar{\phi}^{\pdagger}_{\mathcal{T},0}}{2}+\frac{\bar{\phi}^{\pdagger}_{\mathcal{T},1}}{2}\rightarrow\bar{\phi}^{\pdagger}_{C_4,0},\\
&w^{\pdagger}_{C_4}=1,\,w^{\pdagger}_{\sigma}=1:\notag\\
&\bar{\phi}^{\pdagger}_{\mathcal{T},0}\rightarrow-\bar{\phi}^{\pdagger}_{\sigma},\,\bar{\phi}^{\pdagger}_{C_4,0}-\frac{\bar{\phi}^{\pdagger}_{\mathcal{T},0}}{2}+\frac{\bar{\phi}^{\pdagger}_{\mathcal{T},1}}{2}\rightarrow\bar{\phi}^{\pdagger}_{C_4,0}\,.
\end{align}
Furthermore, we use Eq.~\eqref{eq:id_gauge_relation}(i) and Eq.~\eqref{eq:id_gauge_relation}(j) for $\mathcal{O}=C_4,\sigma$ to summarize the following:
\begin{align}
&w^{\pdagger}_{C_4}=w^{\pdagger}_{\sigma}:\notag\\
 &\bar{\phi}^{\pdagger}_{\sigma,s}=\{0,n^{\pdagger}_{\sigma}\pi\},\,n^{\pdagger}_{\mathcal{T}T}=n,\\
 &\bar{\phi}^{\pdagger}_{C_4,s}=\frac{n\pi}{4}+\frac{p^{\pdagger}_{C_4}\pi}{2}+\{0,n^{\pdagger}_{C_4}\pi\},\\
&w^{\pdagger}_{C_4}\neq w^{\pdagger}_{\sigma}:\notag\\
 &\bar{\phi}^{\pdagger}_{\sigma,s}=\{0,n^{\pdagger}_{\sigma}\pi\},\,n^{\pdagger}_{\sigma T}=n^{\pdagger}_{C_4 T},\\
 &\bar{\phi}^{\pdagger}_{C_4,s}=\frac{n^{\pdagger}_{\mathcal{T}T}\pi}{4}+\frac{p^{\pdagger}_{C_4}\pi}{2}+\{0,n^{\pdagger}_{C_4}\pi\}\,.
\end{align}
Finally, we perform a gauge transformation $W(X,Y,s)=e^{\dot\iota(-1)^{X+Y+1}\frac{n^{\pdagger}_{\mathcal{T}T}}{2}\tau^z}$ to set $\phi^{\pdagger}_{\mathcal{T}}(X,Y,s)=0$. However, $\phi^{\pdagger}_{T_2}$, $\phi^{\pdagger}_{C_4}$, and $\phi^{\pdagger}_{\sigma}$ are also modified as follows:
\begin{align}
 &\phi^{\pdagger}_{T_2}(X,Y,s)\rightarrow\phi^{\pdagger}_{T_2}(X,Y,s)-\frac{n^{\pdagger}_{\mathcal{T}T}\pi}{2},\\
 &\phi^{\pdagger}_{C_4}(X,Y,s)\rightarrow\phi^{\pdagger}_{C_4}(X,Y,s)+((-1)^{w^{\pdagger}_{C_4}+u}X-Y)\frac{n^{\pdagger}_{\mathcal{T}T}\pi}{2},\\
 &\phi^{\pdagger}_{\sigma}(X,Y,s)\rightarrow\phi^{\pdagger}_{\sigma}(X,Y,s)+((-1)^{w^{\pdagger}_{\sigma}+u}-1)\frac{Yn^{\pdagger}_{\mathcal{T}T}\pi}{2}\,.
\end{align}

\section[PSGs of the {\it Ans\"atze} in Appendix~\ref{app:ansatz_u1}]{PSGs of the {\textbf{Ans\"atze}} in Appendix~\ref{app:ansatz_u1}}
\label{app:psg_ansatz}
\begin{itemize}
\item U01: $(w^{\pdagger}_{T_1},w^{\pdagger}_{\mathcal{T}},w^{\pdagger}_{C_4},w^{\pdagger}_{\sigma})=(0100)$, $n^{\pdagger}_{T}=0$, $n^{\pdagger}_{\sigma T_2}=0$, $n^{\pdagger}_{\sigma}=n^{\pdagger}_{C_4}=0,1$.
\item U02: $(w^{\pdagger}_{T_1},w^{\pdagger}_{\mathcal{T}},w^{\pdagger}_{C_4},w^{\pdagger}_{\sigma})=(0100)$, $n^{\pdagger}_{T}=0$, $n^{\pdagger}_{\sigma T_2}=1$, $n^{\pdagger}_{\sigma}=\bar{n}^{\pdagger}_{C_4}=0,1$.
\item U03: $(w^{\pdagger}_{T_1},w^{\pdagger}_{\mathcal{T}},w^{\pdagger}_{C_4},w^{\pdagger}_{\sigma})=(0100)$, $n^{\pdagger}_{T}=1$, $n^{\pdagger}_{\sigma T_2}=0$, $n^{\pdagger}_{\sigma}=n^{\pdagger}_{C_4}=0,1$.
\item U04: $(w^{\pdagger}_{T_1},w^{\pdagger}_{\mathcal{T}},w^{\pdagger}_{C_4},w^{\pdagger}_{\sigma})=(0100)$, $n^{\pdagger}_{T}=1$, $n^{\pdagger}_{\sigma T_2}=1$, $n^{\pdagger}_{\sigma}=\bar{n}^{\pdagger}_{C_4}=0,1$.
\item U05: $(w^{\pdagger}_{T_1},w^{\pdagger}_{\mathcal{T}},w^{\pdagger}_{C_4},w^{\pdagger}_{\sigma})=(0110)$, $n^{\pdagger}_{T}=0$, $n^{\pdagger}_{\sigma T_1}=n^{\pdagger}_{\sigma T_2}=0$, $n^{\pdagger}_{\sigma}=1$.
\item U06: $(w^{\pdagger}_{T_1},w^{\pdagger}_{\mathcal{T}},w^{\pdagger}_{C_4},w^{\pdagger}_{\sigma})=(0110)$, $n^{\pdagger}_{T}=1$, $n^{\pdagger}_{\sigma T_2}={n}^{\pdagger}_{\sigma T_2}=0$, $n^{\pdagger}_{\sigma}=1$.
\item U07: $(w^{\pdagger}_{T_1},w^{\pdagger}_{\mathcal{T}},w^{\pdagger}_{C_4},w^{\pdagger}_{\sigma})=(0110)$, $n^{\pdagger}_{T}=0$, $n^{\pdagger}_{T}=1$, $n^{\pdagger}_{\sigma T_1}=\bar{n}^{\pdagger}_{\sigma T_2}=0$, $n^{\pdagger}_{\sigma}=0$.
\item U08: $(w^{\pdagger}_{T_1},w^{\pdagger}_{\mathcal{T}},w^{\pdagger}_{C_4},w^{\pdagger}_{\sigma})=(0110)$, $n^{\pdagger}_{T}=1$, $n^{\pdagger}_{\sigma T_1}=\bar{n}^{\pdagger}_{\sigma T_2}=0$, $n^{\pdagger}_{\sigma}=0$.
\item U09: $(w^{\pdagger}_{T_1},w^{\pdagger}_{\mathcal{T}},w^{\pdagger}_{C_4},w^{\pdagger}_{\sigma})=(1100)$, $n^{\pdagger}_{\mathcal{T}T_2}=0$, ${n}^{\pdagger}_{\sigma T_2}=0$, $p^{\pdagger}_{C_4}=0,2$, $n^{\pdagger}_{C_4}=n^{\pdagger}_{\sigma}=0,1$.
\item U10: $(w^{\pdagger}_{T_1},w^{\pdagger}_{\mathcal{T}},w^{\pdagger}_{C_4},w^{\pdagger}_{\sigma})=(1100)$, $n^{\pdagger}_{\mathcal{T}T_2}=0$, ${n}^{\pdagger}_{\sigma T_2}=0$, $p^{\pdagger}_{C_4}=1,3$, $\bar{n}^{\pdagger}_{C_4}=n^{\pdagger}_{\sigma}=0,1$.
\item U11: $(w^{\pdagger}_{T_1},w^{\pdagger}_{\mathcal{T}},w^{\pdagger}_{C_4},w^{\pdagger}_{\sigma})=(1100)$, $n^{\pdagger}_{\mathcal{T}T_2}=1$, ${n}^{\pdagger}_{\sigma T_2}=0$, $p^{\pdagger}_{C_4}=1,3$, $n^{\pdagger}_{C_4}=n^{\pdagger}_{\sigma}=0,1$.
\item U12: $(w^{\pdagger}_{T_1},w^{\pdagger}_{\mathcal{T}},w^{\pdagger}_{C_4},w^{\pdagger}_{\sigma})=(1100)$, $n^{\pdagger}_{\mathcal{T}T_2}=1$, ${n}^{\pdagger}_{\sigma T_2}=0$, $p^{\pdagger}_{C_4}=0,2$, $\bar{n}^{\pdagger}_{C_4}=n^{\pdagger}_{\sigma}=0,1$.
\item U13: $(w^{\pdagger}_{T_1},w^{\pdagger}_{\mathcal{T}},w^{\pdagger}_{C_4},w^{\pdagger}_{\sigma})=(1110)$, $n^{\pdagger}_{\mathcal{T}T_2}=0$, ${n}^{\pdagger}_{\sigma T_2}=0$, $p^{\pdagger}_{C_4}=1,3$, $n^{\pdagger}_{C_4}=n^{\pdagger}_{\sigma}=0,1$.
\item U14: $(w^{\pdagger}_{T_1},w^{\pdagger}_{\mathcal{T}},w^{\pdagger}_{C_4},w^{\pdagger}_{\sigma})=(1110)$, $n^{\pdagger}_{\mathcal{T}T_2}=0$, ${n}^{\pdagger}_{\sigma T_2}=0$, $p^{\pdagger}_{C_4}=0,2$, $\bar{n}^{\pdagger}_{C_4}=n^{\pdagger}_{\sigma}=0,1$.
\item U15: $(w^{\pdagger}_{T_1},w^{\pdagger}_{\mathcal{T}},w^{\pdagger}_{C_4},w^{\pdagger}_{\sigma})=(1110)$, $n^{\pdagger}_{\mathcal{T}T_2}=1$, ${n}^{\pdagger}_{\sigma T_2}=0$, $p^{\pdagger}_{C_4}=0,2$, $n^{\pdagger}_{C_4}=n^{\pdagger}_{\sigma}=0,1$.
\item U16: $(w^{\pdagger}_{T_1},w^{\pdagger}_{\mathcal{T}},w^{\pdagger}_{C_4},w^{\pdagger}_{\sigma})=(1110)$, $n^{\pdagger}_{\mathcal{T}T_2}=1$, ${n}^{\pdagger}_{\sigma T_2}=0$, $p^{\pdagger}_{C_4}=1,3$, $\bar{n}^{\pdagger}_{C_4}=n^{\pdagger}_{\sigma}=0,1$.
\item U17: $(w^{\pdagger}_{T_1},w^{\pdagger}_{\mathcal{T}},w^{\pdagger}_{C_4},w^{\pdagger}_{\sigma})=(1000)$, ${n}^{\pdagger}_{C_4T_2}=0,1$, ${n}^{\pdagger}_{\sigma T_2}=0$, $p^{\pdagger}_{C_4}=1,3$, $n^{\pdagger}_{\sigma}=0,1$.
\begin{equation}
u^{\pdagger}_{s}=e^{\dot\iota(\bar{\phi}^{\pdagger}_{C_4}-\frac{(3{n}^{\pdagger}_{C_4T_2}+{n}^{\pdagger}_{\sigma T_2}-2n^{\pdagger}_{\sigma})\pi}{2})}u^{\dagger}_{s} 
\end{equation}
\item U18: $(w^{\pdagger}_{T_1},w^{\pdagger}_{\mathcal{T}},w^{\pdagger}_{C_4},w^{\pdagger}_{\sigma})=(1000)$, ${n}^{\pdagger}_{C_4T_2}=0,1$, ${n}^{\pdagger}_{\sigma T_2}=1$, $p^{\pdagger}_{C_4}=1,3$, $n^{\pdagger}_{\sigma}=0,1$.
\begin{equation}
u^{\pdagger}_{s}=e^{\dot\iota(\bar{\phi}^{\pdagger}_{C_4}-\frac{(3{n}^{\pdagger}_{C_4T_2}+{n}^{\pdagger}_{\sigma T_2}-2n^{\pdagger}_{\sigma})\pi}{2})}u^{\dagger}_{s} 
\end{equation}
\item U19: $(w^{\pdagger}_{T_1},w^{\pdagger}_{\mathcal{T}},w^{\pdagger}_{C_4},w^{\pdagger}_{\sigma})=(0001)$, $\xi^{\pdagger}_{T}=0$,$n^{\pdagger}_{\sigma}=1$.
\item U20: $(w^{\pdagger}_{T_1},w^{\pdagger}_{\mathcal{T}},w^{\pdagger}_{C_4},w^{\pdagger}_{\sigma})=(0001)$, $\xi^{\pdagger}_{T}=\pi$,$n^{\pdagger}_{\sigma}=1$.
\end{itemize}

\section{\texorpdfstring{$\mathbb{Z}_2$}{Z2} PSG derivation}
\label{app:z2_psg_derivation}
Before proceeding with the derivation of the $\mathbb{Z}_2$ PSGs, we introduce the following notation:
\begin{itemize}
    \item We define the sublattice-dependent part of the PSGs as $\bar{W}^{\pdagger}_{\mathcal{O},s}=W^{\pdagger}_{\mathcal{O}}(0,0,s)$\,. 
    \item Any $\eta$-parameter denotes a sign parameter, i.e., $\eta=\pm1$\,.
\end{itemize}
As in the U(1) case, we begin with the gauge choice given by Eq.~\eqref{eq:B4}. Using condition~\eqref{eq:id_gauge_relation}(a), we obtain the following solutions for the projective actions of translations:
\begin{equation}
\label{eq:trans_sol_z2}
W^{\pdagger}_{T_1}(X,Y,s)=\eta^{\pdagger}_{T}\tau^0,\quad W^{\pdagger}_{T_2}(X,Y,s)=\tau^0.\\
\end{equation}
Now, we substitute these solutions into Eqs.~\eqref{eq:id_gauge_relation}(b),(c), which can be recast as:
\begin{align}
&W^{\pdagger}_{C_4}(X,Y,s)=\eta^{\pdagger}_{C_4T_1}\eta^{Y}_{T}W^{\pdagger}_{C_4}(X-1,Y,s),\\
&W^{\pdagger}_{C_4}(X,Y,s)=\eta^{\pdagger}_{C_4T_2}\eta^{X}_{T}W^{\pdagger}_{C_4}(X,Y-1,s),
\end{align}
whose solution can be written as:
\begin{equation}
    \label{eq:c4_sol_z2}
W^{\pdagger}_{C_4}(X,Y,s)=\eta^{X}_{C_4T_1}\eta^{Y}_{C_4T_2}\eta^{XY}_{T}\bar{W}^{\pdagger}_{C_4,s}\,.    
\end{equation}
Similarly, Eqs.~\eqref{eq:id_gauge_relation}(e),(f) can be recast as:
\begin{align}
&W^{\pdagger}_{\sigma}(X,Y,s)=\eta^{\pdagger}_{\sigma T_1}W^{\pdagger}_{\sigma}(X-1,Y,s),\\
&W^{\pdagger}_{\sigma}(X,Y,s)=\eta^{\pdagger}_{\sigma T_2}W^{\pdagger}_{\sigma}(X,Y-1,s),
\end{align}
yielding the following solution for $W^{\pdagger}_{\sigma}$:
\begin{equation}
    \label{eq:sigma_sol_z2}
W^{\pdagger}_{\sigma}(X,Y,s)=\eta^{X}_{\sigma T_1}\eta^{Y}_{\sigma T_2}\bar{W}^{\pdagger}_{\sigma,s}\,.
\end{equation}
Now, as in the U(1) case, we proceed to simplify the above PSG solutions by an appropriate choice of gauge transformations of the form $W(X,Y,s)=\eta^X_x\eta^Y_y\bar{W}^{\pdagger}_{s}$, which does not affect the projective action of translations except for a global sign parameter. However, such a gauge transformation modifies $W^{\pdagger}_{C_4}$ and $W^{\pdagger}_{\sigma}$. With appropriate choices of $\eta^{\pdagger}_{\sigma T_i}$ and $\bar{W}^{\pdagger}_{s}$, one can set:
\begin{equation}\label{eq:local_fixing_z1}
 \eta^{\pdagger}_{C_4T_1}  = +1,\, \bar{W}^{\pdagger}_{C_4,0}=\tau^0,\,\bar{W}^{\pdagger}_{\sigma,0}=e^{\dot\iota\xi^{\pdagger}_{\sigma,0}\tau^z}\,.
\end{equation}
Now, Eq.~\eqref{eq:id_gauge_relation}(d), after using the gauge fixing given by Eq.~\eqref{eq:local_fixing_z1}, yields:
\begin{align}
\eta^{\pdagger}_{C_4}=&\eta^{\pdagger}_{C_4T_2}\bar{W}^{2}_{C_4,1}\notag\\
\Rightarrow &\bar{W}^{\pdagger}_{C_4,1}=\pm\tau^0\,\text{for }\eta^{\pdagger}_{C_4}\eta^{\pdagger}_{C_4T_2}=+1\, ,\label{eq:z2_c4_choice_1}\\
&\bar{W}^{\pdagger}_{C_4,1}=\dot\iota\vec{\alpha}_1\cdot\vec{\tau}\,\text{for }\eta^{\pdagger}_{C_4}\eta^{\pdagger}_{C_4T_2}=-1\, , \label{eq:z2_c4_choice_2}
\end{align}
where $\vec{\alpha}_1\in\mathbb{S}^2$. Similarly, Eq.~\eqref{eq:id_gauge_relation}(g) gives
\begin{align}
\eta^{\pdagger}_{\sigma}=&\bar{W}^{2}_{\sigma,0}=\eta^{\pdagger}_{\sigma T_1}\bar{W}^{2}_{\sigma,1}\notag\\
\Rightarrow &\bar{W}^{\pdagger}_{\sigma,s}=\{\tau^0,\pm\tau^0\}\,\text{for }\eta^{\pdagger}_{\sigma}=\eta^{\pdagger}_{\sigma T_1}=+1\, ,\label{eq:z2_sigma_choice_1}\\
 &\bar{W}^{\pdagger}_{\sigma,s}=\{\tau^0,\dot\iota \vec{\alpha}_2\cdot\vec{\tau}\}\,\text{for }\eta^{\pdagger}_{\sigma}=-\eta^{\pdagger}_{\sigma T_1}=+1\, ,\label{eq:z2_sigma_choice_2}\\
&\bar{W}^{\pdagger}_{\sigma,s}=\{\dot\iota\tau^z,\pm\tau^0\}\,\text{for }\eta^{\pdagger}_{\sigma}=\eta^{\pdagger}_{\sigma T_1}=-1\, ,\label{eq:z2_sigma_choice_3}\\
&\bar{W}^{\pdagger}_{\sigma,s}=\{\dot\iota\tau^z,\dot\iota \vec{\alpha}_2\cdot\vec{\tau}\}\,\text{for }\eta^{\pdagger}_{\sigma}=-\eta^{\pdagger}_{\sigma T_1}=-1\, ,\label{eq:z2_sigma_choice_4}
\end{align}
where $\vec{\alpha}_2\in\mathbb{S}^2$. Furthermore, we are left with Eq.~\eqref{eq:id_gauge_relation}(h), which imposes the following constraints:
\begin{align}
&\eta^{\pdagger}_{C_4T_2}=\eta^{\pdagger}_{\sigma T_1}\eta^{\pdagger}_{\sigma T_2}\,,\label{eq:z2_c4s_choice_1}\\
&\eta^{\pdagger}_{C_4\sigma}=\eta^{\pdagger}_{\sigma T_1}\bar{W}^{\pdagger}_{C_4,s}\bar{W}^{\pdagger}_{\sigma,\bar{s}}\bar{W}^{\pdagger}_{C_4,\bar{s}}\bar{W}^{\pdagger}_{\sigma,s}\,.\label{eq:z2_c4s_choice_2}
\end{align}
Here $\bar{s}=(1+s)\mod 2$. The last condition imposes a further constraint on the choices~\eqref{eq:z2_c4_choice_1}-\eqref{eq:z2_sigma_choice_4}, and the resulting solutions for $\bar{W}^{\pdagger}_{C_4,s},\bar{W}^{\pdagger}_{\sigma,s}$ are summarized as follows.
\begin{align}
& \eta^{\pdagger}_{C_4}\eta^{\pdagger}_{C_4T_2}=+1,\,\eta^{\pdagger}_{\sigma}=\eta^{\pdagger}_{\sigma T_1}=+1:\notag\\
&\bar{W}^{\pdagger}_{C_4,s}=\{\tau^0,\eta^{\pdagger}_{C_4,1}\tau^0\},\,\bar{W}^{\pdagger}_{\sigma,s}=\{\tau^0,\eta^{\pdagger}_{\sigma,1}\tau^0\}\,,\label{eq:z2_fixing_cs_1}\\
& \eta^{\pdagger}_{C_4}\eta^{\pdagger}_{C_4T_2}=+1,\,\eta^{\pdagger}_{\sigma}=-\eta^{\pdagger}_{\sigma T_1}=-1:\notag\\
&\bar{W}^{\pdagger}_{C_4,s}=\{\tau^0,\eta^{\pdagger}_{C_4,1}\tau^0\},\,\bar{W}^{\pdagger}_{\sigma,s}=\{\dot\iota\tau^z,\eta^{\pdagger}_{\sigma,1}\dot\iota\tau^z\}\,,\label{eq:z2_fixing_cs_2}\\
& \eta^{\pdagger}_{C_4}\eta^{\pdagger}_{C_4T_2}=-1,\,\eta^{\pdagger}_{\sigma}=-\eta^{\pdagger}_{\sigma T_1}=+1:\notag\\
&\bar{W}^{\pdagger}_{C_4,s}=\{\tau^0,\eta^{\pdagger}_{C_4,1}\dot\iota\tau^z\},\,\bar{W}^{\pdagger}_{\sigma,s}=\{\tau^0,\eta^{\pdagger}_{\sigma,1}\dot\iota\tau^z\}\,,\label{eq:z2_fixing_cs_3}\\
& \eta^{\pdagger}_{C_4}\eta^{\pdagger}_{C_4T_2}=-1,\,\eta^{\pdagger}_{\sigma}=\eta^{\pdagger}_{\sigma T_1}=-1:\notag\\
&\bar{W}^{\pdagger}_{C_4,s}=\{\tau^0,\eta^{\pdagger}_{C_4,1}\dot\iota\tau^z\},\,\bar{W}^{\pdagger}_{\sigma,s}=\{\dot\iota\tau^z,\eta^{\pdagger}_{\sigma,1}\tau^0\}\,,\label{eq:z2_fixing_cs_4}\\
& \eta^{\pdagger}_{C_4}\eta^{\pdagger}_{C_4T_2}=-1,\,\eta^{\pdagger}_{\sigma}=-\eta^{\pdagger}_{\sigma T_1}=-1:\notag\\
&\bar{W}^{\pdagger}_{C_4,s}=\{\tau^0,\eta^{\pdagger}_{C_4,1}\dot\iota\tau^x\},\,\bar{W}^{\pdagger}_{\sigma,s}=\{\dot\iota\tau^z,\eta^{\pdagger}_{\sigma,1}\dot\iota\tau^y\}\,.\label{eq:z2_fixing_cs_5}
\end{align}
Now we proceed to find the projective solutions for time-reversal symmetry. Eqs.~\eqref{eq:id_gauge_relation}(j) for $\mathcal{O}=T_1,T_3$ can be recast as:
\begin{align}
&W^{\pdagger}_{\mathcal{T}}(X,Y,s)=\eta^{\pdagger}_{\mathcal{T}T_1}W^{\pdagger}_{\mathcal{T}}(X-1,Y,s),\\
&W^{\pdagger}_{\mathcal{T}}(X,Y,s)=\eta^{\pdagger}_{\mathcal{T} T_2}W^{\pdagger}_{\mathcal{T}}(X,Y-1,s),
\end{align}
yielding the following solution for $W^{\pdagger}_{\sigma}$:
\begin{equation}
    \label{eq:time_sol_z2}
W^{\pdagger}_{\mathcal{T}}(X,Y,s)=\eta^{X}_{\mathcal{T} T_1}\eta^{Y}_{\mathcal{T} T_2}\bar{W}^{\pdagger}_{\mathcal{T},s}\,.    
\end{equation}
Furthermore, Eqs.~\eqref{eq:id_gauge_relation}(j) for $\mathcal{O}=C_4,\sigma$ and Eq.~\eqref{eq:id_gauge_relation}(i) impose the following conditions:
\begin{align}
 \eta^{\pdagger}_{\mathcal{T} T_1}&=\eta^{\pdagger}_{\mathcal{T} T_2}\\
\eta^{\pdagger}_{\mathcal{T} C_4}&=\eta^{\delta^{\pdagger}_{s,0}}_{\mathcal{T} T_1}\bar{W}^{\pdagger}_{\mathcal{T},s}\bar{W}^{\pdagger}_{C_4,s}\bar{W}^{-1}_{\mathcal{T},\bar{s}}\bar{W}^{-1}_{C_4,s}\\
\eta^{\pdagger}_{\mathcal{T} \sigma}&=\eta^{\delta^{\pdagger}_{s,1}}_{\mathcal{T} T_1}\bar{W}^{\pdagger}_{\mathcal{T},s}\bar{W}^{\pdagger}_{\sigma,s}\bar{W}^{-1}_{\mathcal{T},{s}}\bar{W}^{-1}_{\sigma,s}\\
\eta^{\pdagger}_{\mathcal{T}}&=\bar{W}^{2}_{\mathcal{T},s}\,.
\end{align}
The next step is to find the possible choices of $\bar{W}^{\pdagger}_{\mathcal{T},s}$ compatible with Eqs.~\eqref{eq:z2_fixing_cs_1}-\eqref{eq:z2_fixing_cs_5}, depending on the signs of $\eta^{\pdagger}_{\mathcal{T} T_1}$ and using the above conditions. The results are summarized in Appendix~\ref{app:projective_symmetry}.

\begin{figure*}[tb]	\includegraphics[width=1.0\linewidth]{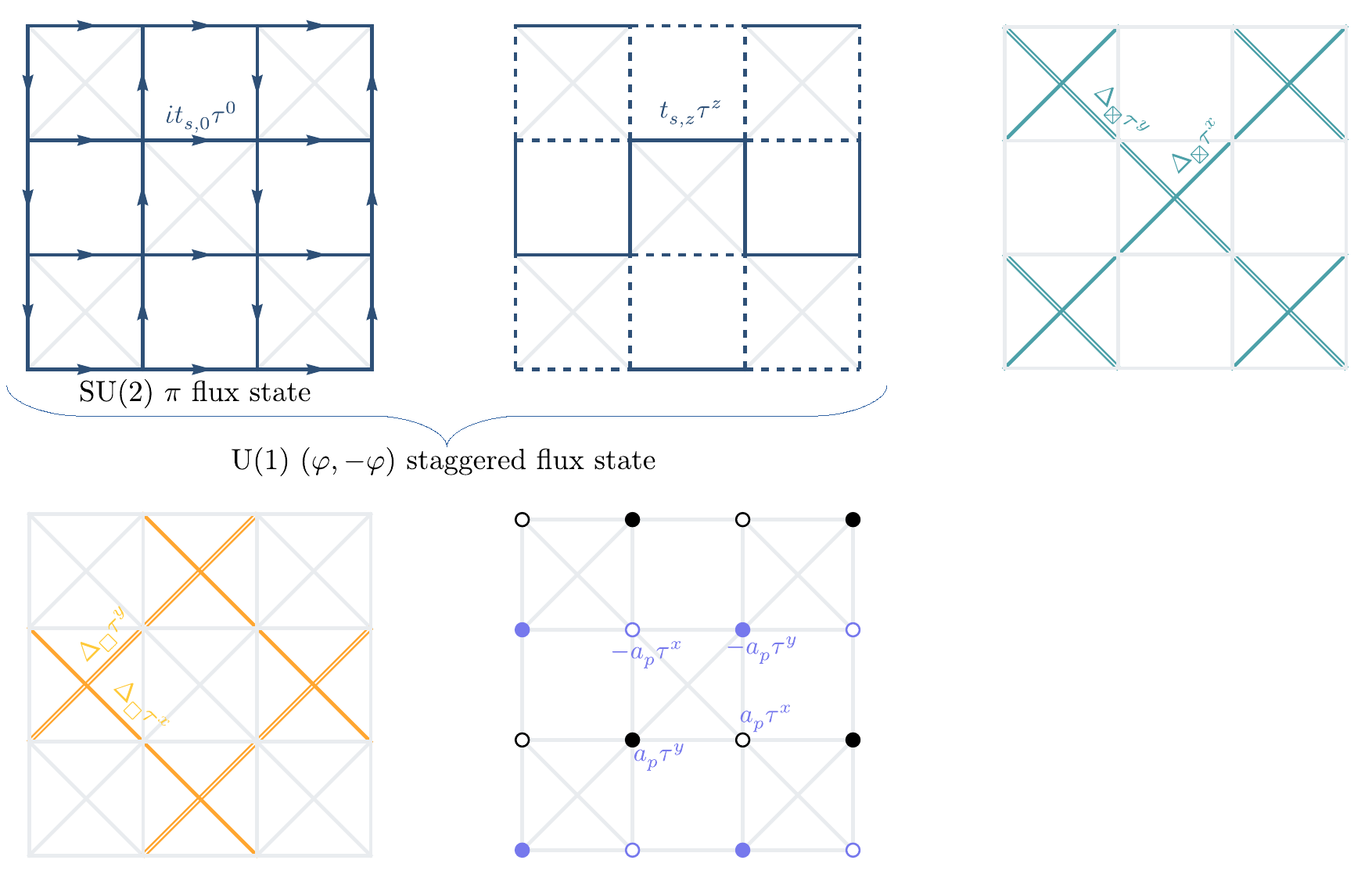}
	\caption{Sign structures of the hopping amplitudes (also for the U(1) {\it Ansatz} labeled by U19) and the pairing amplitudes of the $Z_2$ {\it Ansatz} labeled by Z05. For the staggered-flux structure $(\varphi,-\varphi)$, $t^{}_{s,0}=t\cos\left(\frac{\varphi-\pi}{4}\right)$ and $t^{}_{s,3}=t\sin\left(\frac{\varphi-\pi}{4}\right)$. All dashed links carry an additional negative sign.}
	\label{fig:u800_z3000_ansatz}
\end{figure*} 

\section{Field theory of the Z05 Dirac spin liquid}\label{app:z2_ansatz_qft}
In order to compare our field-theoretic study with previous results on the square~\cite{Shackleton2021,Shackleton:2022zzm} and Shastry--Sutherland~\cite{Feuerpfeil-2026} lattices, we transform the {\it Ansatz} of the Z05 spin liquid into a gauge consistent with~\cite{Feuerpfeil-2026} by a gauge transformation given by $W(\vecr,m_x,m_y)=(-1)^{\vecr}W(m_x,m_y)$, where $W(A,A)=e^{-\dot\iota\frac{\pi}{4}\tau^x}$, $W(B,A)=e^{-\dot\iota\frac{2\pi}{3\sqrt{3}}(\tau^x+\tau^y+\tau^z)}$, $W(B,B)=e^{\dot\iota\frac{pi}{2\sqrt{2}}(\tau^y-\tau^z)}$ and $W(A,B)=e^{\dot\iota\frac{\pi}{3\sqrt{3}}(\tau^x-\tau^y-\tau^z)}$.

We use a $2\times 2$ unit cell labeled by two sublattice degrees of freedom $m_x,m_y=A,B$ and denote $\rho^a,\kappa^a$ as the Pauli matrices acting on $m_x$ and $m_y$; see \figref{fig:lattice}(b).
In this new gauge, the {\it Ansatz} is given by
\begin{widetext}
\begin{align}
&u^{}_{\mbf{r},A,A,\mbf{r},B,A}=u^{}_{\mbf{r},A,B,\mbf{r}+\mbf{\hat{y}},A,A}=u^{}_{\mbf{r}+\mbf{\hat{y}},B,A,\mbf{r},B,B}=u^{}_{\mbf{r},B,B,\mbf{r}+\mbf{\hat{x}},A,B}=
\begin{bmatrix}
-t e^{-\dot\iota\theta} &0\\
0 & t e^{\dot\iota\theta}
\end{bmatrix}=u^{}_s\,,\label{eq:U1_1}\\
&u^{}_{\mbf{r},A,A,\mbf{r},A,B}=u^{}_{\mbf{r},A,B,\mbf{r},B,B}=u^{}_{\mbf{r},B,B,\mbf{r},B,A}=u^{}_{\mbf{r},B,A,\mbf{r}+\mbf{\hat{x}},A,A}=
\begin{bmatrix}
t e^{\dot\iota\theta} &0\\
0 & -t e^{-\dot\iota\theta}
\end{bmatrix}=-u^{\dagger}_s\,,\label{eq:U1_2}\\
&u^{}_{\mbf{r},A,A,\mbf{r},B,B}=u^{}_{\mbf{r},B,B,\mbf{r}+\mbf{\hat{x}}+\mbf{\hat{y}},A,A}=\Delta^{}_{\boxtimes}\sigma^x,\;
u^{}_{\mbf{r}+\mbf{\hat{x}},A,B,\mbf{r}+\mbf{\hat{y}},B,A}=u^{}_{\mbf{r},B,A,\mbf{r},A,B}=\Delta^{}_{\boxtimes} \sigma^y \label{eq:Z2_1}\,,\\
&u^{}_{\mbf{r},B,A,\mbf{r}+\mbf{\hat{x}},A,B}=u^{}_{\mbf{r},A,B,\mbf{r}+\mbf{\hat{y}},B,A}=\Delta_{\square}\sigma ^y\,,\,u^{}_{\mbf{r}+\mbf{\hat{x}},A,A,\mbf{r},B,B}=u^{}_{\mbf{r},B,B,\mbf{r}+\mbf{\hat{y}},A,A}=\Delta_{\square}\sigma^x\,. \label{eq:Z2_end}
\end{align}    
\end{widetext}
Here, $\mathbf{r}=(\vecr,m_x,m_y)=r^{}_{2x}T_{2x}+r^{}_{2y}T_{2y}+r^{}_{m_x,m_y}$, with $T_{2x}=2a\mbf{\hat{x}}$, $T_{2y}=2a\mbf{\hat{y}}$, and $r^{}_{m_x,m_y}$ denoting the positions of the sites within the unit cell. 
All the hopping and pairing mean-field parameters are illustrated schematically in Fig.~\ref{fig:u800_z3000_ansatz}, where the real and imaginary parts of the hopping parameters are given by $t^{}_{s,0}=t \sin\theta$ and $t^{}_{s,z}=t \cos\theta$, respectively.

\subsection{PSGs for the continuum field theory}\label{app:PSGs}
In the following, we present the transformation properties of the checkerboard wallpaper-group elements and the associated projective actions for the {\it Ansatz} under the gauge chosen in \secref{sec:continuum_theory} and \appref{app:z2_ansatz_qft}, in the new coordinate scheme where a four-site unit cell has been defined.
\begin{align}
 T_{1}&:(\vecr,m_x,m_y)\rightarrow(\vecr^{}_{2x}+\delta^{}_{m_x,B},\vecr^{}_{2y}-\delta^{}_{m_y,A},\bar{m}_x,\bar{m}_y),\notag\\
T_{2}&:(\vecr,m_x,m_y)\rightarrow(\vecr^{}_{2x}+\delta^{}_{m_x,B},\vecr^{}_{2y}+\delta^{}_{m_y,B},\bar{m}_x,\bar{m}_y),\notag\\
C_{4}&:(\vecr,m_x,m_y)\rightarrow(-\vecr^{}_{2y},\vecr^{}_{2x},m^{C_4}_x,m^{C_4}_y),\notag\\
\sigma&:(\vecr,m_x,m_y)\rightarrow(\vecr^{}_{2y},\vecr^{}_{2x},m^{\sigma}_x,m^{\sigma}_y).\label{eq:symmetry_checkerboard}
\end{align}
Here, $\bar{m}=B(A)$ for $m=A(B)$.  $(m^{C_4}_x,m^{C_4}_y)=$ $(B,A)$, $(B,B)$, $(A,B)$ and $(A,A)$ for $(m^{}_x,m^{}_y)=$ $(A,A)$, $(B,A)$, $(B,B)$ and $(A,B)$, respectively, while $(m^{\sigma}_x,m^{\sigma}_y)=$ $(A,A)$, $(A,B)$, $(B,B)$ and $(B,A)$ for $(m^{}_x,m^{}_y)=$ $(A,A)$, $(B,A)$, $(B,B)$ and $(A,B)$, respectively.
\subsubsection{\texorpdfstring{$\SU(2)$}{SU(2)} \texorpdfstring{$\pi$}{pi}-flux }
The PSG is given by:
\begin{align}
 & W_{T_{1}}(\vecr,m_x,m_y)=(-1)^{\delta_{m_y,B}}g^{}_{T_1},\notag\\
 & W_{T_{2}}(\vecr,m_x,m_y)=(-1)^{\delta_{m_y,B}}g^{}_{T_2},\notag\\
& W_{C_4}(\vecr,m_x,m_y)=(-1)^{\delta_{(m_x,m_y),(B,A)}}g^{}_{C_4},\notag\\
& W_{\sigma^{}_{}}(\vecr,m_x,m_y)=(-1)^{\delta_{(m_x,m_y),(B,B)}}g^{}_{\sigma^{}_{}}\,,\notag\\
& W_{\mathcal{T}}(\vecr,m_x,m_y)=-(-1)^{\delta_{m_x,m_y}}g^{}_{\mathcal{T}}\,,\label{eq:psg_su2_pi}
\end{align}
where $g^{}_{T_{1}}$, $g^{}_{T_{2}}$, $g^{}_{C_4}$, $g^{}_{\sigma}$ and $g^{}_{\mathcal{T}}$ are global SU(2) matrices. In this gauge, the structure of the {\it Ansatz} is given by:
\begin{equation} \label{eq:SU2_pi-flux}
u_{\vi,\vi+\hatx}=\dot\iota t^{}_{s,0},\;  u_{\vi,\vi+\haty}=(-1)^{i_x}\dot\iota t^{}_{s,0} .
\end{equation}
This {\it Ansatz} also exists with square-lattice symmetry, and the associated PSG is given by:
\begin{align}
& W_{T_{x}}(i_x,i_y)=(-1)^{i_y}g^{}_{T_x},\;W_{T_{y}}(i_x,i_y)=g^{}_{T_y},\notag\\
& W_{\sigma_y}(i_x,i_y)=(-1)^{i_x}g^{}_{\sigma_y},\;W_{\sigma_x}(i_x,i_y)=(-1)^{i_y}g^{}_{\sigma_x},\notag\\
& W_{R_{\pi/2}}(i_x,i_y)=(-1)^{i_x+i_xi_y}g^{}_{C_4},\notag\\
& W_{\mathcal{T}}(i_x,i_y)=(-1)^{i_x+i_y}g^{}_{\mathcal{T}}.\label{eq:psg_su2_pi_squre}
\end{align}
Here $R_{\pi/2}$ generates a $\pi/2$ rotation around $(0,0)$.
\subsubsection{U(1) \texorpdfstring{$(\varphi,-\varphi)$}{}-flux }
The corresponding U(1) PSG can be obtained from \eqnref{eq:psg_su2_pi} by choosing $g^{}_{T_{1}}=e^{i\phi^{}_{T_{1}}\tau^z}$, $g^{}_{T_{2}}=e^{i\phi^{}_{T_{2}}\tau^z}$, $g^{}_{C_4}=e^{i\phi^{}_{C_4}\tau^z}$, $g^{}_{\sigma}=e^{i\phi^{}_{\sigma}\tau^z}i\tau^x$ and $g^{}_{\mathcal{T}}=e^{i\phi^{}_{\mathcal{T}}\tau^z}$ and is given by:
\begin{align}
 & W_{T_{1}}(\vecr,m_x,m_y)=(-1)^{\delta_{m_y,B}}e^{i\phi^{}_{T_1}\tau^z},\notag\\
 & W_{T_{2}}(\vecr,m_x,m_y)=(-1)^{\delta_{m_y,B}}e^{i\phi^{}_{T_2}\tau^z},\notag\\
& W_{C_4}(\vecr,m_x,m_y)=(-1)^{\delta_{(m_x,m_y),(B,A)}}e^{i\phi^{}_{C_4}\tau^z},\notag\\
& W_{\sigma}(\vecr,m_x,m_y)=(-1)^{\delta_{(m_x,m_y),(B,B)}}e^{i\phi^{}_{\sigma}\tau^z}i\tau^x,\notag\\
& W_{\mathcal{T}}(\vecr,m_x,m_y)=-(-1)^{\delta_{m_x,m_y}}e^{i\phi^{}_{\mathcal{T}}\tau^z}.\label{eq:psg_u1}
\end{align}
where $\phi\in[0,2\pi]$. In this gauge, the structure of the {\it Ansatz} is given by:
\begin{align}\label{eq:ansatz_g_2}
    u_{\vi,\vi+\hatx}&=\dot\iota{t}^{}_{s,0}\tau^0-(-1)^{i_x+i_y}{\chi}^{}_{s,z}\tau^z\notag\\ u_{\vi,\vi+\haty}&=(-1)^{i_x}\dot\iota{t}^{}_{s,0}\tau^0+(-1)^{i_y}{\chi}^{}_{s,z}\tau^z. 
\end{align}
In square-lattice symmetry, the PSG is given by:
\begin{align}
& W_{T_{x}}(i_x,i_y)=(-1)^{i_y}e^{i\phi^{}_{T_x}\tau^z}i\tau^x,\;W_{T_{y}}(i_x,i_y)=e^{i\phi^{}_{T_y}\tau^z}i\tau^x,\notag\\
& W_{\sigma_y}(i_x,i_y)=(-1)^{i_x}e^{i\phi^{}_{\sigma_y}\tau^z},\;W_{\sigma_x}(i_x,i_y)=(-1)^{i_y}e^{i\phi^{}_{\sigma_x}\tau^z},\notag\\
& W_{R_{\pi/2}}(i_x,i_y)=(-1)^{i_x+i_xi_y}e^{i\phi^{}_{C_4}\tau^z}i\tau^x,\notag\\
& W_{\mathcal{T}}(i_x,i_y)=(-1)^{i_x+i_y}e^{i\phi^{}_{\mathcal{T}}\tau^z}.\label{eq:psg_u1_squre}
\end{align}

\subsubsection{Z05}
The corresponding $\mathbb{Z}_2$ PSG can be obtained from \eqnref{eq:psg_u1} by choosing $\phi^{}_{T_{1}}=\phi^{}_{T_{2}}=\pi/2$, $\phi^{}_{C_{4}}=\pi/4$, $\phi^{}_{\sigma}=-\pi/2$, and $\phi^{}_{\mathcal{T}}=\pi/2$ and is given by:
\begin{align}
 & W_{T_{1}}(\vecr,m_x,m_y)=(-1)^{\delta_{m_y,B}}i\tau^z,\notag\\
& W_{T_{2}}(\vecr,m_x,m_y))=(-1)^{\delta_{m_y,B}}i\tau^z,\notag\\   
& W_{C_4}(\vecr,m_x,m_y)=(-1)^{\delta_{(m_x,m_y),(B,A)}}\frac{1}{\sqrt{2}}(\tau^0+\dot\iota\tau^z),\notag\\
 & W_{\sigma}(\vecr,m_x,m_y)=(-1)^{\delta_{(m_x,m_y),(BB)}}i\tau^y\,,\notag\\
 & W_{\mathcal{T}}(\vecr,m_x,m_y)=-(-1)^{\delta_{m_x,m_y}}i\tau^z\,.\label{eq:psg_z2_oroinal}
\end{align}

In this gauge, the structure of the {\it Ansatz} is given in Eqs.~\eqref{eq:U1_1}-\eqref{eq:Z2_end}. 

Figure~\ref{fig:u800_z3000_ansatz} illustrates the sign structures of the mean-field hopping and pairing amplitudes of the checkerboard $Z_2$ {\it Ansatz} labeled by $\mathrm{Z05}$ (which is also connected to Z2A$zz$13 on a square grid and to the $\mathrm{Z3000}$ Shastry--Sutherland lattice). Note that, in the illustrated gauge form, the connection to the parent U(1) $(\varphi,-\varphi)$-flux state and the parent SU(2) $\pi$-flux state is explicit. Simply removing the pairing terms leads to the parent state.

\subsection{Lattice realization of Higgs transitions}\label{app:sym_couplings}

\begin{table*}[t]
    \centering
    \begin{tblr}{
          width = \linewidth,
          colspec = {c X[c] X[c] X[c] X[c] X[c] | X[c] X[c] X[c] X[c]},
          cell{2}{7,8,9,10,11} = {bg=colone},
          cell{3}{7,8,9,10,11}  = {bg=coltwo},
          cell{4}{7,8,9,10,11}  = {bg=colthree},
        }\hline\hline
         & $T_x$ & $T_y$ & $\sigma_y$ & $\sigma_x$ & $R_{\pi/2}$ & $T_{1}/T_2$ & $\mathcal{T}$ & $C_4$ &  $ \sigma_{}$\\\hline
         $\Phi_1^a$ & $-$ & $+$ & $-$ & $-$ & $-\Phi_2^a$ & $ -$ & $-$ & $ -\Phi_2^a$ & $ \Phi_2^a$\\
         $\Phi_2^a$ & $+$ & $-$ & $-$ & $-$ & $-\Phi_1^a$ & $ -$ & $-$& $ \Phi_1^a$ & $ \Phi_1^a$\\
         $\Phi_3^a$ & $-$ & $-$ & $+$ & $+$ & $-$ & $ +$  & $+$ & $+ $ & $- $\\\hline
    \end{tblr}
    \caption{Symmetry transformations of the Higgs fields under the square lattice and checkerboard symmetries.}
    \label{tab:higgs_transformations}
\end{table*}

We obtain the continuum version of the perturbations to the U(1) $\mathbf{(\varphi,-\varphi)}$ flux~\footnote{Such state has been denoted by U800 in Ref.~\cite{Maity-2024,Feuerpfeil-2026} and U19 in Fig.~\ref{fig:u1_ansatz}.} spin liquid by expanding around the mean field $u_{\vi\vj}$ in powers of $\theta=\pi/2+\delta \theta$ corresponding to additional real-valued hopping parameters
\begin{equation}
    t^z_{\vi,\vi+\hatx}=-\delta\theta(-1)^{i_x+i_y}\,, \quad t^z_{\vi,\vi+\haty}=\delta\theta(-1)^{i_y}\,.
\end{equation}
These corrections are identical to~\cite{Feuerpfeil-2026} leading to the perturbations
\begin{equation}\label{eq:U1_perturbation}
    \delta \mathcal{L} = -2\delta\theta\Tr[\tau^z\bar{\X}(\mu^y\gamma^yi\partial_x+\mu^y\gamma^xi\partial_y) \X]\,.
\end{equation}
In \eqnref{eq:U1_perturbation}, the Pauli matrix $\tau^z$ is acted on by the $\SU(2)$ gauge symmetry of the $\pi$-flux phase. Gauge invariance hence requires the existence of similar terms with $\tau^x$ and $\tau^y$. Therefore, we express the perturbation in a gauge-independent fashion by introducing an adjoint Higgs field $\Phi_3^a$ with $\SU(2)$ gauge index $a$:
\begin{equation}
    \delta \mathcal{L}=\Phi_3^a\Tr[\tau^a \bar{\X}\mu^y(\gamma^y i\partial_x+\gamma^x i\partial_y) \X]\,.
\end{equation}
The subscript ``3'' is chosen in accordance with~\cite{Shackleton2021,Feuerpfeil-2026}. The continuum version of \eqnref{eq:U1_perturbation} is obtained by condensing $\langle \Phi_3^z \rangle \propto \delta\theta$.

Analogously to~\cite{Feuerpfeil-2026}, by identifying $\Delta_d=\Delta_{d'}=\Delta_{\boxtimes}$ as well as $\Delta_{g,1}=\Delta_{\square}$ and $\Delta_{g,2}=0$, we obtain the perturbation
\begin{equation}
\begin{split}\label{eq:Z2_perturbations}
    \delta \mathcal{L} =(2\Delta_{\boxtimes}+2\Delta_{\square})(&\Tr[\tau^x\bar{\X}\left(\gamma^x\mu^z-\gamma^y\mu^x\right)\X]\\
    +&\Tr[\tau^y\bar{\X}\left(\gamma^x\mu^z+\gamma^y\mu^x\right)\X])
\end{split}
\end{equation}
to the $\mathbb{Z}_2$ spin liquid, where we have ignored additional gradient terms.

Again, we want to express the Lagrangian in a gauge-invariant fashion by introducing two 
Higgs fields $\Phi_{1,2}^{a}$ with the couplings
\begin{equation}
    \Phi_1^a\Tr[\tau^a \bar{\X}\gamma^x\mu^z\X]+\Phi_2^a\Tr[\tau^a \bar{\X}\gamma^y\mu^x\X]\,.
\end{equation}
The continuum version of \eqnref{eq:Z2_perturbations} is then obtained by condensing the Higgs fields as
\begin{equation}
\begin{split}
    \langle \Phi_1\rangle &\propto (2\Delta_{\boxtimes}+2\Delta_{\square})(1,1,0)\,,\\
    \langle \Phi_2\rangle &\propto (2\Delta_{\boxtimes}+2\Delta_{\square})(-1,1,0)\,.\\
\end{split}
\end{equation}

Now, we collect the perturbations to obtain the Lagrangian for the Majorana field $\mathcal{X}$ and three real, adjoint Higgs fields $\Phi_{1,2,3}^a$. We will not explicate the $\SU(2)$ gauge fluctuations, as they can be added via minimal coupling:
\begin{equation}
\begin{split} \label{eq:full_lagrangian_app}
    \mathcal{L}=&i\Tr[\bar{\X}\gamma^\mu \partial_\mu \X]+\Phi_1^a\Tr[\tau^a\bar{\X}\gamma^x\mu^z \X]\\
    &+\Phi_2^a\Tr[\tau^a\bar{\X}\gamma^y\mu^x \X]\\
    &+\Phi_3^a\Tr[\tau^a \bar{\X}\mu^y(\gamma^yi\partial_x+\gamma^xi\partial_y)\X]+V(\Phi)\,.\\
\end{split}
\end{equation}
This continuum Lagrangian is identical to the square and Shastry--Sutherland lattice up to less relevant gradient terms~\cite{Shackleton2021,Feuerpfeil-2026}.

In the form of \eqnref{eq:full_lagrangian_app}, by ignoring the gradient terms, which have a larger scaling dimension and are thus less relevant, the critical theory is identical to the square lattice~\cite{Shackleton2021} and Shastry--Sutherland model~\cite{Feuerpfeil-2026}. As the nearest-neighbor mean-field {\it Ans\"atze} on the checkerboard lattice might lack terms that are symmetry-allowed in the continuum theory, we now analyze all allowed fermion bilinear terms that respect time reversal $\mathcal{T}$, which acts as $\mathcal{T}=i\tau^y K$ and the lattice symmetries as detailed in \appref{app:symmetry}. 

The projective action for the relevant spin liquids is defined in \appref{app:PSGs}. For simplicity, we set all global $\SU(2)$ matrices to the identity.

\subsubsection{Symmetry-allowed terms on the square lattice}
Then, the $\X$ fermions transform nontrivially under time reversal and the square lattice wallpaper group symmetries:
\begin{equation}
\begin{split} \label{eq:square_lattice_symmetries}
    T_x:\X&\rightarrow \mu^x\X\,,\\ 
    T_y: \X &\rightarrow \mu^z \X\,,\\
    \sigma_y: \X &\rightarrow  -\rho^z\mu^z \X(-x,y)\,,\\
    \sigma_x: \X &\rightarrow \rho^x\mu^x \X(x,-y)\,,\\ 
    \mathcal{T}: \X &\rightarrow \rho^y\mu^y \X\,, \quad  i\rightarrow -i\,,\\
    R_{\pi/2}: \X&\rightarrow e^{-i\pi \rho^y/4}e^{i\pi \mu^y/4}\X(-y,x)\,.
\end{split}
\end{equation}
For the Higgs fields $\Phi_{1,2,3}^a$, we deduce the transformation properties in \tabref{tab:higgs_transformations}.

By comparing with the symmetry transformations of all fermion bilinears in~\cite{Thomson_2018}, we see that there are no other symmetry-allowed terms, one could add to these Higgs fields in the Lagrangian \eqnref{eq:full_lagrangian} on the square lattice.

\begin{table}
    \centering
    \begin{tblr}{
              width = 0.98\linewidth,
              colspec = {c X[c] X[c] X[c] X[c]},
            }\hline\hline
            $T^j$ & $T_1/T_2$  & $\mathcal{T}$ & $C_4$ &  $ \sigma$\\\hline
            $\mu^y$ & $ +$ & $ -$ & $ -$ & $ +$ \\
            $\tau^a$ & $ +$  & $ -$ & $+$ & $- $\\
            $\mu^x\tau^a $ & $- $  & $ +$ & $-\mu^z\tau^a$ & $-\mu^z\tau^a$\\
            $\mu^z \tau^a$ & $ -$ & $ +$ & $-\mu^x\tau^a$ & $-\mu^x\tau^a$ \\\hline
        \end{tblr}
    \caption{Transformation properties of $\Tr[\tau^a \bar{\mathcal{X}}T^j \X]$ under the checkerboard symmetries. Here $T^j=\lbrace \mu^y,\tau^a, \mu^x\tau^a,\mu^z\tau^a\rbrace$ are the 10 generators of the emergent $\SO(5)$ symmetry~\cite{Tanaka2005,Ran2006,Wang_2017,Feuerpfeil-2026}.}
    \label{tab:checkerboard_bilinears_1}
\end{table}

\begin{table}
    \centering
    \begin{tblr}{
          width = 0.98\linewidth,
          colspec = {c X[c] X[c] X[c] X[c]},
          row{3} = {bg=colone},
          row{4} = {bg=coltwo},
          row{6} = {bg=colone},
          row{7} = {bg=coltwo},
        }\hline\hline
        $\Gamma^j \gamma^\mu$ & $T_1/T_2$ & $\mathcal{T}$ & $C_4$ &  $ \sigma$\\\hline
        $\mu^x\gamma^0$ & $ -$ & $+$ & $ -\mu^z\gamma^0$ & $ \mu^z\gamma^0$ \\
        $\mu^x\gamma^x$ & $ -$ & $-$ &$ -\mu^z \gamma^y$ &$ \mu^z \gamma^y$ \\
        $\mu^x\gamma^y$ & $ -$ & $-$ &$ \mu^z\gamma^x$ &$ \mu^z\gamma^x$ \\\hline
        $\mu^z\gamma^0$ & $ -$ & $+$ &$ -\mu^x\gamma^0$ &$ \mu^x\gamma^0$ \\
        $\mu^z\gamma^x$ & $ -$ & $-$ &$ -\mu^x\gamma^y$&$ \mu^x\gamma^y$  \\
        $\mu^z\gamma^y$ & $ -$ & $-$ &$ \mu^x\gamma^x$ &$ \mu^x\gamma^x$  \\\hline
        $\mu^y \tau^a\gamma^0$ & $ +$ & $+$ & $ -$& $ -$ \\
        $\mu^y \tau^a\gamma^x$ & $ +$ & $-$ &$ -\mu^y\tau^a\gamma^y$&$ -\mu^y\tau^a\gamma^y$  \\
        $\mu^y \tau^a\gamma^y$ & $ +$ & $-$ &$ \mu^y\tau^a\gamma^x$&$ -\mu^y\tau^a\gamma^x$  \\\hline
    \end{tblr}
    \caption{Transformation properties of $\Tr[\tau^a \bar{\mathcal{X}}\Gamma^j\gamma^\mu \X]$ under the checkerboard symmetries. Here $\Gamma^j=\lbrace \mu^x,\mu^z, \mu^y\tau^a\rbrace$ transforms as a vector under $\SO(5)$~\cite{Feuerpfeil-2026}.}
    \label{tab:checkerboard_bilinears_1_2}
\end{table}

\begin{table*}
    \centering
    \begin{minipage}[t]{0.49\linewidth}
        \centering
        \begin{tblr}{
              width = \linewidth,
              colspec = {c X[c] X[c] X[c] X[c]},
              row{7} = {bg=colthree},
              row{9} = {bg=colthree},
            }\hline\hline
            $\mu^y\gamma^\mu i\partial_\nu$ & $T_{1}/T_2$ & $\mathcal{T}$ & $C_4$ &  $ \sigma$\\\hline
            $\mu^y\gamma^0 i\partial_0$ & $+ $& $+ $ & $-$ & $-$ \\
            $\mu^y\gamma^0 i\partial_x$ & $+ $& $- $ & $-\mu^y\gamma^0i\partial_y $ & $-\mu^y\gamma^0i\partial_y $  \\
            $\mu^y\gamma^0 i\partial_y$ & $+ $& $- $ & $\mu^y\gamma^0i\partial_x $& $-\mu^y\gamma^0i\partial_x $  \\\hline
            $\mu^y\gamma^x i\partial_0$ & $+ $& $- $ & $-\mu^y\gamma^yi\partial_0 $ & $-\mu^y\gamma^yi\partial_0 $ \\
            $\mu^y\gamma^x i\partial_x$ & $+ $& $+ $ & $-\mu^y\gamma^yi\partial_y $ & $-\mu^y\gamma^yi\partial_y $  \\
            $\mu^y\gamma^x i\partial_y$ & $+ $& $+ $ & $\mu^y\gamma^yi\partial_x $  & $-\mu^y\gamma^yi\partial_x $\\\hline
            $\mu^y\gamma^y i\partial_0$ & $+ $& $- $ & $\mu^y\gamma^xi\partial_0 $& $-\mu^y\gamma^xi\partial_0 $  \\
            $\mu^y\gamma^y i\partial_x$ & $+ $& $+ $ & $\mu^y\gamma^xi\partial_y $  & $-\mu^y\gamma^xi\partial_y $ \\
            $\mu^y\gamma^y i\partial_y$ & $+ $& $+ $ & $-\mu^y\gamma^xi\partial_x $ & $-\mu^y\gamma^xi\partial_x $  \\\hline
        \end{tblr}
    \end{minipage}
    \hfill
    \begin{minipage}[t]{0.49\linewidth}
        \centering
        \begin{tblr}{
              width = \linewidth,
              colspec = {c X[c] X[c] X[c] X[c]},
              row{2} = {bg=colthree},
            }\hline\hline
            $\mu^{0,x,y}i\partial_\mu$ & $T_{1}/T_{2}$ & $\mathcal{T}$ & $C_4$ &  $ \sigma_{}$\\ \hline 
            $i\partial_0$ & $ +$& $ +$ & $+ $& $- $  \\
            $i\partial_x$ & $ +$& $ -$ & $i\partial_y $ & $-i\partial_y $  \\
            $i\partial_y$ & $ +$& $ -$ & $-i\partial_x $& $-i\partial_x $  \\\hline
            $\mu^xi\partial_0$ & $ -$ & $- $ & $-\mu^zi\partial_0$& $-\mu^zi\partial_0$ \\
            $\mu^xi\partial_x$ & $ -$ & $ +$ & $-\mu^zi\partial_y$& $-\mu^zi\partial_y$ \\
            $\mu^xi\partial_y$ & $ -$ & $ +$ & $\mu^zi\partial_x$& $-\mu^zi\partial_x$ \\\hline
            $\mu^zi\partial_0$ & $ -$ & $- $ & $-\mu^xi\partial_0$& $-\mu^xi\partial_0$ \\
            $\mu^zi\partial_x$ & $ -$ & $ +$ & $-\mu^xi\partial_y$& $-\mu^xi\partial_y$ \\
            $\mu^zi\partial_y$ & $ -$ & $ +$ & $\mu^xi\partial_x$& $-\mu^xi\partial_x$ \\\hline
        \end{tblr}
    \end{minipage}
    \caption{Transformation properties of $\Tr[\tau^a \bar{\X}i\partial_\mu\X]$, $\Tr[\tau^a\bar{\X}\Gamma^ji\partial_\mu \X]$, and $\Tr[\tau^a \bar{\X}T^j\gamma^\mu i \partial_\nu \X]$ under the checkerboard symmetries. We consider only terms that do not transform under spin~\cite{Thomson_2018}.}
    \label{tab:checkerboard_bilinears_3}
\end{table*}

\subsubsection{Symmetry-allowed terms on the checkerboard lattice}\label{app:sym_checkerboard}

The checkerboard lattice breaks symmetries of the square lattice, see \appref{app:lattice_symmetry}. Thus, we can use the transformation properties of $\X$ in \eqnref{eq:square_lattice_symmetries} to derive the action of the checkerboard point group, which act as
\begin{equation}
\begin{split}
     T_{1}: \X &\rightarrow -i\mu^y\X\,,\\
     T_{2}: \X &\rightarrow -i\mu^y\X\,,\\
     C_{4}: \X &\rightarrow \mu^x e^{-i\pi \rho^y/4}e^{i\pi \mu^y/4}\X(-y,x)\,,\\
    \sigma: \X &\rightarrow -\rho^z\mu^ze^{-i\pi \rho^y/4}e^{i\pi \mu^y/4}\X(y,x)\,,\\
    \mathcal{T}: \X &\rightarrow \rho^y\mu^y \X\,, \quad  i\rightarrow -i\,.
\end{split}
\end{equation}
We tabulated all possible bilinears up to first-order gradient terms in \tabref{tab:checkerboard_bilinears_1} to \tabref{tab:checkerboard_bilinears_3}.

We color code all fermion bilinears, which do not transform under spin and are consistent with the transformation properties of the Higgs fields in \tabref{tab:higgs_transformations}. The additional symmetry-allowed terms are given by 
\begin{equation}
\begin{split}
    \mathcal{L}_{\mathrm{check.}} &\propto \Phi_3^a\Tr[\tau^a\bar{\X}(i\partial_0+\gamma^x\mu^yi\partial_y+\gamma^y\mu^yi\partial_x)\X]\\
    +\Phi_1^a&\Tr[\tau^a\bar{\X}\gamma^x\mu^x\X]+\Phi_2^a\Tr[\tau^a\bar{\X}\gamma^y\mu^z\X]\,,
\end{split}
\end{equation}
For $\Phi_3$, there are only additional gradient terms, whilst for $\Phi_{1/2}$, we have additional zeroth-order bilinears. All the additional symmetry-allowed terms are also allowed on the Shastry--Sutherland lattice~\cite{Feuerpfeil-2026}. As the gradient terms are less relevant, we can ignore them. The zeroth-order bilinears change the coupling between the Higgs fields and Dirac fermions but as was shown in~\cite{Feuerpfeil-2026}, this does not alter the scaling dimensions of the quantum field theory.

\clearpage

\bibliography{references}
\end{document}